\documentclass [10pt,twocolumn]{IEEEtran}
\usepackage{algorithm,algorithmic,amsbsy,amsmath,amssymb,epsfig,bbm,mathrsfs,multirow,amsthm}
\usepackage[T1]{fontenc}
\usepackage[latin9]{inputenc}
\usepackage{amsmath}
\usepackage{fixltx2e}
\usepackage{algorithm}
\usepackage{array,multirow,graphicx}
\usepackage{color}
\usepackage{cite}
\usepackage{float}
\usepackage{makecell}
\usepackage[x11names,dvipsnames,table]{xcolor} 
\usepackage{colortbl}
\usepackage{multirow}
\usepackage{subcaption} 
\usepackage{lipsum}
\definecolor{mygray}{gray}{0.6}
\definecolor{myblue}{rgb}{0.8,0.85,1} 
\usepackage{array}
\newcolumntype{L}[1]{>{\raggedright\let\newline\\\arraybackslash\hspace{0pt}}m{#1}}
\newcolumntype{C}[1]{>{\centering\let\newline\\\arraybackslash\hspace{0pt}}m{#1}}
\newcolumntype{R}[1]{>{\raggedleft\let\newline\\\arraybackslash\hspace{0pt}}m{#1}}

\usepackage{array, tabularx, boldline}
\usepackage{eurosym}
\usepackage{amstext} 
\DeclareRobustCommand{\officialeuro}{%
  \ifmmode\expandafter\text\fi
  {\fontencoding{U}\fontfamily{eurosym}\selectfont e}}

\usepackage{array, tabularx, boldline}
\usepackage{graphicx}
\usepackage{cellspace}
\begin{document}
\title{\huge Applications of Economic and Pricing Models for Resource Management in 5G Wireless Networks: A Survey}

\author{Nguyen Cong Luong, Ping Wang, \textit{Senior Member, IEEE}, Dusit Niyato, \textit{Fellow, IEEE}, Ying-Chang Liang, \textit{Fellow, IEEE}, Fen Hou, \textit{Member, IEEE}, and Zhu Han, \textit{Fellow, IEEE}
\thanks{N.~C.~Luong, P.~Wang, and D.~Niyato are with School of Computer Science and Engineering, Nanyang Technological University, Singapore. E-mails: clnguyen@ntu.edu.sg, wangping@ntu.edu.sg, dniyato@ntu.edu.sg.}
\thanks{Y.-C.~Liang is with School of Electrical and Information
Engineering, The University of Sydney, NSW 2006, Australia, and also with University of Electronic Science and Technology of China, Chengdu, China. E-mail: liangyc@ieee.org.} 
\thanks{F.~Hou is with the Department of Electrical and Computer Engineering,
University of Macau, Macau. E-mail: fenhou@umac.mo.} 
\thanks{Z.~Han is with Electrical and Computer Engineering, University of Houston, Houston, TX, USA. E-mail: zhan2@uh.edu.} 
}

\maketitle
\begin{abstract}
This paper presents a comprehensive literature review on applications of economic and pricing theory for resource management in the evolving fifth generation (5G) wireless networks. The 5G wireless networks are envisioned to overcome existing limitations of cellular networks in terms of data rate, capacity, latency, energy efficiency, spectrum efficiency, coverage, reliability, and cost per information transfer. To achieve the goals, the 5G systems will adopt emerging technologies such as massive Multiple-Input Multiple-Output (MIMO), mmWave communications, and dense Heterogeneous Networks (HetNets). However, 5G involves multiple entities and stakeholders that may have different objectives, e.g., high data rate, low latency, utility maximization, and revenue/profit maximization. This poses a number of challenges to resource management designs of 5G. While the traditional solutions may neither efficient nor applicable, economic and pricing models have been recently developed and adopted as useful tools to achieve the objectives. In this paper, we review economic and pricing approaches proposed to address resource management issues in the 5G wireless networks including user association, spectrum allocation, and interference and power management. Furthermore, we present applications of economic and pricing models for wireless caching and mobile data offloading. Finally, we highlight important challenges, open issues and future research directions of applying economic and pricing models to the 5G wireless networks. 

\end{abstract}

\section{Introduction}
\label{sec:intro}
The increasing proliferation of smart devices together with new emerging multimedia applications leads to the explosive growth of mobile data traffic. According to the latest Visual Network Index (VNI) report by Cisco \cite{VNI_Cisco}, the global mobile data traffic will reach 48.3 Exabytes per month in 2021, up from 7.2 Exabytes per month in 2016. The growth of mobile data demand has already created a significant burden on existing cellular networks which has triggered the investigation of the fifth generation (5G) cellular networks, abbreviated as 5G. Several commissions have launched projects towards 5G. For example,  the European commission has launched more than ten European Union (EU) projects to address the architecture and functionality needs of 5G. Horizon 2020 \cite{Horizon_2020}, the biggest EU Research and Innovation programme, provides funding for the 5G-Public Private Partnership (5G-PPP) to deliver solutions, architectures, technologies and standards for the ubiquitous 5G communications infrastructures. 

The primary technologies proposed for 5G are \cite{hossain20155g} massive Multiple-Input Multiple-Output (MIMO), dense Heterogeneous Networks (HetNets), mmWave communication, full-duplex communication, Device-to-Device (D2D) communication, energy-aware communication and energy harvesting, Cloud-Based Radio Access Networks (C-RANs), the virtualization of network resources, and so on. Compared to 4G cellular networks, 5G is expected to \cite{hossain20155g} (i) improve at least 1000 times of throughput, (ii) support higher network densification, (iii) reduce significantly latency, (iv) improve energy efficiency, and (vi) support a high density of mobile broadband users, D2D, ultra reliable, and massive Machine-Type-Communications (MTC).

However, the adoption of the emerging technologies introduces challenges for the radio resource management such as user association, spectrum allocation, interference and power management. The reasons are (i) the heterogeneity and dense deployment of wireless devices, (ii) the heterogeneous radio resources, (iii) the coverage and traffic load imbalance of Base Stations (BSs), (iv) the high frequency of handovers, (v) the constraints of the fronthaul and backhaul capacities, and (vi) a large number of users and stakeholders with different objectives. The traditional methods, e.g., the system optimization, can provide optimal resource allocation for the entire network. However, they usually require a centralized entity, and the signaling may be excessive for medium- to large-scale networks. Moreover, it is difficult to incorporate economic implication into the solutions. Therefore, the traditional methods may not be suitable for 5G, especially when the rationality of users and stakeholders are important.

\subsection{What and Why Economic and Pricing Approaches}
Since 5G involves multiple entities and stakeholders that may have different objectives, e.g., high data rate, low latency, utility maximization, cost minimization, and profit maximization, economic and pricing approaches have been recently developed and adopted as useful tools to reach the objectives. They allow to model and analyze complex interactions among the entities and stakeholders. Through the interactions, each entity can observe, learn, and predict status/actions of other entities, and then has the best decisions based on equilibrium analysis. In other words, the economic and pricing approaches are inherently suitable for distributed autonomous decision making. Therefore, they are applicable to 5G which consists of a large number of autonomous network entities. Specifically, compared with the traditional approaches, the economic and pricing approaches provide the following advantages:

 
\begin{itemize}
\item Economic and pricing models with their simplicity provide fast and dynamic user association schemes which can adapt to the fast variations of the wireless channels and frequent handovers in 5G. Also, through negotiation mechanisms, economic and pricing models enable the user association to achieve multiple objectives such as throughput maximization, fairness, and load balancing. 

\item The adoption of the emerging technologies enables 5G to provision heterogeneous resources to users. Economic and pricing models such as combinatorial auction allow users to explicitly request bundles of diverse resources to satisfy their dynamic demands and to improve the resource utilization. 

\item The dense deployment of cells increases the spectrum reuse among network operators in 5G. Economic and pricing models allow the network operators to quantify their own revenues or profits before deciding to share spectrum.

 \item The dense and unplanned deployment of wireless devices complicates the dynamics of interferences in 5G. Economic and pricing models such as Stackelberg game can provide distributed and dynamic solutions for the interference and power management with small information exchange and low computational complexity. 

\item The dense deployment of wireless devices along with the huge traffic demand of users increases the burden in terms of bandwidth on backhaul and fronthaul links. Economic and pricing models such as congestion-based pricing or tiered pricing regulate user demands which mitigates the congestion as well as maximizes the resource utilization and revenue for network operators.

\item The burden on the backhaul and fronthaul links can be reduced by using wireless caching. The wireless caching employs network devices close to the users, e.g., small cells, to cache popular contents from remote servers. Economic and pricing models such as contract theory incentivize the selfish owners of the small cells to contribute their resources.
\end{itemize}

Note that the economic and pricing approaches have a common assumption that the stakeholders in the system are rational and react to the strategy for the highest utility. In some cases, the assumption does not hold, e.g., due to limited available information and observation, which confines the applicability of the traditional economic and pricing approaches. Nevertheless, the economic tools dealing with bounded rationality such as evolutionary games can still apply.


\subsection{Contributions of the Paper}

Although there are several surveys related to 5G, they do not focus on economic and pricing approaches, which are emerging as a promising tool. For example, the survey of emerging technologies of 5G was given in \cite{gupta2015survey}, the survey of resource management toward 5G was presented in \cite{olwal2016survey}, and the survey of user association in 5G was given in \cite{liu2016user}. To the best of our knowledge, there is no survey specifically discussing the use of economic and pricing models to address the resource management in 5G. This motivates us to deliver the survey with the comprehensive literature review on the economic and pricing models in 5G.

\begin{figure*}[t!]
\centering
\includegraphics[width=17 cm, height=6 cm]{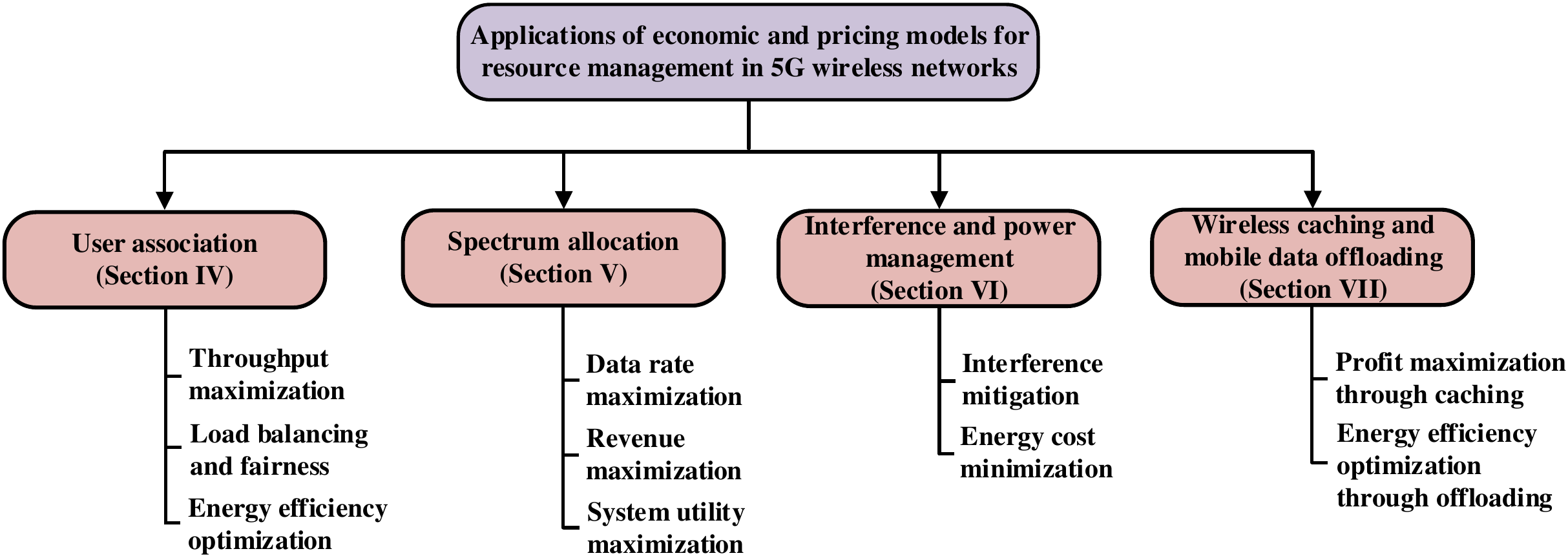}
 \caption{\small A taxonomy of the applications of economic and pricing models for resource management in 5G.}
 \label{Application_pricing_model}
\end{figure*}

For convenience, the related approaches in this survey are classified based on resource management issues and then major objectives as shown in Fig.~\ref{Application_pricing_model}. The resource management issues include user association, spectrum allocation, interference and power management, and wireless caching and mobile data offloading. Furthermore, advantages and disadvantages of each approach are highlighted.

The rest of this paper is organized as follows. Section~\ref{sec:Intro_5G} briefly describes key technologies of the 5G wireless networks. Section~\ref{sec:Intro_Price} presents the fundamentals of economic and pricing models. Section~\ref{sec:User_Association} discusses how to apply economic and pricing models for the user association. Section~\ref{sec:App_SA} reviews applications of economic and pricing models for the spectrum allocation. Applications of economic and pricing models for the interference and power management are given in Section~\ref{sec:App_PM}. Section~\ref{sec:App_Caching_Offloading} considers economic and pricing approaches for wireless caching and mobile data offloading. Important challenges, open issues, and future research directions are outlined in Section~\ref{sec:Open_issues}. Section~\ref{sec:Conclusion} concludes the paper. The list of abbreviations appeared in this paper is given in Table~\ref{tab:table_abb}. 

\begin{table}[b!]
\scriptsize
  \caption{\small Major abbreviations}
  \label{tab:table_abb}\centering
  \begin{tabularx}{8.7cm}{|Sl|X|}
    \hline
  \cellcolor{mygray} \textbf{Abbreviation} &   \cellcolor{mygray} \textbf{Description} \\   
         \hline
  ACA&Ascending Clock Auction \\ 
  \hline
ADMM&Alternating Direction Method of Multiplier\\
         \hline
 BBU/RRH& BaseBand processing Unit/Remote Radio Head\\
     \hline
CAPEX/OPEX &CAPital EXpenditure/OPerational EXpenditure\\
       \hline
FAP/PBS&Femto Access Point/Pico cell Base Station\\  
       \hline
FH/CP&Femto Holder/Content Provider\\
\hline
HetNets&Heterogeneous Networks\\
        \hline
KKT/NUM& Karush-Kuhn-Tucker/Network Utility Maximization \\ 
     \hline
MNO/MVNO&Mobile Network Operator/Mobile Virtual Network Operator\\
       \hline
MBS/SBS & Macro cell Base Station/Small cell Base Station\\
\hline
MUE/FUE/RUE/DUE&Users of MBS/FAPs/RRHs/D2D\\
\hline
MTC/M2M & Machine-Type-Communications/Machine-to-Machine\\
       \hline
IBGA&Increment-Based Greedy Allocation\\
        \hline
   PMP/VCG &Paris Metro Pricing/Vickrey-Clarke-Groves \\
\hline
PO/SO/SP& Primary Operator/Secondary Operator/Service Provider \\
      \hline
RAT/CoMP  & Radio Access Technology/Coordinated Multi-Point\\   
        \hline
        RSSI&Received Signal Strength Indication\\
                \hline
 SINR/RB   &Signal-to-Interference-plus-Noise Ratio/Resource Block \\
        \hline
WDP & Winner Determination Problem\\
 \hline
  \end{tabularx}
\end{table}

\section{Overview of 5G wireless networks}
\label{sec:Intro_5G}
\begin{figure*}[ht]
\centering
\includegraphics[width=\linewidth]{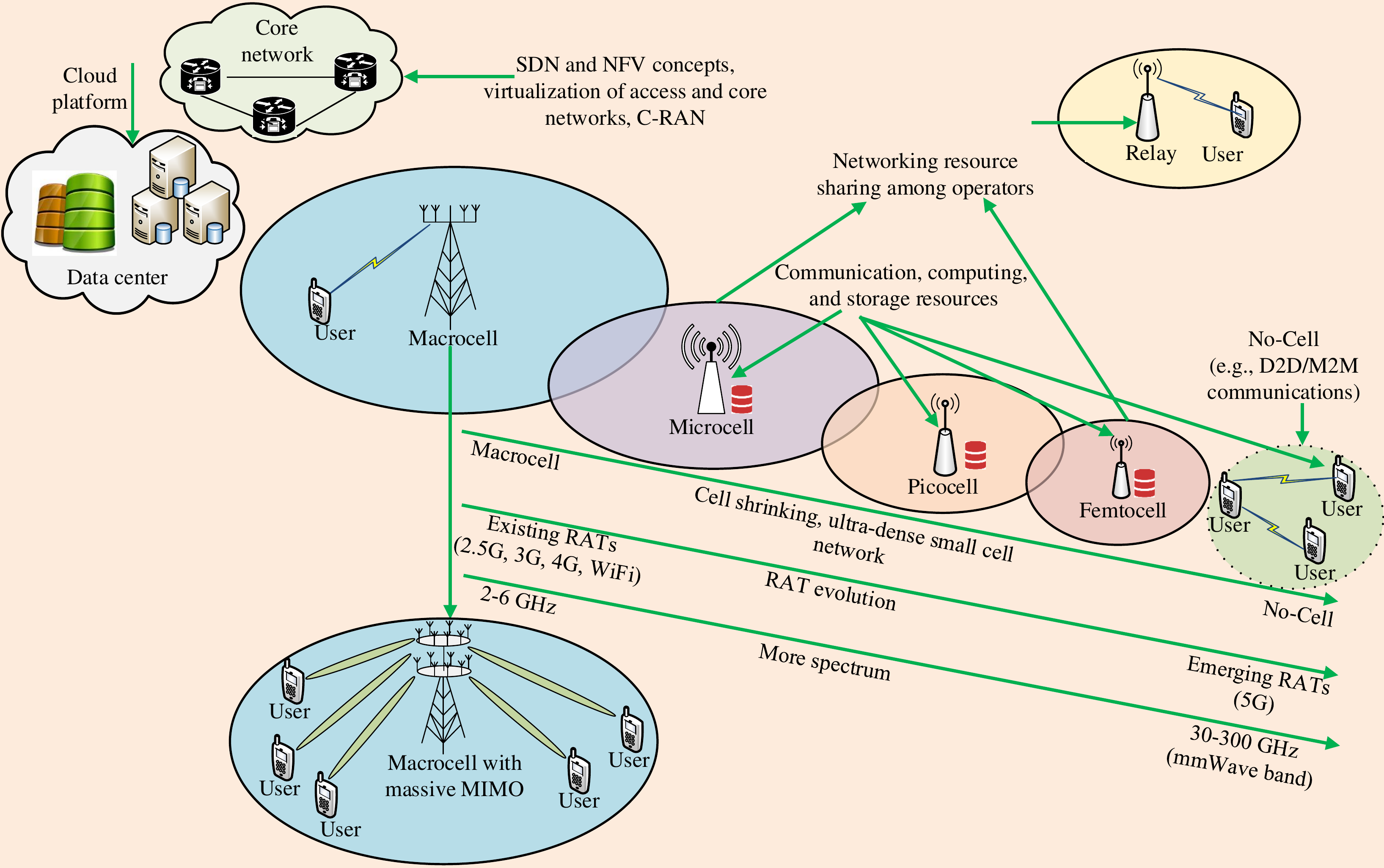}
 \caption{\small Technology trends in 5G era. Arrows indicate the potential technologies deployed in 5G.}
 \label{5G_architecture}
\end{figure*}

 Recent wireless research activities have already considered many technology trends towards 5G as illustrated in Fig.~\ref{5G_architecture}. In this paper, we focus on the technologies related to the function improvements of Base Stations (BSs). The reason is that BSs play an important role in guaranteeing Quality-of-Service (QoS) by providing connections to mobile users through the air interface. Besides, the spectrum used for the connections is the scarcest resource. Especially, more than 80\% of energy consumption of telecommunication networks is for the operation of BSs \cite{demestichas2015intelligent}. 
In following subsections, we discuss four key technologies which have the most significant impact on progressing towards 5G: (i) massive MIMO, (ii) HetNets, (iii) mmWave communications, and (iv) C-RAN. 

\subsection{Massive MIMO}
\label{sec:Intro_5G_Techno_massive_MIMO}
MIMO technology has been widely adopted in many wireless standards including LTE as it significantly improves the capacity and reliability of wireless transmission. However, the conventional MIMO with the limited number of antennas is not scalable \cite{larsson2014massive}. To achieve high multiplexing gains, massive MIMO, also known as large-scale antennas systems, has been proposed. In the massive MIMO, each BS is equipped with a very large number of antennas, e.g., a few hundreds, and the antenna array is typically designed in a two-dimensional grid with an antenna spacing of at least $\lambda_c/2$, where $\lambda_c$ is the wavelength at the intended carrier frequency $f_c$. 

The antenna array allows transmitted power to be concentrated in small regions in the space based on the principle of \textit{coherent superposition} of wavefronts, thus significantly improving the energy efficiency. Besides, extra degrees of freedom can be provided through using inexpensive low-power components such as RF amplifiers. The total power consumption is thus reduced. The massive MIMO system also uses spatial modulation and spatial multiplexing. The spatial modulation encodes data symbols of users based on (i) the signal constellation including PSK/QAM symbols and (ii) the antenna index via an information-driven antenna-switching mechanism \cite{di2014spatial}. Then, the spatial multiplexing simultaneously transmits the independent and separately encoded data signals, so-called \textit{spatial streams}, of users from each transmit antenna of the antenna array. As such, a number of users can be  simultaneously served using the same time-frequency resource, and the spectral efficiency is substantially improved.

The spatial streams of users are considered to be generated by slicing physical resources such as  bandwidth, power, backhaul/fronhaul, infrastructure, and antennas of the massive MIMO system. In fact, to satisfy dynamic demands of users, the massive MIMO system allows the users to explicitly request bundles of physical resources. For example, the users can specify the amount of spectrum and power as well as the number of antennas. However, this imposes an important issue for the resource allocation. The issue is how to allocate the bundles of physical resources to the users to accommodate their dynamic demands while satisfying requirements of efficient resource allocation, e.g., high resource utilization and energy efficiency. To address the issue, optimization-based dynamic resource allocation mechanisms, e.g., \cite{lu2014elastic}, were proposed. Economic and pricing models such as combinatorial auction provide an optimal allocation of resource bundles/combinations to the users with little global information requirement and high economic efficiency.

\subsection{Heterogeneous Networks (HetNets)}
\label{sec:Intro_5G_Techno_HetNet}
HetNets are one key technology which combines different types of cells such as macrocells and small cells, e.g., picocells and femtocells. The macrocells are covered by Macro Base Stations (MBSs), and the small cells are covered by Small cell Base Stations (SBSs) such as Femto Access Points (FAPs), Pico cell Base Stations (PBSs), and Relay Stations (RSs). The coexistence of different cells forms a \textit{multi-tier HetNet} which results in better performance in terms of capacity, coverage, spectral efficiency, and total power consumption. Key features of cells used in HetNets are given in Table~\ref{Small_cell}. As seen, compared with macrocells, the small cells have much smaller coverage. Shrinking the transmission range of the cells leads to \textit{ultra-dense networks}, i.e., $\geq{10^3}$ cells/km$^2$ \cite{kamel2016ultra}, which allows to serve the high density of users, i.e., up $600$ active users/km$^2$. To enhance further spectrum and energy efficiencies, the HetNets exploit the versatile and local area networks such as the D2D and Machine-to-Machine (M2M) communications. The adoption of these networks leads to another tier with the elimination of cells. 

\begin{table}[h!]
\scriptsize
  \caption{\small Types of cell deployed in 5G HetNets.}
  \label{Small_cell}
  \centering
 \begin{tabularx}{8.9cm}{|L{1.1cm}|X|X|X|X|}
    \hline
  \cellcolor{mygray} \textbf{Cells} &  \cellcolor{mygray} \textbf{Transmit power (W)}&  \cellcolor{mygray} \textbf{Coverage (km)}  &   \cellcolor{mygray} \textbf{Deployment scenarios}& \cellcolor{mygray} \textbf{Users} \\ 
   \hline
Femtocell&0.001 to 0.25&0.01 to 0.1&Indoor& Up to 30\\     
      \hline
Picocell&0.25 to 1&0.1 to 0.2& Indoor/outdoor&30 to 100\\ 
    \hline
 Microcell&1 to 10&0.2 to 2&Indoor/outdoor&100 to 2000\\ 
    \hline
Macrocell&10 to 50&8 to 30&Outdoor&>2000\\ 
    \hline
  \end{tabularx}
\end{table}


Apart from multiple tiers, HetNets leverage spectrum of different Radio Access Technologies (RATs), including existing RATs, e.g., Wi-Fi, Evolved High-Speed Packet Access (EHSPA+), LTE, and new 5G RATs. As such, each user can simultaneously transmit/receive data to/from BSs with different RATs, e.g., through the \textit{Coordinated Multi-Point (CoMP)} technology \cite{sawahashi2010coordinated}. Such a multi-RAT environment provides a significant gain in terms of capacity, reliability, coverage, and spectrum efficiency.  

Heterogenours and dense deployment of wireless devices allow 5G to provision radio resources in terms of spectrum, power, and cache storages. However, this raises radio resource management issues such as user association, resource allocation, and interference management. The design of sophisticated radio resource management schemes for the issues is thus needed. To evaluate the resource management schemes, five performance metrics are commonly used as follows:  
\begin{itemize}
\item \textit{Spectrum efficiency:} Spectrum efficiency is defined as the average
achievable data rate per unit bandwidth.
\item \textit{Energy efficiency:} Energy efficiency is the ratio of the total network throughput to the total energy consumption.
\item \textit{Load balancing:} Load balancing is the capability to balance traffic across the entire network. 
\item \textit{Fairness:} Transmission rate fairness among users is typically measured by the Jain's fairness index \cite{sediq2013optimal} as $\frac{\big{(}\sum_{i=1}^{N}r_i\big{)}^2}{N\sum_{i=1}^{N}r_i^2}$, where $N$ is the number of users, and $r_i$ is the rate of user $i$.
\item \textit{Interference:} Two common interferences to be mitigated are \textit{intra-tier interferences} among cells within the same tier, and \textit{inter-tier interferences} among cells in different tiers.
\end{itemize}

To optimize the above performance metrics, traditional approaches for the radio resource management may not be efficient in 5G HetNet environment. For example, the Received Signal Strength Indication (RSSI) \cite{IEEE_802.11ad} is typically used in user association schemes. The scheme is based only on the received signal strength from BSs rather than on the traffic load of the BSs. This might result in unbalanced traffic load among BSs and low network throughput. For the interference management, optimization methods for power control are typically used. The methods usually require central controllers which may lead to the huge signaling and computational overhead. Recently, economic and pricing models have been efficiently adopted to address the above issues. For example, for the interference management, an MBS sets penalty prices according to interferences caused by SBSs, and then the SBSs locally reduce their transmit power to avoid the high payment. Such a simple pricing strategy mitigates the inter-cell interference, enables the MBS to obtain a higher revenue, and reduces the information exchange and computational complexity.

\subsection{Millimeter Wave (mmWave) Communications}
\label{sec:Intro_5G_mmWave}
Spectrum shortage of existing frequency bands, e.g., the 3G and 4G bands, becomes imminent. On the other hand, there is a vast amount of unused or lightly used spectrum in the mmWave band ranging from 30-300 GHz, the wavelenths of which are 1-10 mm. Therefore, mmWave communications have been proposed to be a potential candidate for 5G which can provide multi-gigabit communication services \cite{niu2015survey}.

Compared with existing communication systems using lower carrier frequencies, mmWave communications suffer from high propagation loss and have shorter range due to the high rain attenuation and atmospheric absorption. However, considerable works \cite{niu2015survey} on mmWave propagation confirm that for small distances, the rain attenuation and atmospheric absorption do not create significant additional path loss for mmWaves, particularly at 28 GHz and 38 GHz. As shown in Table~\ref{mmWave_5Gnetwork}, at 28 GHz, the rain attenuation is 0.9 dB over 200m. Also, the attenuation over 200m caused by atmospheric absorption is only 0.012~dB at 28 GHz and 0.016~dB at 38 GHz. 

The above features imply that the mmWave communications can overcome the strong path loss problem when small cells, e.g., picocells and femtocells, with the radius smaller than 200m are deployed. They are thus suitable for 5G. However, exploiting high frequency bands with narrow beams imposes several challenging issues. Specifically, the mmWave communications (i) are sensitive to blockage by obstacles, e.g., walls and human body, (ii) often suffer from the \textit{deafness}, i.e., the misalignment between the main beams of the transmitter and the receiver, and (iii) require a dense deployment of mmWave BSs. These issues lead to frequent handovers and make radio resource management in 5G be more challenging. For example, existing user association schemes are only sub-optimal for mmWave systems \cite{xu2016distributed}. Economic and pricing models with their simplicity provide fast and dynamic association schemes which can adapt to the fast variations of the wireless channels in mmWave systems. 
\begin{table}[h!]
\scriptsize
  \caption{\small Common mmWave bands in 5G and propagation characteristics at precipitation rate of 25 mm/h and the distance of 200m) \cite{niu2015survey}. LOS stands for Line-Of-Sight.}
  \label{mmWave_5Gnetwork}
  \centering
 \begin{tabularx}{8.9cm}{|L{1.2cm}|X|X|X|X|}
    \hline
  \cellcolor{mygray} \textbf{Frequency} &\cellcolor{mygray}\textbf{Propagation}&\cellcolor{mygray} \textbf{Rain}&\cellcolor{mygray}\textbf{Oxygen} &\cellcolor{mygray}\textbf{Available} \\ 
  \cellcolor{mygray} \textbf{band} &\cellcolor{mygray}\textbf{loss (LOS)}&\cellcolor{mygray} \textbf{attenuation}&\cellcolor{mygray}\textbf{absorption} &\cellcolor{mygray}\textbf{bandwidth} \\ 
    \hline
28 GHz&1.8-1.9 & 0.9 dB& 0.04 dB & 500 MHz\\ 
    \hline
38 GHz&1.9-2 &1.4 dB &0.03 dB &1 GHz\\
    \hline
73 GHz& 2 & 2.4 dB&0.09 dB &2 GHz\\
    \hline
  \end{tabularx}
\end{table}

\subsection{Cloud-Radio Access Network (C-RAN)}
\label{sec:Intro_5G_Techno_C_RAN}

C-RAN is a centralized, cloud computing-based architecture for radio access networks \cite{mobile2011c} in which the BaseBand processing Units (BBUs) of conventional BSs are moved to the cloud, i.e., the \textit{BBU pool}, and separated from the radio access units, namely \textit{Remote Radio Heads (RRHs)}. The functions of the BBU pool and RRHs are as follows. 
\begin{itemize}
\item BBU pool performs the centralized baseband processing functions such as coding, modulation, and Fast Fourier Transform (FFT). 
\item RRHs perform digital processing, RF amplification, up/down conversion, filtering, Digital-to-Analog (DA)/Analog-to-Digital (DA) converters, and interface adaptation. The RRHs with antennas transmit radio signals to users in downlink and forward the baseband signals from users to the BBU pool in uplink. The RRHs are connected to the BBU pool via fonthaul links.
\item Fronthaul links can be realized by optical fiber or microwave connections. The optical fibers can provide high bandwidth, i.e., up to 40 Gbps, while the microwave communications have the limited available bandwidth, i.e., a few hundred Mbps. However, the microwave communications are faster and cheaper to deploy than optical fibers. 
\end{itemize}
The BBU assignment for each RRH can be implemented using distributed or centralized approaches. For the distributed approach, each RRH directly connects to its exclusive BBU. This approach is simple and easy to be deployed, but it is not flexible to exploit the advantages of joint signal processing and central controlling in C-RAN. For the centralized approach, several RRHs are served by a BBU, and thus it has many advantages in terms of scalability, network capacity, coverage, interference mitigation, and CAPital EXpenditure (CAPEX)/OPerational EXpenditure (OPEX) reduction. 

In particular, the centralized BBU assignment can be divided into two categories, i.e., partially and fully centralized solutions. For the partially centralized solution, the L1 (Layer 1, PHY) processing is implemented at the RRH which reduces the burden in terms of bandwidth on the fronthaul links. However, this solution is less optimal because resource sharing is considerably limited and advanced features such as CoMP and joint processing Distributed Antenna System (DAS) cannot be efficiently supported. For the fully centralized solution, the functionalities of the L1, L2 (Layer 2, MAC), and L3 (Layer 3, network) are moved to the BBU pool, and the aforementioned advantages of the C-RAN can be achieved. However, the main disadvantage is a high load on fronthaul links due to the In-phase and Quadrature-phase (IQ) data transmission between the BBUs and RRHs. It significantly increases the latency and jitter, especially when a large number of RRHs are deployed. Thus given the constraint of the fronthaul capacity, new radio resource management solutions need to be developed to regulate the resource consumption of RRHs as well as their users. Such solutions can be easily realized by economic and pricing models. For example, the tiered pricing \cite{herrera2007theory} sets high prices to users with high resource demands, and thus reducing their resource consumption proportionally. 

\textbf{Summary:} In this section, we provide a brief overview of key technologies potentially deployed in 5G, e.g., the massive MIMO, HetNets, mmWave, and C-RAN. In each technology, radio resources, resource management issues, and motivation of using economic and pricing models are highlighted. The next section presents some basics and fundamentals of economic and pricing models.

\section{Overview and fundamentals of economic and pricing theories in 5G wireless networks}
\label{sec:Intro_Price}
This section presents the background of economic and pricing models which have been proposed for resource management schemes in 5G. In fact, there are several different pricing models depending on how to set the price. For example, when the price is set based on the profit maximization problem, we have profit-maximization pricing. In particular for our survey, the price is commonly determined using non-cooperative game, Stackelberg game, auctions, and Network Utility Maximization (NUM) problem. The following subsections present fundamentals of the pricing models. 

\subsection{Non-cooperative game}
\label{subsec:Non_cooperative_game}
Non-cooperative game is a game with competition among players. The players are selfish to maximize only their own utilities without forming coalitions with each other. Consider a radio resource market including $N$ competitive MNOs as the sellers which compete for selling spectrum to users, i.e., buyers. The strategy of MNO $i$ is to select spectrum price $p_i$ to maximize its own utility $\pi_i(p_1,\dots,p_N)$. Here, the utility may be the revenue or profit that the MNO receives from selling the spectrum. Let $p_i^*$ be the best strategy, i.e., the best response, of MNO $i$ which maximizes its utility. Then, the set of the best strategies $\mathbf{p^*}=(p_1^*,\dots,p_N^*)$ is the \textit{Nash equilibrium} if no MNO can gain higher utility by changing its own strategy when the strategies of the others remain the same.

Generally, the Nash equilibrium is a stable strategy profile and is the major solution concept of the non-cooperative game. However, it may not always exist and if it exits, it may not be unique. Thus the existence and uniqueness of the Nash equilibrium need to be checked and proved when setting prices based on the game. One common method for proving the existence and uniqueness of the Nash equilibrium is to check if the strategies of the players are non-empty, compact, and convex, and if their utility functions are concave or quasi-concave \cite{luong2017applications}. Different popular methods prove that the game is a potential game or supermodular game. Since the non-cooperative game models the conflict of selfish players, it has been commonly applied in environments with high competition or limited resources. In 5G, it is used for spectrum trading with competitive MNOs as proposed in \cite{gu2016game} or data rate allocation to multimedia applications as proposed in \cite{scott2015multimedia}. Note that the Nash equilibrium is obtained as the solution when all players make their decisions at the same time. However, if at least one player can make decision before the other players, the Stackelberg game \cite{amir1999stackelberg} can be adopted.
\subsection{Stackelberg game}
\label{subsec:Stackelberg_game}
Stackelberg game is a sequential game in which players consist of leaders and followers. The leaders choose their strategies first, and then the followers make corresponding strategies based on the leaders' strategies. 

Consider again the model in Section~\ref{subsec:Non_cooperative_game} with two MNOs, i.e., MNO 1 and MNO 2, as spectrum sellers. The optimization problem of each MNO is to choose its spectrum price $p_i, i \in {1,2},$ so as to maximize its utility $\pi_i (p1,p2), i \in {1,2}$. Assume that MNO 1 decides its pricing strategy before MNO 2 does, then MNO 1 is called the leader, and MNO 2 is called the follower.  The optimization problems of the leader and the follower together form the Stackelberg game. The objective of such a game is to find the Stackelberg equilibrium. 

\textbf{Definition 1.} \textit{Let $p_1^*$ and $p_2^*$ be solutions of the optimization problems of the leader and the follower, respectively. Then, the point ($p_1^*,p_2^*$) is the Stackelberg equilibrium for the Stackelberg game if for any $(p_1,p_2)$ with $p_1\geq 0$ and $p_2\geq 0$, we have $\pi_1(p_1^*,p_2^*) \geq \pi_1(p_1,p_2^*)$ and $\pi_2(p_1^*,p_2^*) \geq \pi_2(p_1^*,p_2)$.}

 To compute the Stackelberg equilibrium, the backward induction method \cite{aumann1995backward} is commonly used. Consider again the above example, for a given $p_1$, the follower solves its problem to find $p_2^*$, and then the leader substitutes $p_2^*$ in the leader problem to find $p_1^*$. Due to the first-move advantage, the leader imposes a favorable solution to itself. Thus the utility of the leader at the Stackelberg equilibrium is guaranteed to be no less than that at the Nash solution. This feature makes the Stackelberg game suitable for the resource management in 5G. For example, it allows an MBS to decide optimal interference prices after measuring transmit power of SBSs as proposed in \cite{lashgari2015distributed} or interferences caused by SBSs as proposed in \cite{liu2016pricing}. However, given the high density of the SBSs, how the MBS observes and then decides the optimal strategies is challenging. 

%

\subsection{Auction}
\label{subsec:Auction}
Auction is an efficient way of allocating resources to buyers which value the resources the most. A number of auctions and their descriptions can be found in \cite{nguyen2017resource} and \cite{luong2017applications}. The following briefly presents the major auctions used in this survey. Note that an auction typically consists of three entities: buyers, i.e., bidders, sellers, and an auctioneer. The auctioneer is an entity that organizes the auction, and it can be the same as the seller.

\subsubsection{Vickrey and Vickrey-Clarke-Groves (VCG) auctions}
\label{subsec:Auction_VCG_auction}
Vickrey and VCG auctions are the sealed-bid auctions in which bidders submit simultaneously their sealed bids to the auctioneer. 
\begin{itemize}
\item{\textit{Vickrey auction:}} In this auction, bidders submit bids, i.e., bidding prices that they are willing to pay for a commodity to the auctioneer. The auctioneer selects the bidder with the highest bid as the winner. The winner pays the second-highest price rather than the highest price that it submitted. Since the winner pays the price less than its expected price, the Vickrey auction motivates bidders to bid their true valuations for the commodity. The auction thus achieves truthfulness. This is an important property because any resource allocation scheme without holding this property may be vulnerable to market manipulation and reduce the revenue for the seller due to the low valuations \cite{luong2017applications}. 

\item{\textit{VCG auction:}} The VCG auction is a generalization of the Vickrey auction for multiple commodities. The VCG auction assigns the commodities to bidders in a socially optimal manner. It then charges each bidder the loss of the social value due to the bidder's getting the commodity. Such a payment strategy enables the VCG auction to be a truthful mechanism. In 5G, this auction can be used to enhance the fronthaul resource utilization in the C-RAN as proposed in \cite{gu2016virtualized}. 
\end{itemize} 

\subsubsection{Combinatorial auction}
\label{subsec:Auction_combinatorial_auction}
Combinatorial auction allows each bidder to bid a bundle of heterogeneous commodities rather than a single commodity \cite{cramton2006combinatorial}. After receiving bids, the auctioneer solves a Winner Determination Problem (WDP) which determines an optimal allocation of the commodities to the bidders under constraints, e.g., the supply constraint. Compared with the traditional auctions such as the Vickrey auction, the combinatorial auction has advantages such as utility maximization for buyers. However, it has one big challenge for solving the WDP. The WDP is generally NP-hard, and there does not exist a polynomial-time algorithm. The Lagrangian relaxation approach \cite{hsieh2010combinatorial} has been commonly used to find approximate solutions for the WDP. In 5G, the combinatorial auction is often used in massive MIMO systems as proposed in \cite{zhu2015virtualization} to allocate heterogeneous resources, i.e., antennas and bandwidth, to users. 

\subsubsection{Ascending Clock Auction (ACA)}
\label{subsec:Auction_ACA}
ACA is a type of multiple-round auctions in which the auctioneer raises the commodity price in each round until the total demand is equal to or less than the supply. Consider a model including one MNO as a spectrum seller and multiple users as bidders, i.e., buyers. Initially, the MNO announces a spectrum price to all the users. Each user submits its spectrum demand so as to maximize the user's utility. If the total demand is equal to or less than the supply, the MNO concludes the auction. Otherwise, the MNO increases the spectrum price and then announces again this price in the next auction round. This process is repeated until the total demand is equal to or less than the supply. Since the utility and the demand of each user decrease with the increase of the auction rounds, the algorithm converges. Due to its simplicity, the ACA has been used in complex markets such as heterogeneous resource allocation in massive MIMO systems as proposed in \cite{ahmadi2016substitutability} or wireless caching in dense HetNets as proposed in \cite{sken2016energy}. However, the ACA is not a truthful mechanism since the users have an incentive to misreport their true demands which can lead to higher utility. 

\subsubsection{Forward, reverse auction, and double auction}
\label{subsec:Auction_forward_reverse_double_auction}
Considering the sides of sellers and buyers, auctions can be classified as follows \cite{nguyen2017resource}:
\begin{itemize}
\item{\textit{Forward and reverse auctions:}} In the forward auction, multiple buyers compete for commodities by submitting their bids, i.e., bidding prices, to one seller. On the contrary, in the reserve auction, multiple sellers compete to sell commodities by submitting their asks, i.e., asking prices, to one buyer. 
\item{\textit{Double auction:}} In the markets with multiple sellers and buyers, the double auction can be used to match the sellers and the buyers. Specifically, the sellers and buyers respectively submit their asks and bids to an auctioneer. Then, the auctioneer sorts the sellers in the ascending order of their asks and the buyers in the descending order of their bids. The auctioneer finds the largest index $l$ at which the ask $p_l^a$ is less than the bid $p_l^b$. The transaction price is determined as $\frac{(p_l^a+p_l^b)}{2}$. The seller receives the transaction price, and the buyer receives spectrum. The process is repeated to determine the remaining seller-buyer pairs, transaction prices, and spectrum allocation. Under certain conditions and settings, the double auction holds desired economic properties such as truthfulness, individual rationality, balanced budget, and economic efficiency.
\end{itemize}

\subsection{Network Utility Maximization (NUM)-based Pricing}
\label{subsec:NUM_problem}
NUM is proposed to allocate resources, e.g., bandwidth and power, to users so as to maximize the total net utility of users given the capacity constraint of the network. Consider a network model with one MNO as the resource seller and multiple users as buyers. Then, the net utility of each user is the difference between the utility associated with its resource allocation and the cost that the user pays the MNO for using the resources. To solve the NUM problem, price update schemes can be used as presented in \cite{nguyen2017resource}. Briefly, by applying the dual decomposition and gradient methods, the MNO iteratively updates resource prices. Also, the users iteratively select their amounts of resources to maximize their own net utilities. The utilities of the users are assumed to be concave functions of their resource allocation, and the NUM problem is a convex optimization problem. Therefore, the algorithm converges to a unique optimal solution for the resource allocation. 

 In 5G, NUM-based pricing is often applied to maximize the total data rate of users. For example, it is used for the aggregated carrier allocation in HetNets as proposed in \cite{shajaiah2015efficient} or for data rate allocation in massive MIMO systems as proposed in \cite{athanasiou2015optimizing}. However, these approaches mostly require utility functions of users to be strictly concave. 

 Apart from the aforementioned pricing models, matching theory and contract theory have been used for the resource management in 5G. In general, the matching theory assigns users to MNOs given their preferences, and the contract theory constructs resource-price bundles under an information asymmetry between users and MNOs.

\textbf{Summary:} In this section, we introduce the basics of economic
and pricing models proposed to address resource management issues in 5G.
Specifically, we provide the definitions, mechanism descriptions,
and rationale behind the use of economic and pricing models
for the resource management. In the subsequent sections, we review economic and pricing approaches for addressing various resource management issues in 5G.

\section{Applications of economic and pricing models for user association}
\label{sec:User_Association}
Before data transmission commences, a user association mechanism is performed to determine which user to be assigned to which BS. The user association mechanism must be designed to optimize metrics such as throughput, load balancing, and energy efficiency. Centralized solutions, e.g., \cite{xu2014qoe}, often require a large amount of signaling and have high computational complexity, which may not be a viable solution for large-scale networks, especially for HetNets. Alternatively, economic and pricing models can provide distributed solutions to optimize the aforementioned metrics with low computational complexity. This section discusses applications of economic and pricing models for the user association in 5G. Refer to \cite{liu2016user}, this section is classified based on metrics which the user association mechanisms aim to achieve, i.e., (i) throughput maximization, (ii) load balancing and fairness, and (iii) energy efficiency optimization, in the following subsections, respectively. Note that each metric can be considered in different technologies, e.g., massive MIMO, mmWave, and HetNets. For example, energy efficiency can be considered to be a major requirement when designing the user association in massive MIMO networks due to the large energy consumption of antennas. On the contrary, given the ultra-dense and unplanned deployment of BSs, throughput maximization and load balancing among BSs/tiers can be set to be the major requirements in the user association in mmWave networks and HetNets, respectively. 
 
\subsection{User Association for Throughput Maximization}
\label{sec:User_Association_Throughput}
A common objective of the network aims to maximize the total throughput of all users in 5G. Distributed auction or distributed algorithms with price update schemes are used to achieve the objective with low complexity.  

\subsubsection{Distributed auction}
\label{sec:User_Association_Throughput_distributed_auction}
\begin{figure}[t!]
 \centering
\includegraphics[width=8.1cm, height=3.8cm]{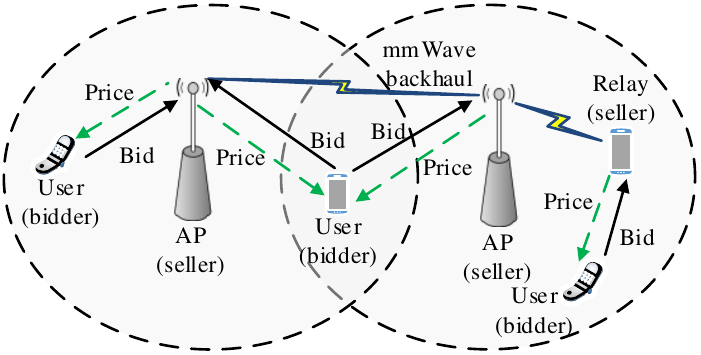}
 \caption{\small User association based on distributed auction in mmWave networks.}
 \label{mmWave_user_relay_association_auction}
\end{figure}

The first work based on the distributed auction was investigated in \cite{shokri2015user} for a mmWave network. The model is shown in Fig.~\ref{mmWave_user_relay_association_auction} in which users act as bidders, i.e., buyers, and BSs are sellers. Initially, the BSs set connection prices to zero and broadcast them to all the users. Each user calculates its utility and payment based on the prices and its required throughput. The user then selects the best BS which yields the user's maximum utility. The user sends its required throughput as a bid and its payment to the BS. The BS selects users with the highest bids and higher payments than the old price. The BS increases its price and then feeds back the new price to the users. The auction terminates when there is only one request user, and this user is associated with the BS. The simulation results show that the proposed scheme can improve network throughput around 12\% compared with the RSSI-based association scheme \cite{IEEE_802.11ad}. In addition to the throughput improvement, the proposed scheme considers the optimization of operating beamwidth to address the \textit{deafness} problem. Accordingly, the network throughput of the proposed scheme increases when the beamwidths are narrower. However, the narrower beamwidths lead to significant alignment overhead since many directions have to be searched. The future work can consider this effect in the optimization of bidding process.


The distributed auction-based approach was also found in \cite{maghsudi2016distributed} to assign each user to each SBS in a HetNet. Here, the SBSs are powered by harvested energy. Different from \cite{shokri2015user}, users in the proposed scheme are considered to be commodities while the SBSs act as bidders, i.e., buyers. A coordinator which ensures that the commodities are sold in a fair manner is the auctioneer or seller. Each SBS selects the number of users to serve depending on its state, i.e., amount, of harvested energy. Since the SBS is uncertain about the state which has to be determined before the user association process, it assigns probabilities to the possible energy states. Then, a multi-round assignment is adopted. At each round, the auctioneer announces arbitrary prices for commodities, i.e., the users. Given the prices and the probabilities of the states, each SBS calculates and submits its demand, i.e., the number of commodities so as to maximize its expected utility. The auctioneer adjusts the price of the commodity using the \textit{Walras' tatonnement} process \cite{mochaourab2015distributed} such that the total demand equals the supply. The commodity allocation and the prices at this point constitute a so-called \textit{Arrow-Debreu equilibrium} \cite{debreu1959theory}. The simulation results show that the proposed scheme outperforms the distance-based assignment \cite{son2011base} in terms of aggregate network throughput. However, how to select the probability distribution function for the states is not specified in the proposed scheme.

\subsubsection{Utility maximization}
\label{sec:User_Association_Throughput_Utility_Max}
\begin{figure}[t!]
 \centering
\includegraphics[width=8.1cm, height=3.8cm]{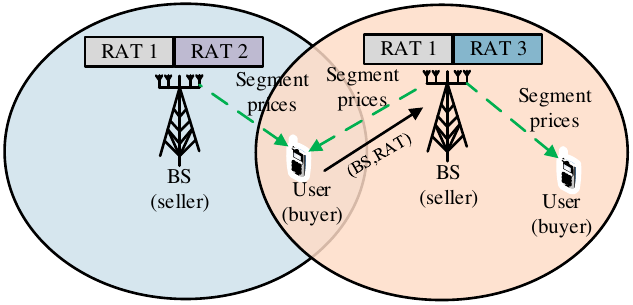}
 \caption{\small  User association based on NUM problem in multi-RAT environments.}
 \label{HetNet_user_association_PMP}
\end{figure}
Consider a HetNet with multi-RAT, the authors in \cite{soldati2015distributed} formulated the NUM problem to jointly associate users with BSs, e.g., MBSs or SBSs, and allocate spectrum segments. The objective is to maximize the sum of the users' utilities. As shown in Fig.~\ref{HetNet_user_association_PMP}, each BS as a seller owns multiple RATs, and each RAT may have multiple spectrum segments to be sold to the users, i.e., buyers. To solve the NUM problem, the Lagrange duality theory is first applied with multipliers as spectrum segment prices that the users pay the BSs. The distributed algorithm with the price update is then executed. Specifically, each BS broadcasts initial prices to its users. Given the prices, each user locally decides on the BS and spectrum segment of the BS to maximize its utility. The BS updates the prices using the projected gradient method \cite{lin2007projected} which ensures that the total traffic load generated by the users equals the maximum capacity of the BS. The algorithm is guaranteed to converge to a near-optimal solution of the original NUM problem. The simulation results show that the proposed scheme improves the network throughput up to 20\% compared with the RSSI-based association scheme. 

By using the projected gradient method, the distributed algorithm in \cite{soldati2015distributed} is guaranteed to converge to the global optimal solution. However, to ensure the convergence, all the prices of the BSs need to be updated at the same time using the same step size. This requires synchronized price updates across the BSs, which is challenging in implementation in large-scale networks. The dual coordinate descent method \cite{bertsekas1999nonlinear} can be an alternative solution as proposed in \cite{shen2014distributed}. The model and the design of the distributed algorithm for the user association in \cite{shen2014distributed} are similar to those in \cite{soldati2015distributed}. However, after applying the Lagrange dual problem, the dual objective function is expressed in a closed form, and the BSs use the dual coordinate descent method to update the prices. As such, the BSs do not need to synchronize their price updates, and the algorithm has faster convergence than that using the projected gradient method. However, the dual coordinate descent method is not guaranteed to converge to a global optimum solution since the closed-form objective function is not differentiable. 

\subsection{User Association for Load Balancing and Fairness}
\label{sec:User_Association_Load_balancing_fairness}
Apart from the throughput improvement, load balancing among BSs and fairness of users need to be considered in the user association. Pricing models such as distributed auction, congestion-based pricing, and Paris metro pricing can well meet the requirements. 

\subsubsection{Distributed auction}
\label{sec:User_Association_Load_balancing_fairness_distributed_auction}
The authors in \cite{xu2016distributed} addressed the joint association and relaying problem in a mmWave network considering the load balancing at BSs. The model involves multiple BSs and multiple users. As shown in Fig.~\ref{mmWave_user_relay_association_auction}, a user can be associated with one BS via one of other users which serve as relays. The joint association and relaying problem generally has no closed-form solution and is NP-complete, but it is convex \cite{bertsekas1998network}. Thus, the problem can be equivalently converted into the \textit{min-cost flow problem} which is then solved by the distributed auction as presented in \cite{shokri2015user}. Note that the bidders are the users, i.e., the buyers, and the sellers are the relays. The proposed scheme is proved to converge quickly when the number of request users for one relay is sufficiently large. The reason is that an increase of the price which is set by the relay is proportional to the number of users requesting to access the relay. Therefore, given a large number of users, one relay will become too expensive in a short time compared with other relays, and there is only few users which can then accept the price. However, a large number of users make the overall convergence speed of the proposed scheme slow. This may not meet the fast variations of the mmWave channel conditions.


\subsubsection{General pricing}
\label{sec:User_Association_Load_balancing_fairness_general_pricing}
A distributed algorithm via Lagrangian dual decomposition for the user association to achieve the load balancing and the fairness in a mmWave network was proposed in \cite{athanasiou2015optimizing}. The model consists of multiple BSs, i.e., sellers, each of which can serve multiple mobile users, i.e., buyers. To achieve load balancing and fairness, a channel utilization metric, i.e., the ratio of the required data rate of the user to the channel capacity, is introduced. The problem is to minimize the maximum BS utilization. Here, the BS utilization is the sum of the channel utilizations. First, the Lagrange dual problem is applied with the multipliers as service prices that the users pay the BSs. Then, the subgradient method \cite{boyd2003subgradient} with the price update is adopted. Generally, each user locally determines its BS so as to minimize its payment, and the BSs communicate with each other to update the prices to regulate users' requests and to balance the channel utilizations among the BSs. The simulation results show that the proposed scheme outperforms the RSSI-based scheme in terms of convergence and fairness among BSs measured by the Jain's fairness index. However, updating prices at each iteration may significantly increase a communication overhead of the proposed scheme. 

\subsubsection{Utility maximization}
\label{sec:User_Association_Load_balancing_fairness_utility_maximization}
\begin{figure}[t!]
 \centering
\includegraphics[width=6.4cm, height=4.6cm]{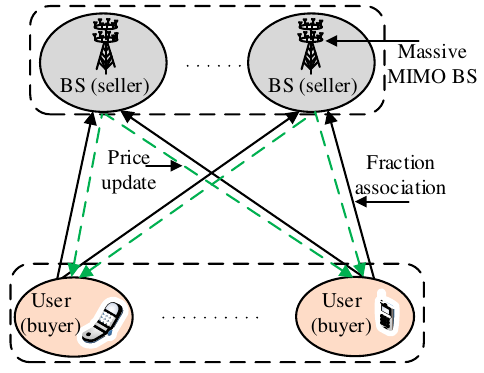}
 \caption{\small Market-based fractional user allocation in a massive MIMO network.}
 \label{MIMO_user_association_NUM}
\end{figure}

The price updating adopted from \cite{athanasiou2015optimizing} can be combined with the NUM problem to achieve fairness and load balancing in a massive MIMO network as proposed in \cite{bethanabhotla2014user}. The model is shown in Fig.~\ref{MIMO_user_association_NUM} which consists of multiple users and multiple BSs. Each of the BSs is equipped with a large number of antennas. Note that each BS, i.e., a seller, can serve multiple users, and each user, i.e., a buyer, can simultaneously be associated with multiple BSs to achieve its desired data rate. The objective is to maximize the total utility of all users. Here, the utility is represented by a logarithmic function which allows the allocation to achieve the proportional fairness. The utility is a concave function, and the problem is convex optimization. The price updating in \cite{athanasiou2015optimizing} is used again to solve the problem. However, different from \cite{athanasiou2015optimizing}, given the BSs' service prices, each user chooses a set of BSs and the corresponding resource fractions so as to maximize its total \textit{bang-per-buck}. The bang-per-buck offered by a BS to a user is defined as the maximization of the ratio of the rate allocated to the user to the price that the user pays the BS. 

\subsubsection{Congestion-based pricing}
\label{sec:User_Association_Load_balancing_fairness_congestion_pricing}
\begin{figure}[t!]
 \centering
\includegraphics[width=5.7cm, height = 5.4cm]{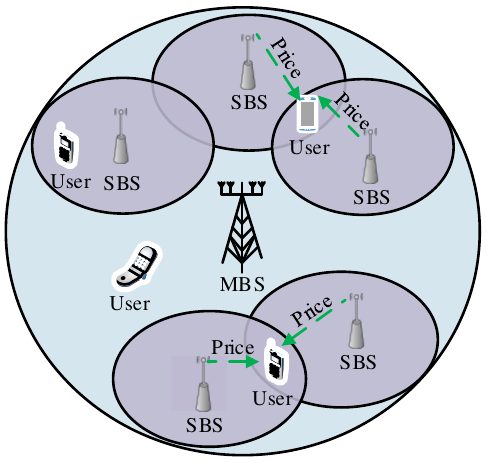}
 \caption{\small User association using price-based distributed algorithm in HetNets.}
 \label{HetNet_user_association_NUM}
\end{figure}

The congestion-based pricing \cite{jin2005dynamic} sets the connection prices depending
on the current network load. Thus it can be used to achieve the load balancing as proposed in \cite{qi2017refunding}. The considered model is a HetNet as shown in Fig.~\ref{HetNet_user_association_NUM} which includes users of one MBS, i.e., MUEs, and multiple SBSs, i.e., femtocells. The problem is to determine the amount of bandwidth which the SBSs allocate to the MUEs so as to maximize the difference between the total utility of the users and the total cost that the MBS pays the SBSs. To solve the problem, the Lagrange dual problem with the distributed algorithm as proposed in \cite{soldati2015distributed} is adopted. However, the multipliers are shadow prices associated with the rate and the bandwidth, and they are updated by the gradient descent method \cite{tseng2009coordinate}. At each iteration, given the shadow and congestion prices of the SBSs, each user selects the best SBS to maximize the ratio of the rate to the sum of the prices offered by the SBS. A more general scenario with multiple MBSs needs to be investigated. However, the interaction between the MBSs and SBSs can be more complicated.  

Apart from small cell networks, M2M networks will be deployed in 5G HetNets which allow a wide range of autonomous devices to communicate wirelessly without human intervention. A massive number of M2M devices create the overload problem which hugely impacts the radio access and core networks of the cellular system. Thus the congestion-based pricing can be used as proposed in \cite{mao2012dynamic}. The model consists of M2M devices as source nodes and routes. The source nodes as buyers buy routing services from the routes, i.e., pairs of nodes, as sellers to forward packets to their destinations. Each route sets a routing price proportionally to the number of source nodes which connect with the route. Given the routing prices, each source node selects the routes with the minimum prices. The routes adjust the prices according to the supply and demand rule to balance the connections among the routes. 


\subsubsection{Paris Metro Pricing (PMP)}
\label{sec:User_Association_Load_balancing_fairness_Paris_pricing}
PMP is used in Paris metro to give passengers the ability to select less congested wagons \cite{odlyzko1999paris}. For Internet service, the PMP is applied to set different prices for services based on QoSs \cite{ros2004mathematical}. For example, channels with higher prices would be less congested than those with lower prices. The authors in \cite{passasparis} employed the PMP for the user association in the multi-RAT environment of a HetNet to balance the traffic among RATs. The model consists of one MNO as a seller and multiple users as buyers. The MNO owns different RATs, and the users decide which RAT they prefer to access depending on the RAT's current access price and the users' congestion sensitivity. First, the MNO assigns randomly the users to its RATs. Then, it collects and calculates system information including the available capacity and the number of connected users to each RAT as well as the access price for the RAT. In particular, the access price of the RAT depends on the availability of the RAT's capacity. The MNO then broadcasts the system information to the users, each of which chooses the RAT which maximizes its utility. Then, the MNO accepts or denies the user depending on the available capacity of the RAT. The steps are repeated until no user desires to switch to another RAT. However, how to prove the stability of the algorithm is not presented. 
\subsubsection{Forward auction}
\label{sec:User_Association_Load_balancing_fairness_General_auction}
In the context of a multi-tier HetNet including a large number of BSs, i.e., MBSs and SBSs, the authors in \cite{sun2015joint} adopted the forward auction for the user association to achieve the max-min fairness, i.e., maximizing the minimum SINR received at each user. The users are bidders, and the BSs are sellers. Each BS has its association price that a user pays if the user is associated to the BS. Each unassigned user finds the BS which maximizes the user's utility, i.e., the channel gain between the user and the BS minus the association price. The user also calculates its bid, i.e., the difference between the largest utility and the second largest utility, and submits it to the BS. Upon receiving the bids, the BS selects the user with the highest bid as the winner. The BS cancels the previous association and adds a new association with the winner. Then, the BS increases its association price by the winner's bid as its new price. Note that the winner is not guaranteed for a long-term association, e.g., when the BS finds a new user with a higher bid. This may discourage users to participate in the auction. 


\subsection{User Association for Energy Efficiency Optimization}
\label{sec:User_Association_Energy_efficiency}
Energy efficiency is also a common requirement for the user association in 5G due to the tremendous number of antennas in massive MIMO networks. The pricing models based on the NUM problem and repeated game are developed for the energy efficiency improvement. 

\subsubsection{Utility maximization}
\label{sec:User_Association_Energy_efficiency_non_cooperative}
The authors in \cite{liu2015distributed} developed a user association algorithm in a massive MIMO network considering the energy efficiency, fairness, and QoS provision. The model consists of multiple BSs and users. The energy efficiency of each user when associated with a BS is defined as the ratio of the downlink data rate of the user to the energy consumed by the BS. The problem is to find a user association solution which maximizes the total utility of all users subject to their minimum SINR requirements. The utility of the user is a logarithmic function of the energy efficiency, which is concave. To solve the problem, the Lagrange dual theory is first applied with the multipliers as the dissatisfactory factors of the users and the service prices that the users pay the BSs. Similar to \cite{athanasiou2015optimizing}, the subgradient method is adopted to update the dissatisfactory factors and the service prices. Then, each user selects a BS so as to maximize the difference between its achievable rate and the service price. As shown in the simulation results, the proposed scheme achieves much higher energy efficiency than that of the Reference Signal Received Power (RSRP)-based user association \cite{qian2013performance}. However, the energy efficiency of the proposed scheme decreases with an increasing number of antennas due to more energy consumption. 

\subsubsection{Repeated game}
\label{sec:User_Association_Energy_efficiency_repeated_game}
 The repeated game can be used to provide the distributed user association to improve the energy efficiency in a HetNet with massive MIMO as proposed in \cite{feng2016boost}. The repeated game allows players to repeat a so-called \textit{stage game} over multiple periods to achieve their objectives \cite{hoang2015applications}. The players in the network model include users and BSs, i.e., multiple SBSs and one MBS equipped with a large number of antennas. The users act as buyers, and the BSs act as sellers. The utility of each user is a function of its achievable rate, and the utility of each BS is the difference between the total price paid by its connected users and the cost of power consumption. First, each user/SBS forms a descending order preference list in terms of its utilities gained from all the SBSs/users. Then, each user bids for the top SBS in its preference list. The SBS puts users with the highest bids in the waiting list and rejects other users. The bidding procedure is repeated for the rejected users. The SBS then chooses to be turned ON for the users' demands when the total payment from its users is larger than the power cost. Otherwise, the SBS is switched OFF, and the users in its waiting list will connect to the MBS. The ON-OFF switching strategy aims to improve the energy efficiency of the overall network. As shown in the simulation results, the proposed scheme improves significantly the energy efficiency compared with the ACTIVE-IDLE switching procedure proposed in \cite{ashraf2010improving}. 
 
In practice, the assumption in which the utility of each user is a function of only its achievable rate as mentioned in \cite{feng2016boost} may not be reasonable since the users need to pay service prices to the BSs. The authors in \cite{xu2015user} introduced the cost, i.e., the service price, in the users' utility functions when applying the repeated game for the distributed user association. The players in the game include users as buyers and a service provider which owns BSs as the seller. Initially, the service provider sets a service price of each BS for each user and broadcasts the prices to the users. The prices are determined so as to maximize the total payment that the service provider receives from the users. Given the prices, each user chooses its BS to maximize its utility. This process is repeated until both the service provider and the users are satisfied with the prices at which the total payment received by the service provider and individual users' utilities are maximized.

\subsubsection{Cost minimization}
\label{sec:User_Association_Energy_efficiency_cost_minimization}
D2D networks are also a tier of 5G HetNets in which each D2D user as a source can transmit its data to its destination by using different relay nodes. The problem is to select relay nodes to minimize the energy consumption cost of the source. The authors in \cite{abdellatif2017concurrent} addressed this problem by using the cost minimization problem. Initially, given a forwarding price of each relay node, the source determines the optimal fractions of data for the relay node to minimize its total cost including the energy consumption and monetary cost. Here, the monetary cost is determined based on the forwarding price of the relay node. Note that if this price is too high, the source may use other relay nodes. Thus the relay node needs to set an appropriate price, e.g., reducing the price while maintaining a positive profit. In particular, the price adjustment is inversely proportional to the fraction of demand of the source. Given the new price, the source solves its problem again. The process is iterated until the optimal data allocation does not change. However, how to solve the source's problem is not specified. 

\textbf{Summary:} In this section, we have reviewed the applications of economic and pricing models for the user association in 5G. The objective is to achieve the throughput maximization, load balancing, fairness, and energy efficiency optimization. The reviewed approaches are summarized along with the references in Table~\ref{table_mmWave_communications}. Additionally, a summary of advantages and disadvantages of major approaches is shown in Table~\ref{table_sum_advantage_user_association}. We observe that the utility maximization problem with price updating schemes has been primarily used for the user association. In fact, with the surge of data traffic and limited spectrum
resources, high spectrum efficiency is also a major requirement of 5G. The next section thus discusses how to apply the economic and pricing models for spectrum allocation to enhance the spectrum efficiency.
\begin{table*}[h!]
\caption{Applications of economic and pricing models for user association in 5G (MWN: Millimeter Wave Network, MMN: Massive MIMO Network, HN: Heterogeneous Network).}
\label{table_mmWave_communications}
\scriptsize
\begin{centering}
\begin{tabular}{|>{\centering\arraybackslash}m{0.6cm}|>{\centering\arraybackslash}m{0.4cm}|>{\centering\arraybackslash}m{1.4cm}|>{\centering\arraybackslash}m{1cm}|>{\centering\arraybackslash}m{0.6cm}|>{\centering\arraybackslash}m{1cm}|>{\centering\arraybackslash}m{7.7cm}|>{\centering\arraybackslash}m{1.5cm}|>{\centering\arraybackslash}m{0.7cm}|}
\hline
\multirow{2}{*} {\textbf{}} & \multirow{2}{*} {\textbf{Ref.}} & \multirow{2}{*} {\textbf{Pricing model}} & \multicolumn{3}{c|} {\textbf{Market structure}} & \multirow{2}{*} {\textbf{Mechanism}}& \multirow{2}{*} {\textbf{Solution}} & \multirow{2}{*} {\textbf{Network}} \tabularnewline
\cline{4-6}
 & & & \textbf{Seller} & \textbf{Buyer} & \textbf{Item} & &&\tabularnewline
\hline
\hline
\parbox[t]{2mm}{\multirow{9}{*}{\rotatebox[origin=c]{90}{ \hspace{0 cm} Throughput}}}
\parbox[t]{2mm}{\multirow{9}{*}{\rotatebox[origin=c]{90}{ \hspace{0 cm}maximization}}}

&\cite{shokri2015user}&Distributed auction& Access points&Users& Transmission rate& Given the access points' service prices, each user selects the best access point. The access point selects users with higher payments. &Optimal solution  &MWN\tabularnewline \cline{2-9}

&\cite{maghsudi2016distributed} &Distributed auction&Coordinator&SBSs&Users&Based on the demand of SBSs, the coordinator iteratively adjusts the price of each item, i.e., each user, using the \textit{Walras' tatonnement} process. &Arrow-Debreu equilibrium&HN\tabularnewline \cline{2-9}

&\cite{soldati2015distributed} &Utility maximization&MBSs and SBSs&Users&Spectrum& Each user selects an MBS or an SBS, and spectrum segment. Then, the sellers update the prices by using the projected gradient method. &Optimal solution &HN\tabularnewline \cline{2-9}

&\cite{shen2014distributed} &Utility maximization&MBSs and SBSs&Users&Spectrum segments& Same as \cite{soldati2015distributed}, but  the dual coordinate descent method is used to update the prices.&Optimal solution &HN\tabularnewline \cline{2-9}

\hline
\parbox[t]{2mm}{\multirow{9}{*}{\rotatebox[origin=c]{90}{ \hspace{-1cm} Load balancing}}}
\parbox[t]{2mm}{\multirow{9}{*}{\rotatebox[origin=c]{90}{ \hspace{-1cm} and fairness}}}

& \cite{athanasiou2015optimizing}&General pricing& Access points&Users&Transmission rate&Given the users' requests, the access points update service prices by using the projected subgradient method. &Optimal solution  &MWN\tabularnewline \cline{2-9}

&\cite{bethanabhotla2014user} &Utility maximization& BSs&Users& Data rates&Each user selects a set of BSs to maximize its total bang-per-buck. Then, the BSs adjust the resource prices using the subgradient method. &Optimal solution&MMN\tabularnewline \cline{2-9}

&\cite{qi2017refunding} &Congestion-based pricing&SBSs&MUEs&Bandwidth& Same as \cite{soldati2015distributed}, but the SBSs update the prices, i.e., shadow prices and congestion prices, using the gradient descent method. &Optimal solution &HN\tabularnewline \cline{2-9}

&\cite{mao2012dynamic} &Congestion-based pricing&M2M devices&Relay nodes&Routing services& Relay nodes update routing prices according to the demand and supply rule. &Market
equilibrium &HN\tabularnewline \cline{2-9}

&\cite{passasparis} &Paris metro pricing&MNOs&Users&Access services&Each user decides to access to RAT of the MNO depending on the RAT's access price and the user's congestion sensitivity. &Stable solution &HN\tabularnewline \cline{2-9}

&\cite{sun2015joint} &Forward auction&MBSs and SBSs&Users&Access services&Given the users' bids, each MBS or SBS selects the buyer with the highest bid as the winner.&Optimal solution &HN\tabularnewline \cline{2-9}
\hline
\parbox[t]{2mm}{\multirow{9}{*}{\rotatebox[origin=c]{90}{ \hspace{-0.5cm} Energy efficiency}}}
\parbox[t]{2mm}{\multirow{9}{*}{\rotatebox[origin=c]{90}{ \hspace{-0.5cm} optimization}}}

&\cite{liu2015distributed}&Utility maximization& BSs&Users& Data rates&Same as \cite{bethanabhotla2014user}, but the Lagrange dual theory is applied to solve the optimization problem.  &Optimal solution&MMN\tabularnewline \cline{2-9}

&\cite{feng2016boost}&Repeated game& BSs&Users& Data rates&Each user repeats the selection of its preferred BS while each BS repeats the selection of its preferred users until the users in the waiting list of each BS do not change anymore.&Nash equilibrium&HN\tabularnewline \cline{2-9}

&\cite{xu2015user}&Repeated game& Service provider&Users& Data rates& Service provider iteratively sets prices for its BSs, and users iteratively select BSs until both the service provider and the users are satisfied. &Nash equilibrium&HN\tabularnewline \cline{2-9}

&\cite{abdellatif2017concurrent}&Cost minimization&Relay nodes&Source node& Forwarding services& Given the relay nodes' prices, the source node determines the optimal fractions of data for each relay node by solving the source node's cost minimization problem. &Optimal solution&HN\tabularnewline \cline{2-9}
\hline
\end{tabular}
\par\end{centering}
\end{table*}

\begin{table*}[h]
\caption{A summary of advantages and disadvantages of major approaches for user association in 5G.}
\label{table_sum_advantage_user_association}
\scriptsize

\begin{centering}
\begin{tabular}{|>{\centering\arraybackslash}m{2cm}|>{\centering\arraybackslash}m{7.8cm}|>{\centering\arraybackslash}m{6cm}|}
\hline
\cellcolor{myblue} &\cellcolor{myblue} &\cellcolor{myblue} \tabularnewline
\cellcolor{myblue} \multirow{-2}{*} {\textbf{Major approaches}} &\cellcolor{myblue} \multirow{-2}{*} {\textbf{Advantages}} &\cellcolor{myblue} \multirow{-2}{*}{\textbf{Disadvantages}} \tabularnewline
\hline
\hline
\cite{maghsudi2016distributed} &\begin{itemize} \item Achieve low computational complexity \end{itemize} & \begin{itemize}  \item Dot not specify the distribution function for the states \end{itemize}\tabularnewline \cline{2-3}
\hline
\cite{bethanabhotla2014user} &\begin{itemize} \item Support multiple users and multiple BSs \end{itemize} & \begin{itemize}  \item Require synchronized
price updates across the BSs \end{itemize}\tabularnewline \cline{2-3}
\hline
 \cite{athanasiou2015optimizing}&\begin{itemize} \item Support multiple BSs and multiple user and have fast convergence\end{itemize} & \begin{itemize}  \item Require communication among BSs for price updates\end{itemize}\tabularnewline \cline{2-3}
\hline
\cite{feng2016boost}&\begin{itemize} \item Support multiple users and multiple BSs\end{itemize} & \begin{itemize}  \item Have high communication overhead \end{itemize}\tabularnewline \cline{2-3}
\hline
\end{tabular}
\par\end{centering}
\end{table*}
\section{Applications of economic and pricing models for spectrum allocation}
\label{sec:App_SA}
5G will support a large number of mobile users with ubiquitous real-time services such as video streaming and online gaming. The services are generally delay sensitive and consume a large amount of spectrum. Thus spectrum allocation schemes need to be developed to maximize the spectrum efficiency, i.e., the transmitted data rate over a given bandwidth. Traditional spectrum allocation schemes, e.g., \cite{peng2006utilization}, can improve the overall data rate, but they often require central servers and the cooperation among the users. This may not be feasible in large-scale networks of 5G. Economic and pricing mechanisms are proposed not only to provide an incentive to different entities in spectrum allocation, but also to introduce distributed solutions in which the spectrum is efficiently assigned to the users based only on local observations. This section reviews the applications of economic and pricing models for the spectrum allocation in 5G. Similar to Section \ref{sec:User_Association}, this section is organized based on objectives which the spectrum allocation mechanisms aim to achieve. The major objectives include (i) data rate maximization for users, (ii) revenue/profit maximization which is to maximize the revenue/profit for network operators, i.e., MNOs, and (iii) system utility/social welfare maximization for both the users and the network operators, in the following subsections, respectively. Note that the different objectives lead to the use of different pricing models. For example, the auctions are commonly used to maximize the revenue for the MNOs while the Stakeleberg game is mostly adopted to maximize the overall system utility. 
\subsection{Data Rate Maximization}
\label{sec:App_SA_SE}
This section reviews the applications of the economic and pricing models for the spectrum/rate allocation in 5G, i.e., mmWave networks, massive MIMO networks, C-RANs, and HetNets. Different pricing models can be applied to different technologies. For example, in mmWave networks, the mmWave band can be aggregated with existing bands, e.g., 3G band, and then the NUM problem can be used for the rate allocation to maximize the sum of data rates of all users. In massive MIMO networks, antennas and spectrum can be sliced into spatial streams which are then allocated to multiple users. Auctions can be used for the stream allocation to maximize the users' data rates while guaranteeing the truthfulness. 


\subsubsection{MmWave networks}
With the high throughput, mmWave communications can support users' multimedia applications and provide wireless backhaul links between BSs. Pricing models are developed for different network scenarios. For example, if the users are selfish, then the non-cooperative game can be used. 

\textbf{Non-cooperative game:}
The authors in \cite{scott2015multimedia} addressed the data rate allocation for multimedia users so as to maximize their individual application data rates. The users are selfish, and the non-cooperative game among them is applied. The model consists of users as players, i.e., buyers, and a network controller as the seller. The strategy of each user is to request data rate to maximize its utility. The utility of the players is a function of the allocated data rate and the price offered by the controller. Generally, if there are few users competing for the available resource, the price is low. Otherwise, there is high competition between the users, and the price for requesting high data rate is increased. This results in decrease of the users' utilities. The desired outcome of the game is the Nash equilibrium in terms of the optimal data rates of the users. However, there is no method for checking and proving the existence and uniqueness of the Nash equilibrium for this proposed scheme. 

In fact, the bargaining game \cite{muthoo1999bargaining} can be used for a fair data rate allocation. The bargaining game allocates data rates optimally to users so as to maximize the product of the users' utilities. However, this game requires the users to cooperate with each other which may be hard to achieve in practice. 


\textbf{Utility maximization:}
To achieve larger spectrum bandwidth, the mmWave band, e.g., 38 GHz, can be aggregated with existing bands, e.g., 3G band and 4G band. In this context, the authors in \cite{shajaiah2015efficient} investigated the aggregated carrier allocation to users to maximize their rates. The model consists of one BS and multiple users. First, the users as buyers submit their data rate requests to maximize their utilities. Then, the BS as the seller determines user groups, each of which is a set of users located in the coverage area of a carrier. The BS allocates the aggregated carriers to user groups in a descending order of the carrier frequencies. Specifically, the BS formulates the resource allocation of each carrier to the user group as the NUM problem. In the problem, the utility of each user is represented by the logarithm of the sigmoidal-like utility function of the total aggregated rate obtained from the allocated carriers. This function is strictly concave, and the globally optimal solution is obtained using the Lagrange method with the multiplier as the bandwidth price that the users pay the BS for receiving the rate. The simulation results show that the proposed scheme always converges to the optimal rates given the different available resources of carriers. Moreover, when the available resource of each carrier is low, the resource price paid by the users is high. The pricing strategy thus regulates user's demand and improves network resource efficiency. 

\textbf{Matching game:}
\begin{figure}[t!]
 \centering
\includegraphics[width=5.1cm, height = 3.8cm]{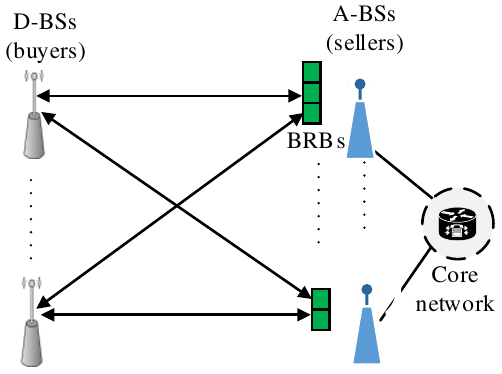}
 \caption{\small Matching between Demanding BSs (D-BSs) and Anchored
BSs (A-BSs) for Backhaul Resource Blocks (BRBs).}
 \label{mmWave_backhaul_network_matching}
\end{figure} 
The mmWave communications can provide wireless backhaul links between BSs as shown in Fig.~\ref{mmWave_backhaul_network_matching}. In this network model, the \textit{Demanding BSs} (D-BSs) are connected to the core network via the \textit{Anchored BSs} (A-BSs). Thus to serve their users, the D-BSs as buyers need to buy mmWave spectrum in terms of Backhaul Resource Blocks (BRBs) from the A-BSs serving as the sellers. The problem is to determine the number of BRBs of the A-BSs for each D-BS to minimize the D-BS's total cost while meeting the data rate requirements for the D-BS's users. 

In such a multi-seller multi-buyer scenario, the matching game \cite{jorswieck2011stable} is used to solve the problem as proposed in \cite{semiari2015matching}. First, each A-BS sets prices for its BRBs. Then, each D-BS ranks all BRBs in a descending order of utilities obtained from receiving BRBs. Here, the utility is the difference between the achievable rate and the BRB price. The D-BS selects for its preferred BRBs with the highest utilities. At each A-BS, if a preferred BRB receives more than one request from D-BSs, the A-BS selects the D-BS which maximizes the preferred BRB's utility as the winner. When the selected D-BS receives more BRBs, its aggregated rate and total cost also increase accordingly. This process is repeated until for every D-BS, either its aggregated rate is larger than its demand or its total cost is larger than its budget. The simulation results show that the proposed scheme can improve the average rate around 30\% compared with the best-effort approach adopted from \cite{gao2012deep}. However, the complexity of the proposed scheme is higher, especially when the number of D-BSs or BRBs is large.

\subsubsection{Massive MIMO networks}
\label{sec:App_SA_SE_MMM}

\begin{figure}[t!]
 \centering
\includegraphics[width=7cm, height=6.7cm]{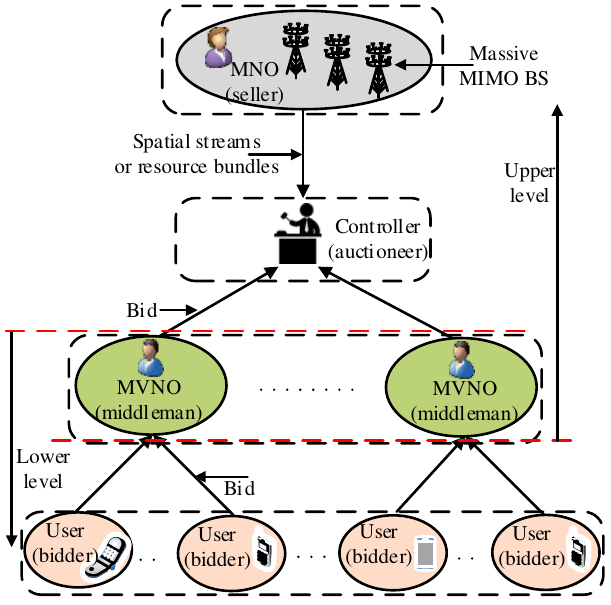}
 \caption{\small Spatial stream allocation based on auction schemes in a massive MIMO network.}
 \label{Auction_based_resource_MIMO}
\end{figure}
In massive MIMO networks, physical resources of BSs such as antennas and spectrum can be sliced into spatial streams using the spatial multiplexing. In this context, the authors in \cite{ahmadivirtualization} addressed the allocation of spatial streams of one MNO to others MNOs, called \textit{Mobile Virtual Network Operators (MVNOs)}, to maximize the data rates for the MVNOs. To incentivize the MVNOs to submit their actual valuation, the VCG auction is used. The model is shown in Fig.~\ref{Auction_based_resource_MIMO} in which the MNO is the seller and the MVNOs are the bidders. First, based on the number of its users, the MVNO computes a bid, i.e., the number of spatial streams to maximize its utility, i.e., the difference between the value and the price that the MVNO pays the MNO. Here, the value is proportional to the average achievable rate, and the price information is taken from the last auction. The winners and the payments are then determined according to the VCG auction. The simulation results show that the proposed scheme improves significantly the average data rate compared with the equally-divided spectrum allocation. However, compared with the optimal assignment, the proposed scheme has a slight performance loss because of the inaccurate estimate of the stream price that is taken from the last auction. Learning algorithms can be used to improve the performance. 

\subsubsection{C-RANs}
\label{sec:App_SA_SE_C_RAN}
 The C-RANs have the constraint on the fronthaul capacity. Subject to this constraint, the authors in \cite{ha2016resource} formulated an optimization problem which determines the transmission rates and the numbers of quantization data bits for all users to maximize the total transmission rate. In this model, the users as buyers consume the fronthaul bandwidth and cloud resources of a cloud provider serving as the seller. The optimization problem is a non-linear integer problem, and a \textit{pricing parameter} is introduced. The pricing parameter is actually the price of fronthaul bandwidth that the users pay the cloud provider. On one hand, the pricing parameter incentivizes the users to efficiently utilize resources. On the other hand, by updating the pricing parameter iteratively based on the binary searching method, the optimization problem is divided into independent subproblems which can be locally solved by the RRHs. The optimal transmission rates for the users are achieved by applying the Karush-Kuhn-Tucker (KKT) condition and the Lagrangian method. However, when the cloud resource is too low, the optimization problem in the proposed scheme may be infeasible.

\subsubsection{HetNets}
\label{sec:App_SA_SE_HetNets}
Rate maximization is also a major requirement in HetNets, especially in M2M networks which provide emergency services, e.g., road accident warning, with low latency. Economic and pricing models can provide rate allocation schemes to maximize the total rate of the M2M services under the network constraints. 

\textbf{Utility maximization:} Such an approach can be found in \cite{huang2013utility} which adopts the NUM problem to allocate data rates from an MBS as a seller to M2M services as buyers. The NUM problem is essentially similar to the utility maximization problem as discussed in the previous approaches, e.g.,\cite{shajaiah2015efficient}. However, the utility of each M2M service is calculated proportionally to the delay and the price that the M2M service must pay for the allocated rate. Therefore, the optimization problem is to minimize the sum of utility functions. By using the second derivative, the objective utility function is proved to be concave, and there is a unique optimal solution for the rate allocation and the price. However, using the Newton's method to obtain the optimal solution can result in slow convergence. 

\textbf{VCG auction:} 
\begin{figure}[t!]
 \centering
\includegraphics[width=7.4cm, height=2.7cm]{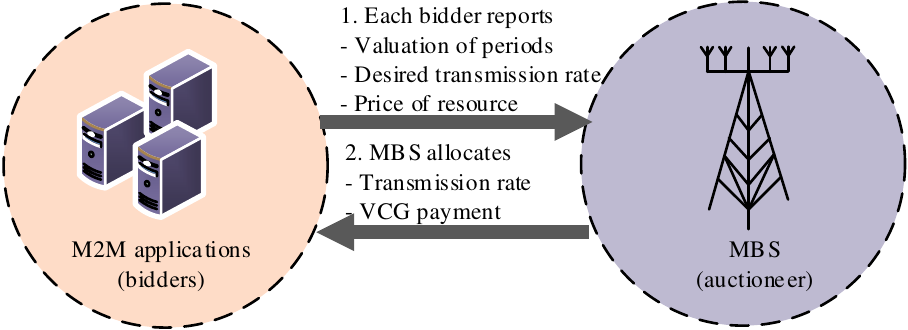}
 \caption{\small Rate allocation for M2M applications based on auction.}
 \label{M2M_communication_auction}
\end{figure}
In practice, different M2M applications may have different valuations of allocated data rates on the same period. For example, at the traffic peak hours, the valuation of an M2M application in a vehicular network may be much higher than that of an M2M application to monitor environments. The auction schemes can be adopted to allocate the data rates to the M2M applications with the highest valuations. The approach based on the VCG auction is presented in \cite{lin2014multi}. The model is shown in Fig.~\ref{M2M_communication_auction} in which each M2M application as a buyer submits its bid to an MBS serving as an auctioneer. The bid includes parameters such as the M2M application's valuation, data rate requirement, and the price that the application is willing to pay. The MBS selects the winners that maximize the social welfare and charges each winner according to the VCG payment strategy for a truth-telling valuation report. The simulation results show that the transmission success probability can reach up to 97\%. However, the assumption that the data rate of each M2M application is constant in each time slot is not so realistic.

To improve further the transmission success probability of the M2M applications, their data rate demands should be predicted. The authors in \cite{lin2015auction} introduced an estimation stage through the maximum likelihood method before conducting the VCG auction. The simulation results show that the transmission success probability can reach up to 99\%, i.e., improved by additionally 2\% compared with that in \cite{lin2014multi}. 

\textbf{Option pricing:} The M2M applications can reserve data rates by using the option pricing \cite{cox1979option} as proposed in \cite{lee2012optimal}. In the scheme, each M2M application as a buyer reserves the resource, i.e., data rate, at a pre-specified price and a pre-defined time via a call option contract. The contract will protect the buyer against a higher resource price in the future, and also provide for the buyer the right to purchase the requested data rate even when the buyer loses in the current auction round. Moreover, a convex cost function is introduced which allows the buyers to choose the strike price, i.e., the price at which the call option can be exercised, to minimize their reservation costs while obtaining the QoS in terms of guaranteed data rate. Under the proposed scheme, the data rates can be allocated to the M2M applications with the maximum utilization. The Vickrey auction is then applied for the rate allocation to achieve the truthfulness. 

\subsection{Revenue Maximization}
\label{sec:App_SA_RM}
To stimulate network operators to share spectrum, improving their revenue is a major motivation of the spectrum management. This subsection discusses the applications of economic and pricing models for the spectrum management to maximize the network operators' revenue in 5G. Different economic and pricing models can be applied to different technologies, e.g., mmWave, massive MIMO, and HetNets. For example, in mmWave networks which have the large bandwidth to serve a number of users, the economic terms ``network size'' or ``network effect''  are introduced to improve and quantify the revenue of the network operators. In massive MIMO networks, auctions are adopted to allocate the bundle of spectrum and antennas to enhance the resource efficiency while guaranteeing the highest revenue of the network operators.
\subsubsection{MmWave networks}
\label{sec:App_SA_RM_GP}
Economic and pricing approaches for the mmWave spectrum sharing between the network operators are reviewed in this section.

\textbf{General pricing:}
The authors in \cite{gupta2016gains} addressed the issue of mmWave spectrum sharing between a Primary Operator (PO) and a Secondary Operator (SO). The PO which owns an exclusive-use license of a certain spectrum band sells a license of the same band to the SO under a certain restriction. The restriction means that the SO's network serves its users without violating an interference threshold at the PO's network. The SO pays the PO a license price which is linear in the sum rate that the SO receives. Note that this sum rate depends on the interference threshold set by the PO. Increasing the threshold allows the SO's network to increase its sum rate, but this also decreases the sum rate of the PO's network. In this case, theoretically, the revenue of the PO decreases due to the reduction of the sum rate for its own users. However, the simulation results show that the PO's revenue does not always decrease with an increase in the threshold. The reason can be that the payment which the PO receives from the SO is more than that the PO loses from its own users. The simulation results also show that there exists an interference threshold that maximizes the PO's revenue, meaning that the PO has an incentive to share its spectrum. However, a general system model with multiple POs and SOs needs to be investigated. 

\textbf{Non-cooperative game:}
Similar to \cite{gupta2016gains}, the authors in \cite{fund2016spectrum} considered mmWave spectrum sharing between the two Service Providers (SPs). The objective is to quantify the SPs' profits with or without sharing the spectrum. Here, each SP's profit is a function of the number of its users, i.e., the network size, the service quality level that the SP chooses, and the price that the SP sets. In particular, the network size is determined by introducing a so-called \textit{network effect} into the utility of each user. The SPs are competitive, and the non-cooperative game is adopted in which the SPs simultaneously set prices and choose service quality levels to maximize their profits. It is proved that if the ratio of service quality levels between the two SPs is larger than a certain threshold, there is a unique Nash equilibrium of the game at which both SPs set prices higher than their marginal costs and earn a non-zero profit. The profits of the SPs at the Nash equilibrium are finally determined. The simulation results show that the SPs' profits with spectrum sharing are significantly improved compared with those of the case without the sharing. However, in the real competitive markets, the SP which chooses the high service level may prefer to share spectrum only when the intensity of the network effect is small.

\subsubsection{Massive MIMO networks}
\label{sec:App_SA_RM_NG}
In massive MIMO networks, the MNO trades both spectrum and antennas of a BS as a bundle to the users to satisfy their dynamic QoS requirements. Auctions allow the users to express preferences over bundles of physical resources, i.e., the spectrum and antennas, while guaranteeing the highest revenue for the MNO.  

\textbf{Ascending Clock Auction (ACA):}
The ACA as presented in Section~\ref{subsec:Auction_ACA} is often used for the resource allocation due to its simplicity which allows bidders to quickly discover the resource prices in complex markets. The authors in \cite{ahmadi2016substitutability} adopted the ACA for the allocation of the resource bundles. The model is shown in Fig.~\ref{Auction_based_resource_MIMO} which includes one MNO as the seller, multiple MVNOs as bidders, and one auctioneer. First, the auctioneer broadcasts the spectrum price and the antenna price to the MVNOs. Given the prices, the MVNOs determine and submit their bids to the auctioneer. Each bid of an MVNO includes the amount of spectrum and the number of antennas that satisfy a minimum rate requirement of its users while minimizing the cost. At the auctioneer, if there is an excess demand for the spectrum in the current round, the auctioneer increases the posted spectrum price in the next round. This process is repeated until the spectrum supply equals the spectrum demand. The auctioneer then selects bids through formulating and solving the WDP in which bids are selected so as to maximize the auctioneer's revenue. The simulation results show that a lower antenna price leads to the higher overall revenue from selling the spectrum. The reason is from the complementarity of the antennas and the spectrum. In particular, a low antenna price leads to the increase of the spectrum demand and the revenue. However, the proposed scheme has high computational complexity due to the complexity of the branch-on-bids algorithm in the WDP.

The ACA approach can also be found in \cite{mcmenamy2016enhanced}, but the spectrum is channelized into blocks each with 5MHz to be compatible with the existing standards, e.g., the 4G LTE. Moreover, the Licensed Shared Access (LSA) is used to improve the spectrum utilization for the MVNOs. The LSA is the European spectrum-sharing framework which allows MVNOs as LSA licensees to share the allocated spectrum with the MVNOs' users, called \textit{incumbents}, on an exclusive basis \cite{irnich2013spectrum}. This means that each MVNO has the rights to access the spectrum that is unused by an incumbent at certain locations and times. The information about the spectrum and its use by the incumbent is updated by the LSA
repository. The ACA approach from \cite{ahmadi2016substitutability} is then applied to assign bundles of resources, i.e., the numbers of antennas and blocks, to the MVNOs. 

\textbf{Utility maximization:}
To gain revenue for the MNO and to reduce the payments of the users, the presence of the MVNOs which are the intermediaries may not be required. The MNO can thus allocate directly bundles of physical resources at each massive MIMO BS to the users as proposed in \cite{jumba2015resource}. The users are first divided into groups, each of which has a minimum required rate. To maximize the total utility of user groups, the NUM problem is formulated. Each group's utility is the difference between the total achievable rate of users in the group and the total price that the users in the group pay the MNO for receiving the resource bundles. The problem is then solved by applying the Lagrange multiplier method and the KKT conditions. In particular, the power and the number of antennas allocated to each user in each iteration is inversely proportional to the power price and the antenna price, respectively. The simulation results show that when these prices increase, the total rate decreases. Moreover, the effect of increasing the antenna price is more considerable than that of the power price. The reason is that in the massive MIMO network, the maximum allowable number of allocated antennas for each user can be much larger than that of allowed power units for the user. 

\subsubsection{C-RAN}
\label{sec:App_SA_RM_TP}
\begin{figure}[t!]
 \centering
\includegraphics[width=6.5cm, height=4.3cm]{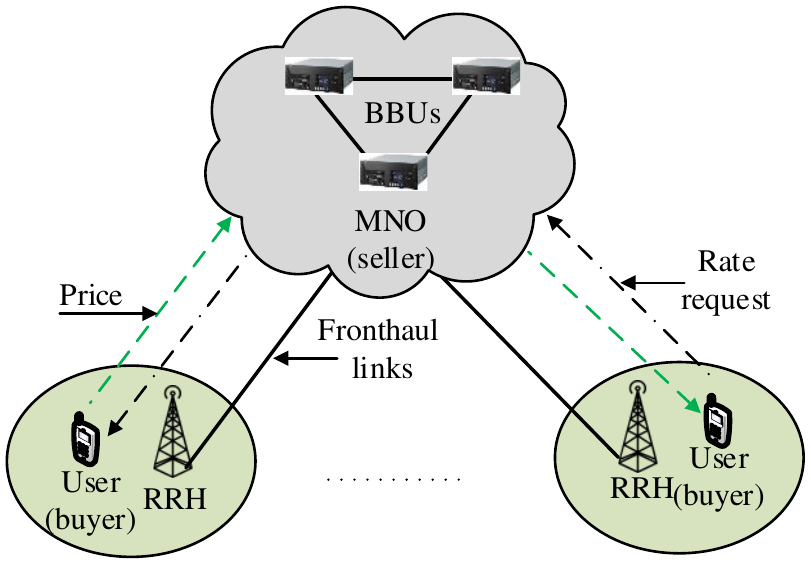}
 \caption{\small  Tiered pricing-based channel allocation in C-RANs.}
 \label{CRAN_V_RAN}
\end{figure}
A large number of RRHs are deployed in the C-RAN, and the load balancing among the RRHs is an important requirement in addition to the revenue maximization of the MNOs. Pricing models such as tiered pricing can be used to meet both requirements.

\textbf{Tiered pricing:} The authors in \cite{niu2016dynamic} adopted the tiered pricing for the spectrum, i.e., channel, allocation in the C-RAN as shown in Fig.~\ref{CRAN_V_RAN}. First, the users as buyers send their QoS requirements to the MNO as the seller. Based on the QoS information received, the MNO determines the prices and then the channel allocation for the users to maximize its revenue. Here, the price is set using a piecewise function \cite{herrera2007theory} in which the price for the user is low/high if the user requires a range of low/high data rates. This results in a balanced resource allocation among the users. The Increment-Based Greedy Allocation (IBGA) \cite{zhu2009objective} is then applied to allocate the channels to the RRHs one by one. Specifically, in each iteration, the MNO assigns the channel to the RRH which provides the highest increment of revenue. The allocation of the channel continues until no more RRH can use it. Afterwards, the user association issue for each RRH is addressed based on the minimum rate guarantee for the RRH's users. The simulation results show that the proposed scheme can achieve the same performance as the standard high-complexity scheme, i.e., the branch and bound method \cite{beraldi2002branch}, in terms of system throughput. However, how the MNO's revenue improves is not demonstrated. 


\textbf{Knapsack problem:}
To satisfy the processing demands of a massive number of users, multiple clouds can be deployed for the C-RAN. The problem is how to assign users to the clouds to maximize the profits of cloud providers. To solve the problem, the authors in \cite{dahrouj2015distributed} adopted the knapsack problem \cite{chu1998genetic}, i.e., a combinatorial optimization for efficiently allocating resources, e.g., spectrum or cloud, to users given the cost budget. Each cloud provider pays its users penalty costs if the QoS performance cannot be guaranteed. Therefore, the net profit function of the cloud provider serving a user is the difference between the price that the user pays and the penalty cost. The optimization problem is then solved by using the full polynomial time approximation scheme \cite{martello1990knapsack} to find an optimal set of users for the cloud provider. In practice, the cloud provider can increase the penalty cost for attracting more users and iteratively execute the algorithm to find a new optimal set of users. Some users in the previous optimal set may remain in the new one. These users show their mutual interests in the cloud provider. Thus the cloud provider may set the penalty costs for them to zero before executing again the algorithm to maximize its net profit. However, the high computational complexity is incurred. 

\subsubsection{HetNets}
\label{sec:App_SA_RM_GG}
A HetNet consists of multiple tiers, and one of the important issues is how to allocate the spectrum to the tiers to maximize the MNOs' revenue. Pricing models based on game theory and auctions are used to make appropriate investment decisions for the MNOs. 

\textbf{General game:}
\begin{figure}[t!]
 \centering
\includegraphics[width=5.8cm, height=5.6cm]{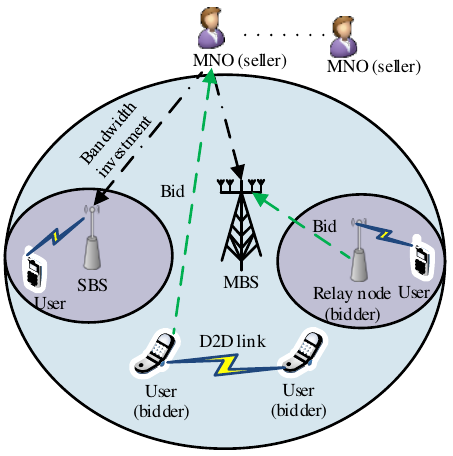}
 \caption{\small Pricing approaches for spectrum allocation in HetNets}
 \label{HetNet_Spectrum_allocation}
\end{figure}
A two-stage game for the spectrum allocation in the HetNet can be found in \cite{chen2015bandwidth}. The network model is shown in Fig.~\ref{HetNet_Spectrum_allocation} which involves multiple MNOs as sellers and multiple users as buyers. Each MNO owns one MBS and one SBS. For each MNO, the first stage of the game determines the amount of bandwidth for trading at its MBS and SBS, and the second stage determines access prices for the MBS and SBS to maximize the MNO's revenue. Given the access prices, each user selects the MNO to maximize the user's utility, i.e., the valuation of its achievable rate minus the access price. The access prices are iteratively adjusted by the MNOs so that the total demand equals the total supply. The simulation results with two MNOs show that with limited bandwidth, both MNOs only allocate bandwidth to SBSs. The reason is that when the bandwidth is limited, the access prices charged to MBSs and SBSs are very high. The MNOs would thus invest all bandwidth to the SBSs which may result in more data rate and hence higher revenue. However, when the MNOs have more bandwidth, investing only in SBSs would significantly decrease the access prices for the SBSs and reduce revenue of the MNOs. Thus the MNOs will invest the bandwidth to both MBSs and SBSs. These results are useful for network operators to make appropriate investment decisions. However, how to adjust the access prices as well as to prove the existence of equilibriums of the game is not specified. 

\textbf{Forward auction:}
When one MNO decides to allocate the bandwidth to multiple SBSs, the forward auction can be adopted to enable the MNO to achieve the highest revenue as proposed in \cite{chen2017auction}. In this model, one MNO is seller, i.e., the auctioneer, and multiple Femto Holders (FHs) which own SBSs are buyers. Based on bandwidth requests of its subscribers, each FH calculates the bandwidth demand and the rental price to maximize its utility, i.e., profit. The FHs then submit the information of bandwidth demands and rental prices to the MNO. Based on the information, the MNO determines the winning FHs which maximize its revenue by solving the knapsack problem. Assume that the FHs submit the same rental price, the simulation results show that there exists a rental price at which their profit is maximized. However, the profit maximization may not be held in the long term since the MNO as the monopolist can re-determine the winners to maximize its own revenue rather than the FHs' profit. 

To gain more revenue, the MNO can also trade unused spectrum to D2D users. In this context, the authors in \cite{swainspectrum} applied the forward auction with multiple rounds for the spectrum sharing between the MNO, i.e., the seller, and the D2D users, i.e., the bidders. In each round, each D2D user submits its bid to specify the amount of bandwidth, the sub-band, and the price that the D2D user is willing to pay. The D2D user with the highest price is the winner. This process is repeated until all bandwidth is sold. To minimize the interference caused by the D2D users, the MNO also allocates a transmit power on each sub-band when allocating the sub-band to each D2D user. 

Different from \cite{swainspectrum}, the authors in \cite{koseoglu2015smart} considered the spectrum allocation for the M2M networks which will support a large number of M2M devices. This may create a significant congestion at the radio access of the MBSs. Thus the congestion problem should be considered in addition to the revenue. The congestion-based pricing can be adopted to address the problem. The model consists of one MBS as a seller which allocates access channels to the M2M users as buyers. The MBS adjusts the access price based on the channel status, i.e., underutilized or overutilized, in the previous access period. Generally, the step sizes for adjusting price in different access periods may not be fixed to avoid the fluctuations of the channel utilization. For example, if the channel is found to be underutilized in two consecutive access periods, the MBS increases the price by a fixed step size. If the channel is overutilized in the previous period but underutilized in the current period, the step size is reduced to its half to prevent instability. Such a pricing scheme alleviates the congestion as well as the low access delay for the users. However, the revenue maximization for the MBS may not be guaranteed. 

Instead of addressing the congestion problem, the authors in \cite{koseoglu2016pricing} sets the access price in each period to maximize the MBS's total revenue. The total revenue is determined based on the probability of successful access of each user and the access price. The simulation results show that the pricing scheme proposed in \cite{koseoglu2016pricing} improves the MBS's revenue up 50\% compared with that in \cite{koseoglu2015smart}. 

In addition to the SBSs and D2D/M2M users, the relay nodes are deployed in HetNets to improve the coverage at cell edges of the network. The authors in \cite{zhang2017auction} adopted the forward auction to allocate the spectrum of one MBS to the relay nodes. In this model, the relay nodes are the bidders, the MBS is the seller, and a controller is the auctioneer. The relay nodes submit their bids to the controller. Each bid specifies the number of RBs that the relay node requests and the price that the user is willing to pay the MBS. Upon receiving the bids, the controller determines the winners via calculating the ratio of the price to the number of requested RBs of each bidder. The bidder with the highest ratio is selected as the winner, and the winner selection for the remaining unallocated bidders is repeated until there is no RB at the MBS. Such a winner selection process guarantees the highest revenue for the MNO and resource efficiency. Then, using a binary search algorithm, the controller calculates \textit{critical prices} that the winners pay the MBS. Since the critical price is less than its bidding price, the proposed scheme guarantees the truthfulness. However, the interferences between the MBS-relay link and the relay-user link because of using the in-band relaying solution are not considered in the proposed scheme.  

\subsection{System Utility Maximization}
\label{sec:App_SA_SUM}
The aforementioned sections, i.e., Sections \ref{sec:App_SA_SE} and \ref{sec:App_SA_RM}, discuss the applications of the economic and pricing models for the spectrum allocation to improve either data rates for the users or revenue for the network operators. To incentivize both the users and the network operators to participate in a resource market, their utilities, e.g., in terms of data rates and revenue, need to be simultaneously improved. Given the rationality of stakeholders in 5G, economic and pricing models can handle efficiently the conflicting objectives of all stakeholders while minimizing the information exchange. This section reviews the applications of economic and pricing models for the spectrum allocation for the \textit{system utility maximization}. Here, the system utility maximization refers to maximizing the utilities of stakeholders or the sum of their utilities, i.e., the social welfare. 


\subsubsection{HetNets}
\label{sec:App_SA_SUM_HN}
\begin{figure}[t!]
 \centering
\includegraphics[width=4cm, height = 6.2cm]{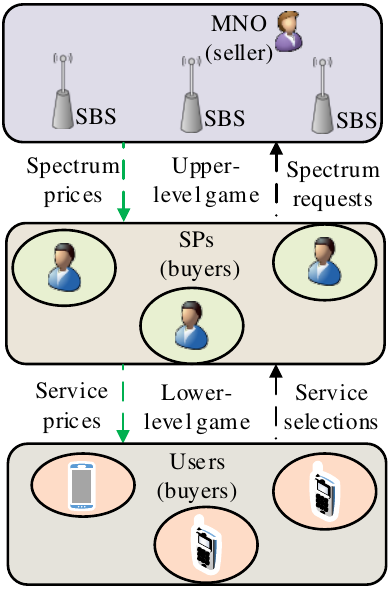}
 \caption{\small Spectrum allocation in an ultra-dense small cell network based on hierarchical game.}
 \label{User_Association_Hierarchical_Game}
\end{figure}
The authors in \cite{chang2015service} adopted a two-level game to maximize the utilities of stakeholders in a HetNet. As shown in Fig.~\ref{User_Association_Hierarchical_Game}, the SPs lease spectrum at SBSs owned by one MNO to provide network services to the SPs' users. The two-level game is based on the Stackelberg game described as follows.

In the lower-level game, the SPs are the leaders, and the users are the followers. Each user chooses a service from an SP to maximize the user's utility, and then the SP autonomously sets prices for its services so as to maximize its utility, i.e., its profit. Note that due to a large number of users, modeling the strategic interaction among individual users is not feasible. Thus the competition among users is modeled as a non-atomic game \cite{aumann2015values}, and the SPs' competition is modeled as a non-cooperative game. The equilibrium solution for the users is obtained by applying the Krasnoselskii algorithm \cite{berinde2007iterative}, and that for the SPs is obtained as the Nash equilibrium. In the upper-level game, the Stackelberg game is used again with the MNO as the leader and the SPs as the followers. Based on the MNO's spectrum prices and users' service requests, the SPs decide spectrum leasing policies to maximize their own utilities. Given the leasing decisions of SPs, the MNO determines the spectrum prices to maximize its utility, i.e., profit. The optimal strategies of the SPs and the MNO constitute the Stackelberg equilibrium. This is the first work that investigates the interactions among the users, SPs, and MNO jointly. The case with multiple MNOs needs to be considered in the future work. However, it is challenging to prove the convexity of the SPs' best responses in such a case.

Similar to \cite{chang2015service}, the authors in \cite{sunauction} adopted the Stackelberg game to maximize the utilities of stakeholders in a multi-tier HetNet. The stakeholders includes one FH as a seller and two competitive network operators, i.e., the MNO and MVNO, as buyers. The FH owns multiple SBSs which provide RBs to the users of the MNO and MVNO. Before the Stackelberg game is implemented, a reverse auction is applied which allows the network operators to select the SBSs with the lowest payment. Since some SBSs may be selected by both the network operators, the Stackelberg game is used to allocate the RBs at the selected SBSs to the network operators. Here, the network operators are the leaders, and the FH is the follower. First, the leaders determine the prices that they pay the FH to maximize the leaders' profits. Then, the FH determines the number of the RBs at the selected SBSs for each leader to maximize the FH's utility, i.e., revenue. The simulation results show that the total utility of stakeholders, i.e., the network operators and the FH, of the proposed scheme is much higher than that of the scheme without sharing the resources. Additionally, the total price that the network operators pay the FH decreases when the SBS density is high. The reason is due to the high competition among the SBSs. However, this may not be an important issue in practice because SBSs owned by the same FH do not compete with each other. 
 
\begin{figure}[t!]
 \centering
\includegraphics[width=6cm, height = 5.6cm]{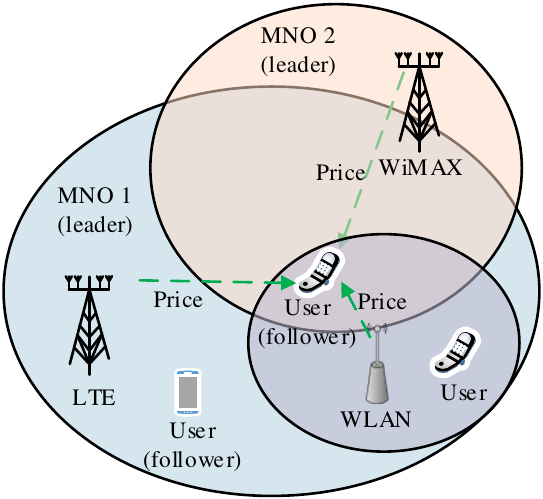}
 \caption{\small Stackelberg game-based spectrum allocation in HetNets.}
 \label{multi_tier_bandwidth_allocation_Stackelberg}
\end{figure}

The Stackelberg game can be used to maximize the utilities of MNOs and users in a multi-RAT environment of the HetNets as proposed in \cite{zhang2015utility}. The model is shown in Fig.~\ref{multi_tier_bandwidth_allocation_Stackelberg} in which the MNOs are the leaders, and the users are the followers. Each MNO as a seller owns multiple RATs, and each user as a buyer can access simultaneously to more than one RAT. Given MNOs' bandwidth prices, each user determines the amount of bandwidth of the chosen RATs to maximize its own utility. The Lagrange function and the projected gradient method are applied to achieve the optimal solutions for the users. Given the users' best responses, the MNOs compete with each other on the bandwidth prices to maximize their utilities, i.e., revenues. The optimal prices for each MNO is obtained by taking the first derivative of its utility function. The simulation results show that the utility of the MNO raises as it increases the bandwidth price. However, when the price is too high, the MNO's revenue decreases since the users tend to access the RATs of other MNOs. In fact, the competition level among the MNOs may vary in different areas depending on the number of MNOs in the areas, which can be taken into account in the future work. 

Different from \cite{zhang2015utility}, the users in \cite{yun2015energy} are assumed to cooperate with each other. Accordingly, given the bandwidth prices of the MNO, the users determine their optimal data rates to maximize the sum of their utilities. The problem of the users is proved to be convex, and the Lagrange function and the KKT conditions are applied to find optimal solutions for the users. Given the optimal data rates, the MNO determines bandwidth prices for the RATs to maximize its utility, i.e., revenue, using the greedy pricing algorithm \cite{neely2009optimal}. This pricing strategy aims to maximize the social welfare while ensuring that all participants make non-negative profit. However, the computational time for finding the optimal prices depends heavily on the number of RATs, and it can be unacceptably long with many RATs in the network.

\subsubsection{C-RAN}
\label{sec:App_SA_SUM_CRAN}
C-RANs often have limited fronthaul bandwidth, and the problem is how one MNO (i) selects users/SPs which value the resource the most and (ii) allocates the resource to them to achieve the highest social welfare. The VCG auction is used to solve the problem as proposed in \cite{gu2016virtualized}. First, each SP, i.e., a bidder, submits a bid to the MNO, i.e., the auctioneer. The bid specifies the price that the SP is willing to pay and the certain amount of resources, i.e., the spectrum at RRHs and the fronthaul bandwidth. Upon receiving the bids, the MNO formulates the WDP to maximize the total utility of the MNO and the SPs. The MNO's utility is its total payment, and the SP's utility is the difference between the true valuation of its selected price and its actual payment. The WDP is solved by the decomposition technique in \cite{lavi2011truthful}, and the prices charged to the winners are determined based on the VCG payment policy. The simulation results show that the proposed scheme outperforms the greedy algorithm in terms of social welfare. Moreover, the social welfare increases as the number of winners and the capacity ratio are large. However, the large capacity ratio results in reduced revenue of the MNO due to abundant resources in the C-RAN system. 

After receiving the resources from the MNO, the SPs allocate them to users. This is a hierarchical model with two levels as shown in Fig.~\ref{User_Association_Hierarchical_Game}. In this scenario, the authors in \cite{morcos2017two} adopted two auction schemes at the two levels for Resource Block (RB) allocation. Generally, the higher-level auction uses the share auction \cite{huang2010game} which allocates the RBs of the MNO to the SPs by using the proportional fair allocation. Accordingly, the number of RBs that one SP receives is proportional to the price that the SP is willing to pay. Then, each SP allocates the RBs to its own users using the VCG auction. The greedy algorithm is applied to determine the winning users and the corresponding prices that the users pay. Using the greedy algorithm and the share auction enables the lower- and higher-level auctions to achieve the high computational efficiency. However, as shown in the simulation results, the uniqueness of the equilibrium in the higher-level auction is not always guaranteed due to the lack of concavity of the utility, i.e., profit, function of the SPs. 

\subsubsection{Massive MIMO networks}
\label{sec:App_SA_SUM_MMM}
In massive MIMO networks, the combinatorial auction is efficiently used to allocate a bundle of resources, i.e., antennas and spectrum, to users. To maximize the social welfare, the combinatorial auction can be combined with the VCG payment policy. 

\textbf{Combinatorial auction:} 
The integration of the two approaches, i.e., the combinatorial auction and the VCG auction, can be found in \cite{zhu2015virtualization}. The network model consists of one MNO, multiple MVNOs, and multiple users. The auction is implemented in two levels as shown in Fig.~\ref{Auction_based_resource_MIMO}. In the upper level, each MVNO as a bidder submits a bid including a bundle of subchannels and antennas as well as its valuation to the MNO. The MNO determines the winning bidders so as to maximize the sum of bids. The dynamic programming algorithm \cite{bertsekas1995dynamic} is applied to obtain an optimal solution. After receiving the resources from the MNO, in the lower level, each MVNO as the auctioneer executes the combinatorial auction to allocate bundles of resources to the MVNO's users, i.e., bidders. The process of the combinatorial auction is similar to that in the upper level. The VCG payment policy is adopted in both levels to charge the winning bidders. The simulation results show that the proposed scheme outperforms the fixed sharing scheme in terms of social welfare and average subchannel utilization. However, using the dynamic programing algorithm leads to high computational complexity. Lower complexity algorithms such as the polynomial-time greedy algorithm \cite{zaman2013combinatorial} can be used. However, the greedy algorithm only achieves an approximate optimal solution based on which the VCG payment policy will be no longer incentive compatible.

\textbf{Double auction:}
When there are multiple MNOs, the double auction can be used as proposed in \cite{zhangdouble}. In this model, the MNOs as sellers sell data rates to MVNOs as buyers. Different from the classical double auction as presented in Section~\ref{subsec:Auction_forward_reverse_double_auction}, the deal price of each seller-buyer pair is a pre-negotiated price between them. The problem is to find optimal data rates requested by the MVNOs and those offered by the MNOs to maximize the social welfare. First, the KKT conditions are applied to the problem to obtain the optimal dual variables, called \textit{shadow prices}. The shadow prices ensure the desirable economic properties as well as the total welfare. Based on the shadow prices and the pre-negotiated prices, each MVNO determines an optimal data rate for each of its user, and each MNO also calculates the optimal data rate that it offers to the user. The MNOs and MVNOs then submit their asks and bids including the optimal data rates to a broker, i.e., a Software-Defined Networking (SDN) controller. The broker checks some convergence conditions, e.g., the requested data rate is less than the offered data rate. If the conditions are not satisfied, the broker updates the shadow prices using the subgradient descent method, and subsequently the MVNOs and the MNOs update and submit again their bids/asks. This process is repeated until the convergence conditions are met. The simulation results show that the proposed scheme outperforms the random allocation in terms of social welfare. However, how to determine the pre-negotiated prices for seller-buyer pairs is not explained, which is practically challenging in the large-scale 5G.

\begin{table*}
\caption{Applications of economic and pricing models for spectrum allocation in 5G (MWN: Millimeter Wave Network, CRAN: Cloud-Radio Access Network, MMN: Massive MIMO Network, HN: Heterogeneous Network).}
\label{table_spectrum_allocation}
\scriptsize
\begin{centering}
\begin{tabular}{|>{\centering\arraybackslash}m{0.6cm}|>{\centering\arraybackslash}m{0.4cm}|>{\centering\arraybackslash}m{1.4cm}|>{\centering\arraybackslash}m{1cm}|>{\centering\arraybackslash}m{1cm}|>{\centering\arraybackslash}m{1.1cm}|>{\centering\arraybackslash}m{7cm}|>{\centering\arraybackslash}m{1.6cm}|>{\centering\arraybackslash}m{0.8cm}|}
\hline
\multirow{2}{*} {\textbf{}} & \multirow{2}{*} {\textbf{Ref.}} & \multirow{2}{*} {\textbf{Pricing model}} & \multicolumn{3}{c|} {\textbf{Market structure}} & \multirow{2}{*} {\textbf{Mechanism}}& \multirow{2}{*} {\textbf{Solution}} & \multirow{2}{*} {\textbf{Network}} \tabularnewline
\cline{4-6}
 & & & \textbf{Seller} & \textbf{Buyer} & \textbf{Item} & &&\tabularnewline
\hline
\hline
\parbox[t]{2mm}{\multirow{9}{*}{\rotatebox[origin=c]{90}{ \hspace{-3.3cm} Data rate}}}
\parbox[t]{2mm}{\multirow{9}{*}{\rotatebox[origin=c]{90}{ \hspace{-3.3cm}maximization}}}

&\cite{shajaiah2015efficient} &Utility maximization&BS&Users&Aggregated carriers&The rate allocation problem is formulated as the NUM problem. The global optimal solution is obtained by applying the Lagrange method. &Optimal solution&MWN  \tabularnewline \cline{2-9}

&\cite{scott2015multimedia} &Non-cooperative game&Network controller&Users&Transmission rate& Given the resource price offered by the network controller, the Nash equilibrium transmissions rates for the users are determined.&Nash equilibrium&MWN \tabularnewline \cline{2-9}

&\cite{semiari2015matching} &Matching theory&Anchored-BSs&Demanding-BSs&Backhaul resource blocks& Given resource prices offered by the anchored-BSs, a two-sided assignment scheme is applied to assign each anchored-BS to each demanding-BS.&Competitive
equilibrium& MWN \tabularnewline \cline{2-9}

&\cite{ahmadivirtualization}& VCG auction& MNO&MVNOs& Spatial streams&MVNOs compute demands and submits it to the MNO. The MNO selects the winners and charges them according to the VCG auction.&Bayesian Nash equilibrium&  MMN\tabularnewline \cline{2-9}

&\cite{ha2016resource}& General pricing& Cloud provider&Users& Fronthaul bandwidth and cloud resources& The optimal transmissions rates for the users are determined by using the penalty method, binary searching method, and KKT condition.&Optimal solution& CRAN \tabularnewline \cline{2-9}

&\cite{huang2013utility}& Utility maximization& MBS&M2M services& Data rates& The optimal transmission rates and price for the services are determined by applying the Newton's method.&Optimal solution& HN \tabularnewline \cline{2-9}

&\cite{lin2014multi}&VCG auction& MBS&M2M applications& Data rates& The winners and their charges are determined based on the VCG auction.&Bayesian Nash
equilibrium& HN \tabularnewline \cline{2-9}

&\cite{lee2012optimal}&Vicrkey auction and option pricing& MBS&M2M applications& Data rates& M2M applications reverse data rates via the option pricing and Vicrey auction is adopted for the rate allocation.&Nash
equilibrium& HN \tabularnewline \cline{2-9}

\hline
\parbox[t]{2mm}{\multirow{9}{*}{\rotatebox[origin=c]{90}{ \hspace{-4cm} Revenue}}}
\parbox[t]{2mm}{\multirow{9}{*}{\rotatebox[origin=c]{90}{ \hspace{-4cm} maximization}}}

&\cite{gupta2016gains}&General pricing& Primary operator&Secondary operator&Spectrum&Secondary operator receives the sum rate from primary operator and pays the primary operator a price which is linear in the sum rate. &Optimal solution& MMW \tabularnewline \cline{2-9}

&\cite{fund2016spectrum}&General pricing&Service providers&Users&Spectrum&Service providers quantify their profits based on the network effect, the users' willingness to pay, the price and the inherent quality offered by the service providers. &Nash equilibrium&MMW  \tabularnewline \cline{2-9}

&\cite{ahmadi2016substitutability}& Ascending clock
auction& MNO&MVNOs& Spectrum and antennas&MVNOs submit their bids for the resources. The MNO adjusts the resource prices according to the law of supply and demand. The winner selection problem is then solved by using the branch-on-bids. &Walrasian
equilibrium & MMN \tabularnewline \cline{2-9}

&\cite{mcmenamy2016enhanced}& Ascending clock
auction& MNO&MVNOs& Spectrum and antennas&Same as \cite{ahmadi2016substitutability}, but the LSA framework is used to allow the MVNOs to access the spectrum that is unused by their users. &Walrasian
equilibrium & MMN \tabularnewline \cline{2-9}

&\cite{jumba2015resource}& General pricing& MNO&Users& Sub-carrier and antennas&Users are divided into groups, and the problem is to maximize the total utility of all groups. The Lagrange multiplier and the KKT conditions are applied to solve the problem. &Optimal solution & MMN \tabularnewline \cline{2-9}

&\cite{niu2016dynamic}& Piecewise function-based pricing& MNO&Users&Channels& MNO sets prices for users using the piecewise function. Then, the MNO formulates its revenue maximization problem which is solved by the IBGA algorithm.&Optimal solution&CRAN \tabularnewline \cline{2-9}


&\cite{chen2017auction} &Forward auction&MNO&FHs&Resource blocks&MNO determines the winning FHs by solving the knapsack problem. Then, the winners pay the MNO the submitted prices.&Optimal solution &HN\tabularnewline \cline{2-9}

&\cite{swainspectrum} &Forward auction&MNO&D2D users&Bandwidth&The MNO selects the D2D user with the highest bid as the winner. This process is repeated for the remaining D2D users until all bandwidth is sold.&Optimal solution &HN\tabularnewline \cline{2-9}

&\cite{koseoglu2016pricing}&General pricing&MBS&M2M users&Access channels&MBS sets an access price so as to maximize its revenue. &Optimal solution &HN\tabularnewline \cline{2-9}

&\cite{zhang2017auction} &Forward auction&MBS&Relay nodes&Resource blocks&Winning relay nodes are determined based on their bids. Then, the binary search algorithm is used to determine the charges for the winners.&Optimal solution &HN\tabularnewline \cline{2-9}

\hline
\parbox[t]{2mm}{\multirow{9}{*}{\rotatebox[origin=c]{90}{ \hspace{-5cm} System utility}}}
\parbox[t]{2mm}{\multirow{9}{*}{\rotatebox[origin=c]{90}{ \hspace{-5cm} maximization}}}

&\cite{chang2015service} &Hierarchical game&MNO&Service providers and users&Network services and spectrum&The Stackelberg game at the lower-level game determines the optimal service prices for the service providers and the optimal service selections for the users. The Stackelberg game at the upper-level game determines the optimal spectrum leasing policies for the service providers and the optimal spectrum prices for the MNO.&  Stackelberg equilibrium& HN\tabularnewline \cline{2-9}

& \cite{sunauction} &Stackelberg game&FH&MNO and MVNO&Resource blocks& Each buyer, i.e., MNO or MVNO, selects a set of SBSs of the buyer using the reverse auction. The resource block allocation to the buyers is then based on the Stackelberg game. &Stackelberg equilibrium & HN \tabularnewline \cline{2-9}

&\cite{zhang2015utility} &Stackelberg game&MNOs&Users&Bandwidth& The Lagrange function and the projected gradient method are applied to determine the optimal amounts of bandwidth for the users. Then, optimal bandwidth prices for the MNOs are obtained by taking the first derivation of their utilities. &Stackelberg equilibrium& HN\tabularnewline \cline{2-9}

&\cite{yun2015energy} &Stackelberg game&MNO&Users&Bandwidth& The Lagrange function and the KKT conditions are applied to find the optimal data rates for the users. Then, the greedy pricing algorithm is adopted to update the bandwidth prices for the MNO. &Stackelberg equilibrium& HN\tabularnewline \cline{2-9}

&\cite{gu2016virtualized}& VCG auction& MNO&Service providers& Fronthaul bandwidth and radio spectrum& Given the service providers' bids, the MNO solves the WDP and payment using the decomposition technique.&Optimal solution&CRAN \tabularnewline \cline{2-9}

&\cite{morcos2017two}& Share auction and VCG auction& MNO&SPs and users& Resource blocks& MNO adopts the share auction to allocate the MNO's RBs to the SPs. Each SP applies the VCG with the greedy algorithm to allocate the RBs to the users.&Nash equilibrium &CRAN \tabularnewline \cline{2-9}

&\cite{zhu2015virtualization}& Combinatorial auction& MNO&MVNOs and users& Subchannels and antennas&The combinatorial auction is adopted in both upper and lower levels to allocate bundles of resources to the MVNOs and their users, respectively. In both levels, the dynamic programming algorithm is used to solve the WDP, and the VCG auction is employed to charge the winners. &Optimal solution & MMN \tabularnewline \cline{2-9}

&\cite{zhangdouble}& Double auction& MNOs&MVNOs& Data rates&MNOs and MVNOs calculate and submit their bids to an SDN controller. The SDN controller checks the convergence condition to either match each MVNO and each MNO or update the shadow prices using the subgradient descent method. &Competitive
equilibrium&MMN\tabularnewline \cline{2-9}

\hline
\end{tabular}
\par\end{centering}
\end{table*}

\begin{table*}[h]
\caption{A summary of advantages and disadvantages of major approaches for spectrum allocation in 5G.}
\label{table_sum_advantage_spectrum_allocation}
\scriptsize

\begin{centering}
\begin{tabular}{|>{\centering\arraybackslash}m{2cm}|>{\centering\arraybackslash}m{8.4cm}|>{\centering\arraybackslash}m{6cm}|}
\hline
\cellcolor{myblue} &\cellcolor{myblue} &\cellcolor{myblue} \tabularnewline
\cellcolor{myblue} \multirow{-2}{*} {\textbf{Major approaches}} &\cellcolor{myblue} \multirow{-2}{*} {\textbf{Advantages}} &\cellcolor{myblue} \multirow{-2}{*}{\textbf{Disadvantages}} \tabularnewline
\hline
\hline
\cite{semiari2015matching}&\begin{itemize} \item Support multiple BSs \end{itemize} & \begin{itemize}  \item Have high computational complexity \end{itemize}\tabularnewline \cline{2-3}
\hline
\cite{ahmadi2016substitutability} &\begin{itemize} \item Allow to allocate bundles of heterogeneous resources \end{itemize} & \begin{itemize}  \item  Have high computational complexity and support only one MNO\end{itemize}\tabularnewline \cline{2-3}
\hline
\cite{fund2016spectrum} &\begin{itemize} \item Consider network effect \end{itemize} & \begin{itemize}  \item  Support only two Service Providers (SPs) \end{itemize}\tabularnewline \cline{2-3}
\hline
\cite{zhu2015virtualization} &\begin{itemize} \item Analyze interactions among multiple users, multiple MVNOs, and MNO \end{itemize} & \begin{itemize}  \item  Have high computational complexity \end{itemize}\tabularnewline \cline{2-3}
\hline
\cite{zhangdouble} &\begin{itemize} \item Achieve win-win solution and support multiple MNOs and multiple MVNOs  \end{itemize} & \begin{itemize}  \item  Have slow convergence \end{itemize}\tabularnewline \cline{2-3}
\hline
\end{tabular}
\par\end{centering}
\end{table*}

\textbf{Summary:} In this section, we have discussed the applications of economic and pricing models for the spectrum allocation in 5G. The objective is to maximize the spectrum efficiency, revenue, and the total system utility. The reviewed approaches are summarized along with the references in Table~\ref{table_spectrum_allocation}. Furthermore, a summary of advantages and disadvantages of major approaches is given in Table~\ref{table_sum_advantage_spectrum_allocation}. We observe that the different objectives lead to the use of different pricing models. For example, the auctions are commonly used to maximize the revenue while the Stakeleberg game is mostly adopted to maximize the overall system utility. Given the ultra-dense BS deployment, the interference mitigation is also a major requirement of 5G. The next section discusses how to apply the economic and pricing models for power management in 5G to mitigate the interference while guaranteeing the QoS.  

\section{Applications of economic and pricing models for interference and power management}
\label{sec:App_PM}
Due to the ultra-dense deployment of heterogeneous BSs and devices, one of the biggest challenges of power management in 5G is to minimize inter-cell and inter/intra-tier interferences. The optimization methods for power control in conventional radio networks such as fractional programming \cite{zhou2016energy} usually require centralized controllers. This leads to substantial signaling and computational overhead to 5G. Alternatively, economic and pricing models can be used in the power management to mitigate interference with low computational complexity while improving revenue. For example, simply setting appropriate prices according to the interference levels was shown to achieve decent outcomes with minimal overhead. This section thus discusses applications of economic and pricing models for power management. Specifically, subsection \ref{sec:App_PA_IM} reviews economic and pricing approaches to mitigate interference, and subsection \ref{sec:App_PA_PCM} reviews economic and pricing approaches to minimize power consumption cost. 



\subsection{Interference Mitigation}
\label{sec:App_PA_IM}
In this section, the economic and pricing approaches for the interference mitigation in HetNets are reviewed. The common idea is that each BS measures locally the interference levels caused by its neighboring BSs. Based on the interference levels, the BS charges prices to the neighboring BSs. The neighboring BSs then adjust their transmit power correspondingly. The actions, i.e., setting the price and adjusting the transmit power, are done sequentially, and thus the Stackelberg game is commonly used. 

\subsubsection{Stackelberg game}
\label{sec:App_PA_IM_HetNets_SG}
\begin{figure}[t!]
 \centering
\includegraphics[width=8.7cm, height = 6.2cm]{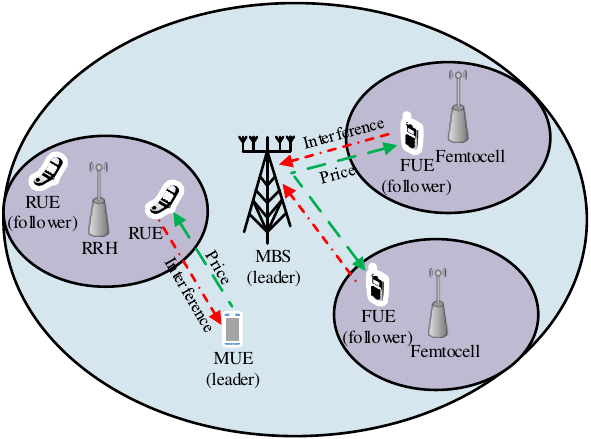}
 \caption{\small Stackelberg game-based interference management in HetNets.}
 \label{Interference_penalty_pricing_multi_tier}
\end{figure}
The first work can be found in \cite{lashgari2015distributed} which adopted the Stackelberg game for interference management in a multi-tier HetNet. The network model is shown in Fig.~\ref{Interference_penalty_pricing_multi_tier} in which the MBS is the leader, and UEs of femtocells, i.e., FUEs, are the followers. Initially, the MBS sets interference prices as penalty costs that the FUEs pay for causing interferences to users of the MBS \cite{wang2017multi}. Based on the prices, each FUE determines its transmit power to maximize its utility. The optimization problem of the FUE is convex, and KKT conditions are applied to obtain the optimal power. Given the optimal power, the MBS determines optimal interference prices to maximize its utility, i.e., its payment. The problem of the MBS is also convex, and the Lagrange method and the complementary slackness condition are adopted to determine the optimal prices. Generally, the price charged to each FUE is inversely proportional to its transmit power. As shown in the simulation, a high price reduces significantly the transmit power of the FUE. Thus the proposed scheme is useful for interference control. 

The same approach can be found in \cite{wang2016interference}, but the followers are SBSs instead of the FUEs. Moreover, the channel allocation is jointly considered. Specifically, given the interference price offered by the MBS and the number of allocated channels, each SBS determines its transmit power to maximize its utility defined similarly to that in \cite{lashgari2015distributed}. Given the SBSs' optimal transmit power, the MBS first determines the optimal number of channels allocated to each SBS and then the optimal interference price using the subgradient algorithm. More specifically, the MBS initially sets the interference price at a high value. Such a high price reduces the transmit power of all SBSs and also the MBS's utility. Then, the price decreases with a step in each round until the utility of the MBS achieves the maximum. The simulation results show that the proposed scheme outperforms the algorithm without the power control scheme in terms of average utility of the SBSs. The reason is that the MBS charges a high interference price to the SBSs if they do not perform power control. 

The model and problem in \cite{wang2016interference} are also found in \cite{chen2017dormancy}. However, the exact potential game \cite{marden2009cooperative} is adopted to determine the optimal transmit power of the SBSs. The potential game is a subclass of strategic normal games in which players' objective functions are perfectly be aligned with one potential function. The exact potential game is used since it always converges to a unique Nash equilibrium. 
 
The aforementioned approaches, i.e., \cite{wang2016interference} and \cite{chen2017dormancy}, assume that the channel gain information between the SBSs and their associated users is known by all SBSs. In practice, only the probability distribution of the information is typically available. Therefore, for a given price offered by the MBS, i.e., the leader, the authors in \cite{duong2016stackelberg} formulated a Bayesian game with incomplete information among the SBSs, i.e., followers. The fictitious play algorithm \cite{rabinovich2013computing} is applied to the follower game to obtain the Bayesian Nash equilibrium, i.e., the SBSs' optimal power. Then, a learning algorithm is adopted which allows the MBS to adjust the interference price to maximize its average profit. Generally, the fictitious play algorithm for the follower game does not converge. However, the simulation results show that the proposed algorithm can converge very fast in the case that the initial price is set close to the optimal value.

In fact, since transmit power of SBSs is far smaller than that of the MBS, the cross-tier interference from an MBS to SBSs needs to be studied. This scenario was found in \cite{wang2014low} in which one MBS adjusts its transmit power according to the interference prices offered by picocells. The MBS is thus the follower, and the picocells are the leaders. Given the interference prices of the picocells, the MBS determines the optimal power allocation on its sub-carriers to maximize its utility. The MBS's problem is solved using the Lagrangian function and the KKT condition. Then, each picocell determines the optimal power on its sub-carrier and optimal interference price to maximize its utility. The Lagrangian function and the ellipsoid method \cite{grotschel1981ellipsoid} are applied to solve the picocell's problem. The picocell replies the information, i.e., power and price, to the MBS such that the MBS updates until the Stackelberg equilibrium is reached. Such a process does not require complete network knowledge, and thus reducing signaling overhead. 

The issues such as interference management, power allocation, and channel allocation in \cite{wang2016interference} were also considered in \cite{yuan2015towards}. However, in the considered model, the leader is the MNO instead of the MBS, and the followers are Unlicensed Users (UUs) of femtocells instead of the SBSs. The optimization problems of both the MNO and the UUs are then solved using the standard convex optimization approaches. 


UEs of RRHs, i.e., RUEs, in C-RANs may cause interferences to the macro cell networks. In this context, the authors in \cite{gu2016game} considered the interference mitigation between macro cell networks and C-RANs. As shown in Fig.~\ref{Interference_penalty_pricing_multi_tier}, the MUE is the leader which sets the interference prices, and the RUEs as the followers decide their transmit power. Given the MUE's interference prices, the non-cooperative game is adopted to model the interaction among the follower RUEs. The reason is that while attempting to transmit their data, each RUE may cause interference to other RUEs and reduce their utilities. The iterative water-filling algorithm \cite{yu2004iterative} is used to obtain RUEs' optimal power. The MUE then applies the Lagrange method and the KKT conditions to determine the optimal interference prices. The simulation results show that the RUEs tend to reduce the transmit power to stable values after only few iterations. This means that the proposed scheme can quickly converge and solve the interference problem.
 
Generally, due to the first-move advantage, the pricing model based on the Stackelberg game is efficient for the interference management in HetNets. However, 5G has very high density of BSs/devices, and thus how one BS predicts strategies of other BSs/devices and makes optimal strategies is challenging.
\subsubsection{General pricing}
\label{sec:App_PA_IM_HetNets_GP}
In fact, MBSs can set pre-defined interference limits/thresholds for SBSs. Subject to the interference limits, then the SBSs determine their transmit power to maximize the sum-rate of all the SBSs. The authors in \cite{zheng2016optimal} adopted a distributed algorithm based on pricing strategy to address the problem in a HetNet with multiple SBSs and MBSs. The algorithm is essentially similar to that in \cite{wang2016interference}. The difference is that each SBS computes its transmit power based on interference prices of multiple MBSs. Additionally, each MBS locally updates its price depending on the difference between the interference caused by the SBS to the MBS and the MBS's interference limit. If the interference is greater than the interference limit, the MBS increases the price. Otherwise, the price is set to zero. The process repeats until the interference experienced by the MBS remains unchanged. It is proved that by choosing the step-size sequence according to the decreasing rule in \cite{scutari2014decomposition} for the price update, the proposed algorithm converges to a unique optimal power allocation solution. The simulation results show that the transmission rate of each SBS increases as the interference limit of each MBS increases. However, the increased interference limit leads to the high interference experienced by the MBSs' users, thus reducing the MBSs' transmission rate. 

A distributed algorithm based on pricing strategy was also adopted in \cite{ho2016distributed} for the interference control for femtocells in a HetNet. First, each femtocell measures locally its SINR, utility function, and effective interference caused by other femtocells and MBSs. Femtocell $i$ then calculates and broadcasts an interference price to other femtocells and MBSs.  It simultaneously updates its transmit power which is inversely proportional to the sum of interference prices set by other femtocells and MBSs. Note that before updating the current price/transmit power, femtocell $i$ observes and takes into account the last power/price updates of other femtocells and MBSs. Such an updating scheme is called \textit{gradient play} \cite{chen2010random}. As shown in the simulation results, the proposed scheme has much lower average transmit power than that of the distributed power control without the price updating. However, the proposed scheme has slower convergence speed due to simultaneously updating price and power as well as synchronizing the step for the updating. 

The same algorithm for the interference  management was found in \cite{wang2016distributed}, but the proximal point method \cite{facchinei2007finite} is adopted for updating the prices and power at the femtocells. More specifically, at each iteration, the Variational Inequality (VI) theory \cite{scutari2010convex} is adopted to determine the generalized Nash equilibrium for the femtocells. Then, the current price and power are updated based on the generalized Nash equilibrium obtained in the previous iteration. 


Different from \cite{zheng2016optimal} and \cite{ho2016distributed}, the authors in \cite{pischella2016resource} developed the allocation of transmit power to the RBs of D2D transmitters to maximize the total utility of the D2D transmitters subject to the interference at an MBS. The Lagrange method is first adopted to the problem, and then the power and price updating algorithm using the gradient projection is applied. Here, two dual prices are updated to each D2D transmitter in which the prices are paid to the MBS. Generally, the prices increase if the total interference caused by all D2D transmitters is greater than a pre-defined interference level at the MBS. Then, the D2D transmitters update their transmit power which is inversely proportional to the dual prices. The simulation results show that the D2D pairs can achieve very high data rates while generating low interference level at the MBS. Moreover, the proposed algorithm's complexity is polynomial in the number of RBs and D2D pairs. This makes the proposed algorithm feasible for practical implementations in 5G. 

In fact, the problem in \cite{pischella2016resource} can be addressed by a distributed auction with multiple iterations as proposed in \cite{hasan2015distributed}. In this model, the MBS is the auctioneer, and multiple underlay users, i.e., SUEs and DUEs, are bidders. At each iteration, each user selects a bid for the resources including the RB and power level to maximize the user's utility. Given users' bids, the MBS calculates the aggregated interference of each RB and forms an assignment vector of all users. The MBS then broadcasts these information to all users. Each user checks if the aggregated interference of its selected RB is greater than an interference threshold on the RB. If so, the user needs to select another RB and power level. This process is repeated until the assignment vector remains unchanged for two successive iterations. The simulation results show that
the proposed algorithm performs close to the achievable data rate of the exhaustive search with less complexity. However, since each user locally decides its resource assignment, the truthfulness of the auction needs to be studied.


\subsubsection{Non-cooperative game}
\label{sec:App_PA_IM_HetNets_NG}
\begin{figure}[t!]
 \centering
\includegraphics[width=5.6cm, height = 5.5cm]{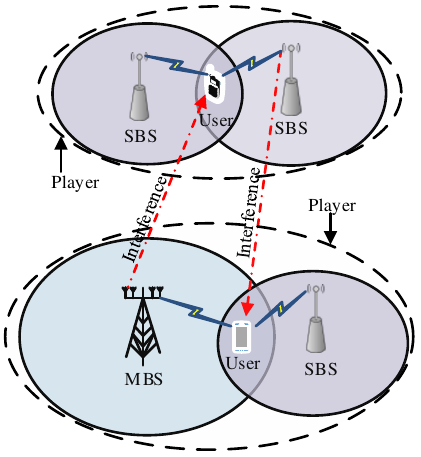}
 \caption{\small Non-cooperative game-based interference management in HetNets which use Coordinated Multi-Point (CoMP).}
 \label{interference_penalty_non_cooperative}
\end{figure}
To improve the throughput of each user, the Coordinated Multi-Point (CoMP) \cite{sawahashi2010coordinated} can be used which allows the user to be simultaneously served by a set of BSs, i.e., MBSs and SBSs. As shown in Fig.~\ref{interference_penalty_non_cooperative}, the power received by the user is the total power transmitted from the BSs in the set. While attempting to serve one user, a set of BSs may cause interference to users in the other sets of BSs. The non-cooperative game can be adopted as proposed in \cite{xu2015energy} to address the interference among the sets of BSs in CoMP and to maximize the system energy efficiency. In this game, the player is the whole set of BSs serving a particular user. In other words, the BSs in the same set act collectively as a single player. Each set determines the total transmit power to serve its user to maximize its utility. The utility is a function of the energy efficiency of the user and the price, i.e., the penalty cost, that the set of BSs pays when the BSs in this set cause interference to the users of other sets. It is proved that there exists a unique Nash equilibrium, i.e., the minimum total transmit power of sets, which yields the highest utilities of the sets of BSs. However, by using CoMP, the proposed scheme has high communication overhead because of more signaling among BSs in the set.  

\subsection{Energy Cost Minimization}
\label{sec:App_PA_PCM}
To transmit a large amount of data traffic, 5G consumes significant amount of energy. Renewable energy sources, i.e., solar and wind, are viable solutions to reduce the energy cost. However, these energy sources have limited capacity and their supply is random which may degrade performance of the networks. The MNOs thus need to purchase the energy from Power Suppliers (PSs). With energy trading, MNOs aim to minimize their cost of energy while the PSs aim to improve their profit. This section discusses the applications of economic and pricing models for the energy trading in 5G technologies, i.e., mmWave networks, HetNets, and C-RANs. Different economic and pricing models are applied to different technologies to minimize the energy consumption cost. For example, the mmWave networks can support the backhaul links which require high QoS, and thus the Stackelberg game and the time-depending pricing are jointly used to maximize the PSs' profit while minimizing the energy cost and guaranteeing the QoS for the MNOs. The C-RANs consume a large amount of energy at BBUs and RRHs, and the cost minimization problem can be used to minimize the energy costs at the BBUs and RRHs. 

\subsubsection{MmWave networks}
\label{sec:App_PA_PCM_MMW}
Consider the mmWave backhaul network as shown in Fig.~\ref{mmWave_backhaul_network_matching}, the authors in \cite{li2016decentralized} studied the energy trading between one MNO as a buyer and multiple PSs as sellers. The MNO purchases energy from the PSs for its backhaul nodes to forward data traffic in multiple epochs. The Stackelberg game is applied to minimize the energy cost for the MNO and maximize the PSs' profits. First, the MNO as a leader sets price of power supply for each PS in different epochs using the time-depending pricing \cite{ha2012tube}. The time-depending pricing is used due to the fact that the MNO has different QoS levels to support its data traffic forwarding in different epochs according to the energy supply from the PS. Given the prices, the non-cooperative game is formulated among the PSs in which they simultaneously determine optimal energy storage levels so as to maximize their expected profits over epochs. The PSs need to determine the optimal storage levels since overstocked energy will incur extra storage costs for them. Based on the PSs' optimal strategies, the MNO determines prices of power supply so as to maximize its expected profit. Here, the MNO's profit is inversely proportional to the prices that the MNO pays to the PSs over epochs. The simulation results show that the system profit, i.e., the sum of profits of the MNO and the PSs, obtained from the proposed scheme is close to that obtained from the centralized solution in which the MNO decides on the power storage levels of all PSs. However, the system profit of the proposed scheme substantially decreases when the number of backhaul nodes increases due to their significant energy consumption. 

\subsubsection{HetNets}
\label{sec:App_PA_PCM_HetNet}
Small cell networks in HetNets with energy harvesting can make the networks greener. The SBSs in these networks can share their energy with each other rather than buying the energy from PSs to reduce the energy cost. However, the SBSs may belong to different MNOs, and thus the economic and pricing models are used to motivate MNOs of the SBSs with surplus harvested energy to share and sell their energy to those with energy deficit. 

In a multi-seller multi-buyer market, the authors in \cite{reyhanian2017game} adopted the double auction for the energy trading. The model consists of SBSs with energy surplus as sellers, SBSs with energy deficit as buyers, and one Central Authority (CA) as the auctioneer. First, each SBS submits (i) the amount of total energy in its battery and (ii) the amount of energy that it wants to sell or to buy to the CA. Based on these information, the CA calculates asks and bids for each unit of energy to sell and to buy, respectively. The CA sorts the asks in an ascending order and the bids in a descending order. The seller-buyer matching is then determined similarly to that in the classical double auction with some modifications. First, the deal price is a random price which can follow any probability distribution such that the algorithm converges. Setting the random price independent of the information submitted by the SBSs is to avoid misreporting from them. Second, the matching process considers locations of the SBSs to minimize the load on the electrical grid. The CA then calculates the prices paid to the winning SBSs and the prices charged to the winning SBSs based on their valuations of the amount of energy. The simulation results show that the proposed scheme outperforms the matching game-based algorithm \cite{bando2012many} in the electrical grid usage reduction. The reason is that when performing the matching process, the proposed scheme considers more the distance between SBSs rather than the utility functions, i.e., usage costs, as in the matching game-based algorithm. 



\subsubsection{C-RANs}
\label{sec:App_PA_PCM_CRANs}
\begin{figure}[t!]
 \centering
\includegraphics[width=\linewidth]{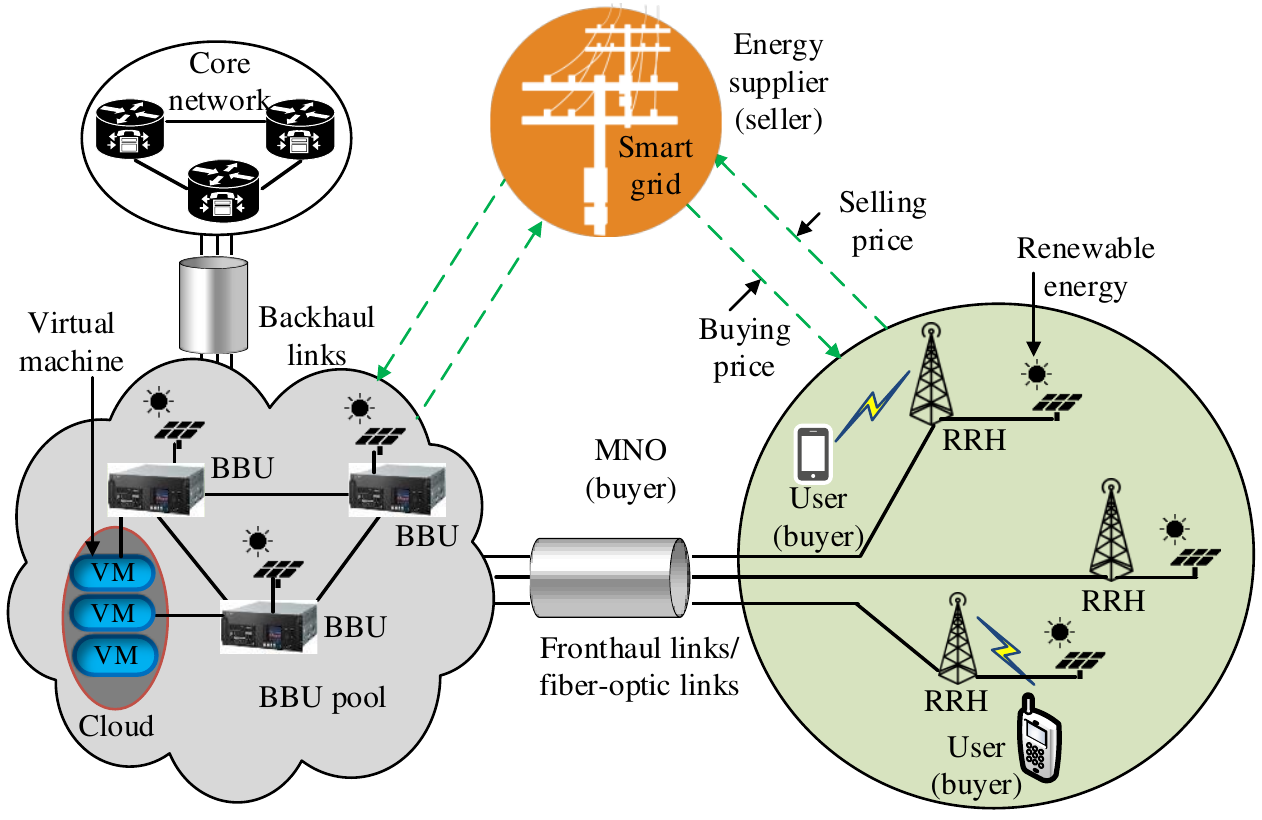}
 \caption{Energy trading in C-RAN for cost minimization.}
 \label{CRAN_Cost_minimization}
\end{figure}

RRHs and BBUs in C-RANs can also be equipped with renewable energy sources as shown in Fig.~\ref{CRAN_Cost_minimization}. Then, the MNO which owns the C-RANs may buy/sell energy from/to the PSs depending on the availability of the renewable energy. In particular, if the renewable energy at each RRH/BBU is more than the needed energy for data transmission, the MNO sells the surplus energy to the PSs. Otherwise, the MNO needs to buy energy for the RRH from the PSs.

In this context, the authors in \cite{alameer2016joint} analyzed the energy trading among one MNO and one PS by formulating the cost minimization problem. Specifically, the problem is to determine the resource allocation to minimize the sum of the cost functions at the RRHs and at the BBUs. Here, the resources to be allocated include (i) the energy bought from the PS for the RRHs and the BBUs, (ii) the achievable rates for the MNO's users, and (iii) the assigned computation capabilities of VMs in the BBUs to serve users. The cost function at each RRH/BBU is the difference between the price for purchasing energy from the PS and the price received for selling renewable energy to the PS. The problem is then solved by the Alternating Direction Method of Multipliers (ADMM) algorithm \cite{boyd2011distributed}. The simulation results show that the cost function at each RRH or BBU reduces with the decrease of the data rate. Moreover, there is a balance point at which the cost function is zero. This means that the MNO uses all the harvested energy to satisfy the user requirements without buying/selling energy from/to the PS. However, the prices used in the energy trading are given a priori. A dynamic pricing strategy needs to be considered in the future work. 

In practice, the fronthaul links between the RRHs and the BBUs may belong to a cloud provider. Therefore, the MNO needs to incorporate the cost of using the fronthaul links into the cost functions. Thus the authors in \cite{ha2015coordinated} formulated a problem which determines transmit power for all RRHs to minimize (i) the total transmit power from the RRHs to users and (ii) the total cost that the MNO pays the cloud provider for using the fronthaul links. In particular, the total cost is calculated based on prices of fronthaul bandwidth associated with the RRHs. The optimization problem is first relaxed using the concave approximation and then solved using the gradient method. It is shown that the total transmit power of the proposed scheme is much lower than that of the iterative link removal algorithm in \cite{zhao2013coordinated}. However, the computational complexity of the proposed scheme significantly increases when the cluster size, i.e., the maximum number of RRHs serving one user, is large.

\begin{table*}
\caption{Applications of economic and pricing models for interference and power management (MWN: Millimeter Wave Network, CRAN: Cloud Radio Access Network, HN: Heterogeneous Network).}
\label{table_power_management}
\scriptsize
\begin{centering}
\begin{tabular}{|>{\centering\arraybackslash}m{0.6cm}|>{\centering\arraybackslash}m{0.4cm}|>{\centering\arraybackslash}m{1.4cm}|>{\centering\arraybackslash}m{0.9cm}|>{\centering\arraybackslash}m{1.1cm}|>{\centering\arraybackslash}m{0.9cm}|>{\centering\arraybackslash}m{7.6cm}|>{\centering\arraybackslash}m{1.4cm}|>{\centering\arraybackslash}m{0.8cm}|}
\hline
\multirow{2}{*} {\textbf{}} & \multirow{2}{*} {\textbf{Ref.}} & \multirow{2}{*} {\textbf{Pricing model}} & \multicolumn{3}{c|} {\textbf{Market structure}} & \multirow{2}{*} {\textbf{Mechanism}}& \multirow{2}{*} {\textbf{Solution}} & \multirow{2}{*} {\textbf{Network}} \tabularnewline
\cline{4-6} & & & \textbf{Seller} & \textbf{Buyer} & \textbf{Item} & &&\tabularnewline
\hline
\hline
\parbox[t]{2mm}{\multirow{9}{*}{\rotatebox[origin=c]{90}{ \hspace{-7cm} Interference}}}
\parbox[t]{2mm}{\multirow{9}{*}{\rotatebox[origin=c]{90}{ \hspace{-7cm}mitigation}}}

&\cite{lashgari2015distributed} &Stackelberg game&MBS&FUEs&Power&The KKT conditions are used to determine the optimal power for the FUEs. The Lagrange method and the complementary slackness condition are applied to determine the optimal interference prices for the MBS.&Stackelberg equilibrium&HN\tabularnewline \cline{2-9}

&\cite{wang2016interference} &Stackelberg game&MBS&SBSs&Channels and power&SBSs determine their optimal transmit power using the first derivative. Then, the MBS determines the optimal number of channels and interference price for the SBSs based on the subgradient algorithm. &Stackelberg equilibrium&HN\tabularnewline \cline{2-9}

&\cite{chen2017dormancy} &Stackelberg game&MBS&SBSs&Power& Same as \cite{wang2016interference}, but the exact potential game is used to determine the SBSs' optimal transmit power. &Stackelberg equilibrium&HN\tabularnewline \cline{2-9}

&\cite{duong2016stackelberg} &Stackelberg Bayesian game&MBS&SBSs&Power& The Bayesian game is used to determine the SBSs' optimal transmit power, and a learning algorithm is adopted to determine the optimal price. &Stackelberg equilibrium&HN\tabularnewline \cline{2-9}

&\cite{wang2014low} &Stackelberg game&Picocells&MBS&Power&The KKT conditions are used to determine the optimal power of the MBS and picocells, and the ellipsoid method is used to calculate the optimal interference prices of picocells. &Stackelberg equilibrium&HN\tabularnewline \cline{2-9}

& \cite{yuan2015towards} &Stackelberg game&MNO&Unlicensed users&Sub-bands and power&Same as \cite{lashgari2015distributed}, but the optimal power for the unlicensed users and the optimal interference prices for the MNO are determined using the standard convex optimization approaches.&Stackelberg equilibrium& HN\tabularnewline \cline{2-9}

&\cite{gu2016game} &Stackelberg game&MUE&RUEs&Power& The optimal power for the RUEs are determined using the iterative water-filling algorithm. Then, the optimal interference prices for the MUE are determined using the Lagrange method and the KKT conditions.&Stackelberg equilibrium&HN\tabularnewline \cline{2-9}

&\cite{zheng2016optimal} &General pricing&MBSs&SBSs&Power&MBSs control transmit power of the SBSs by adjusting the interference prices according to the decreasing rule in \cite{scutari2014decomposition}. &Optimal solution&HN\tabularnewline \cline{2-9}

&\cite{ho2016distributed} &General pricing&Femtocells&Femtocells and MBSs&Power&Femtocells calculate and broadcast interference prices. Each femtocell then updates its transmit power and price by using the gradient play scheme. &Nash equilibrium&HN\tabularnewline \cline{2-9}


&\cite{pischella2016resource} &General pricing&MBS&D2D transmitters&Power&The gradient projection algorithm is applied to update power and dual prices for the D2D transmitters. &Nash equilibrium&HN\tabularnewline \cline{2-9}

& \cite{hasan2015distributed}&Distributed auction&MBS&Underlay users&Resource blocks and power&Each underlay user locally decides its resource assignment if the aggregated interference caused by the resource is smaller than an interference threshold.&Nash equilibrium&HN\tabularnewline \cline{2-9}

&\cite{sawahashi2010coordinated} &Non-cooperative game&BS sets&BS sets&Power&The optimal transmit power for sets of buying BSs are determined via the Nash equilibrium. &Nash equilibrium&HN\tabularnewline \cline{2-9}

\hline
\parbox[t]{2mm}{\multirow{9}{*}{\rotatebox[origin=c]{90}{ \hspace{0cm} Energy cost}}}
\parbox[t]{2mm}{\multirow{9}{*}{\rotatebox[origin=c]{90}{ \hspace{0cm} minimization}}}

&\cite{li2016decentralized} &Stackelberg game&Renewable power suppliers&MNO&Renewable energy& MNO sets power prices over epochs using the time-depending pricing. Then, the power suppliers determine optimal energy storage levels via the non-cooperative game.&Stackelberg equilibrium& MWN\tabularnewline \cline{2-9}

& \cite{reyhanian2017game} &Double auction&SBSs&SBSs&Renewable energy&The central authority calculates asks for the selling SBSs and bids for the buying SBSs. Then, it matches the buyers and the sellers based on a random price. &Nash equilibrium &HN\tabularnewline \cline{2-9}

&\cite{alameer2016joint} &Cost minimization& Power supplier&MNO&Power&MNO determines the amount of energy for trading to minimize its total cost. The ADMM algorithm is then applied to solve the problem. &Optimal solution &CRAN\tabularnewline \cline{2-9}

&\cite{ha2015coordinated} &General pricing& Cloud provider&MNO&Fronhaul capacity&Given the fronthaul bandwidth prices, the concave approximation and the gradient method are used to determine the transmit power for all RRHs. &Optimal solution&CRAN\tabularnewline \cline{2-9}

\hline
\end{tabular}
\par\end{centering}
\end{table*}

\begin{table*}[h]
\caption{A summary of advantages and disadvantages of major approaches for interference and power management.}
\label{table_power_management_advantage}
\scriptsize
\begin{centering}
\begin{tabular}{|>{\centering\arraybackslash}m{2cm}|>{\centering\arraybackslash}m{7.2cm}|>{\centering\arraybackslash}m{7cm}|}
\hline
\cellcolor{myblue} &\cellcolor{myblue} &\cellcolor{myblue} \tabularnewline
\cellcolor{myblue} \multirow{-2}{*} {\textbf{Major approaches}} &\cellcolor{myblue} \multirow{-2}{*} {\textbf{Advantages}} &\cellcolor{myblue} \multirow{-2}{*}{\textbf{Disadvantages}} \tabularnewline
\hline
\hline
\cite{wang2016interference} &\begin{itemize} \item Consider jointly interference management and channel allocation \end{itemize} & \begin{itemize}  \item Need to know the channel gain information between the SBS and their users \end{itemize}\tabularnewline \cline{2-3}
\hline
\cite{duong2016stackelberg} &\begin{itemize} \item Have fast convergence \end{itemize} & \begin{itemize}  \item Support only one MBS \end{itemize}\tabularnewline \cline{2-3}
\hline
 \cite{hasan2015distributed}&\begin{itemize} \item Have low complexity \end{itemize} & \begin{itemize}  \item Support only one MBS and do not consider truthfulness\end{itemize}\tabularnewline \cline{2-3}
\hline
\cite{sawahashi2010coordinated}&\begin{itemize} \item Support multiple MBSs and SBSs and CoMP technology \end{itemize} & \begin{itemize}  \item Have high communication overhead\end{itemize}\tabularnewline \cline{2-3}
\hline
 \cite{reyhanian2017game}&\begin{itemize} \item Analyze interactions among multiple SBSs \end{itemize} & \begin{itemize}  \item Need a Central Authority (CA) and have slow convergence \end{itemize}\tabularnewline \cline{2-3}
\hline
\end{tabular}
\par\end{centering}
\end{table*}

\textbf{Summary:} This section has discussed the applications of economic and pricing models for the interference and power management in 5G. The objective is to minimize the intra-tier and inter-tier interferences as well as the energy consumption cost. A summary of the related works is given in Table~\ref{table_power_management}, and a summary of advantages and disadvantages of major works in Table~\ref{table_power_management_advantage}. As seen, the economic and pricing models are mostly used to mitigate the interference caused by the SBSs or their users to MBSs. Although the SBS deployment poses the interference management as a challenge, the SBSs can provide mobile data offloading services and wireless caching schemes to offload as well as to reduce traffic from MBSs. The next section discusses the economic and pricing models for the mobile data offloading and wireless caching in 5G HetNets to improve the profits for the owners of both SBSs and MBSs.

\section{Applications of economic and pricing models for wireless caching and mobile data offloading}
\label{sec:App_Caching_Offloading}
Two important services provided by SBSs in 5G HetNets are wireless caching and mobile data offloading. The wireless caching refers to caching contents such as popular videos or web information in the storage/memory of SBSs. The mobile data offloading allows to use SBSs to reduce the traffic being carried on MBSs. These services can (i) reduce the transmission latency of content requests from users, (ii) mitigate the redundant transmissions of popular contents over backhaul links, (iii) achieve higher energy efficiency, and (iv) significantly improve network capacity as well as coverage. However, the SBSs and MBSs may belong to different owners. Besides, the SBSs typically have limited resources, i.e., power, storage, and bandwidth, as well as less security. Economic and pricing models can be applied to improve profits of the owners of the SBSs, and thus incentivizing the owners to provide the wireless caching and offloading services. The following subsections review the economic and pricing approaches for wireless caching and mobile data offloading. 
\subsection{Profit Maximization Through Caching}
\label{sec:App_Caching_Offloading_Caching}
This section reviews economic and pricing models for wireless caching in 5G HetNets. A common scenario involves Content Providers (CPs), e.g., Youtube, which act as buyers and MNOs which act as sellers. The CPs lease the storage space at SBSs of MNOs to store files and serve the CPs' users. The CPs pay certain prices to the MNOs for leasing the storage space. Different economic and pricing models can be used depending on specific objectives. For example, to maximize profits for both MNOs and CPs which have a hierarchical interaction, the Stackelberg game is used. To either maximize the profits for MNOs or minimize the caching payments for CPs, single-side auctions are adopted. 
\subsubsection{Stackelberg game}
\label{sec:App_Small_Cell_Caching_stack}
\begin{figure}[t!]
 \centering
\includegraphics[width=7.9cm, height = 5.3cm]{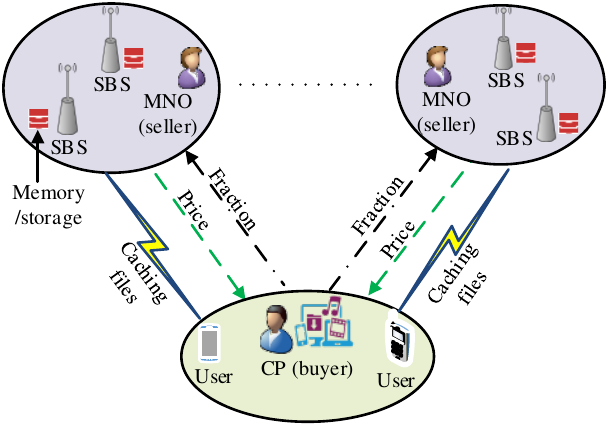}
 \caption{\small Wireless caching based on Stackelberg game in HetNets where CP stands for Content Provider.}
 \label{Caching_Stackelberg}
\end{figure}
The authors in \cite{li2016efficient} applied the Stackelberg game for the wireless caching to improve the profits of one CP as a follower and multiple MNOs as leaders. Such Stackelberg game with multiple leaders but single follower is practically not so common. The model is shown in Fig.~\ref{Caching_Stackelberg} in which each MNO as a seller owns multiple SBSs, and the CP as the buyer decides to rent a fraction of cache storages of SBSs based on the MNO's offered price. First, given the MNOs' prices, the CP adopts the conventional water-filling algorithm \cite{yu2004iterative} to determine an SBS utilization fraction vector to maximize its profit. The profit of the CP is the difference between the total revenue from providing downloading services to its users and the total price for renting the SBSs. Based on the CP's optimal renting fractions of SBSs, each MNO determines a common leasing price for its SBSs using the KKT conditions so as to maximize the MNO's profit. Due to the limited budget of the CP, the MNOs compete with each other in the non-cooperative game. Thus each MNO needs to update its leasing price after the other MNOs change the leasing prices. The optimization process is repeated to reach the convergence. The simulation results show that the profits of the CP and the MNOs increase as the budget increases. Especially, there is a budget value at which the CP's profit is maximum. However, how the CP sets this budget value is not explained. Also, the proof of existence and uniqueness of the Stackelberg equilibrium is not given. 


The Stackelberg game for the wireless caching was also studied in \cite{shen2016stackelberg}. However, the considered model consists of multiple CPs as followers, i.e., buyers, and a single MNO as a leader, i.e., a seller. Based on the caching prices announced by the MNO, the CP responds with a quantity of cached files to maximize the CP's profit, i.e., the difference between its revenue and the price that it pays the MNO. Due to the limited storage capacity of the SBSs, the competition among the CPs is formulated as a non-cooperative game. The Cramer's rule is applied to obtain the optimal quantity of cached files for each CP. Given the CPs' optimal responses, the MNO optimizes the price to maximize its profit, i.e., the difference between its total price received from the CPs and the total cost of caching the CPs' files. The optimal price is obtained using the first derivative of the profit function. The simulation results show that the MNO's profit improves up to 50\% of that in the case when the MNO chooses arbitrary prices. However, the profits of the CPs decrease when the number of CPs increases. This is because of that more CPs result in higher competition on the cache storage, and the price charged to the CPs is higher. 

\begin{figure}[t!]
 \centering
\includegraphics[width=6cm, height=4.6cm]{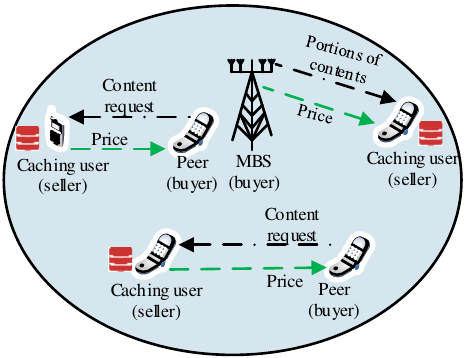}
 \caption{\small Wireless caching based on Stackelberg game in D2D networks.}
 \label{wireless_caching_D2D_Stackelberg}
\end{figure}
Due to the scarcity of licensed spectrum, using the SBSs for wireless caching may not be effective. D2D users can cache video contents of the CPs to reduce caching costs at SBSs. Such a D2D user is called \textit{caching user} which precaches certain amount of contents from the CPs. Then, the caching user can transfer the content to other users, namely \textit{peers}.  However, the design of incentive mechanisms is challenging due to the heterogeneous preference and selfish nature of the caching user.

To address the challenge, the authors in \cite{Zhang2016social} adopted the Stackelberg game to model the interaction between a caching user as a leader, i.e., a seller, and its peers as the followers, i.e., buyers. Given the prices offered by the caching user, each peer determines an amount of content to maximize its utility. The peer's utility depends on its valuation from receiving the content, network effect, congestion effect, and price that the peer pays the caching user. In particular, the network effect is defined in Section~\ref{sec:App_SA_RM_GP}. The optimal amount of content to be transferred for each peer is obtained by taking the first derivative of its utility function. Given the optimal amounts requested by the peers, the caching user chooses prices per unit of content to maximize its profit. Generally, the optimal prices depend on the network and congestion effects. As shown in the simulation results, the caching user receives high profit if the network effect increases since higher prices can be set to the peers. However, when there exists only the congestion effect, the profit is zero since the peers cannot obtain their requested contents from the caching user due to the network congestion.

  
In fact, caching users can assist the MBS to transmit their precached popular files to the MBS's subscribers. To maximize the profit for the caching users while guaranteeing a non-negative utility for the MBS, the Stackelberg game is used as proposed in \cite{liu2016d2d}. In this model, the caching users initiate transactions by providing prices for transmitting the files. Thus the caching users are leaders, i.e., sellers, and the MBS is the follower, i.e., the buyer. First, the MBS broadcasts the file request from its subscribers to all caching users. Each caching user replies to the MBS with a price per unit of power for transmitting the file. The optimal price is determined so as to maximize the caching user's profit. Given the prices, the MBS calculates the optimal power for each caching user to maximize the MBS's utility, i.e., the difference between the gain from the sum rate of transmitting the file from the caching users and the prices that the MBS pays them. Generally, the optimal power that the MBS buys from a caching user is affected by not only the caching user's optimal price, but also optimal power of other caching users. As shown in the simulation results, when the MBS buys more power, the profits of the caching users and the utility of the MBS both increase. However, the utility of the MBS then decreases as the power increases due to its higher payment.  

The same model is also found in \cite{chen2016caching}, but the caching users are the followers, and the MBS is the leader. The MBS, i.e., a buyer, requests each caching user, i.e., a seller, to cache contents of the caching user's neighbors. As shown in Fig.~\ref{wireless_caching_D2D_Stackelberg}, given the caching price offered by the MBS, each caching user determines caching strategies, i.e., the portions of contents that the caching user serves their neighbors, to maximize its utility. The utility of the caching user is the difference between the total price that the caching user receives from the MBS and the caching user's total delay cost. Based on the Schauder fixed-point theorem \cite{rosen1965existence}, it is proved that there exists at least one Nash equilibrium for the sub-game among the caching users. Given the optimal caching strategies, the MBS determines the caching price to minimize its total cost. The iterative gradient algorithm is then used to find the Stackelberg equilibrium of the game. In practice, due to their mobility, the neighbors of each caching user may vary which affects the caching strategies of the caching user. This issue needs to be considered in the future work. 

\subsubsection{Auction}
\label{sec:App_Small_Cell_Caching_auction}
Due to the limited cache storage capacity of the SBSs, CPs have to evaluate their own contents and compete for cache storages of the MNO. Auction schemes such as the iterative auctions can be used for the wireless caching to maximize the profits of both sides. 

Such an approach can be found in \cite{you2017auction} which adopts the Ascending Clock Auction (ACA) for the wireless caching. The model consists of one MNO as a seller and multiple CPs as buyers. The MNO first announces a common leasing price for its SBSs, and each CP computes the faction of the SBSs' storages so as to maximize its profit. The CPs then submit their optimal fractions to the MNO. If the total fraction demand is greater than the resources that the MNO can supply, the MNO will increase the price in the next iteration to curtail the demand. Otherwise, the CPs cannot afford renting more SBSs, and the auction is terminated. Since the renting price is an increasing function with iteration, and the fraction demand is a decreasing function of the price, the auction will converge. In fact, since each CP locally decides its resource allocation, the CP may cheat the MNO by submitting a distorted demand. For example, the CP may claim less fraction of the SBSs than that needed to obtain a lower price from the MNO. To solve this issue, the MNO determines the SBS allocation to each CP based on the other CPs' bids rather than the one from itself. Then, the payment for each CP in each iteration is determined by computing a \textit{cumulative clinch}, i.e., the fraction of SBSs that the CP is
guaranteed to win in the iteration. As shown in the simulation results, each CP achieves its maximum profit only when it does not cheat the MNO.

\begin{figure}[t!]
 \centering
\includegraphics[width=7cm, height = 4.7cm]{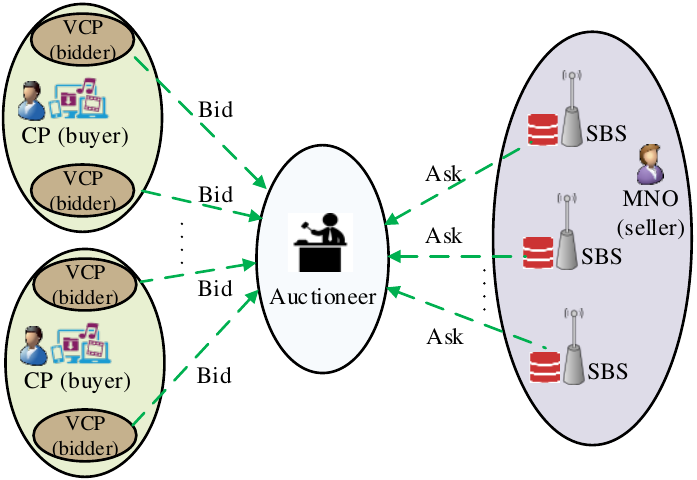}
 \caption{\small Wireless caching based on double auction in HetNets where VCP stands for Virtual Content Provider.}
 \label{Caching_Double_Auction}
\end{figure}
To improve the revenue for both the MNO and the CPs as well as some desired economic properties, the double auction can be used as proposed in \cite{xu2016self}. The model is shown in Fig.~\ref{Caching_Double_Auction} with one MNO as the seller and CPs as buyers. Each CP requires cache storage space for its files at different SBSs. Depending on the popularity of the files, the CP may have different valuations of the storage space at different SBSs. Thus the CP can be represented by a set of Virtual CPs (VCPs), each of which as a bidder is associated with an SBS. The VCP submits its bid to an auctioneer which specifies the price that the VCP is willing to pay the MNO for the storage space at the corresponding SBS. Also, the MNO submits its asks to the auctioneer which specifies the price of cache storage space at the corresponding SBS. Matching the VCPs' bids and the MNO's asks as well as payment determination for them are similar to the matching and payment of the classical double auction. Generally, reference \cite{xu2016self} is the pioneer work which uses the concept of VCP. However, other issues such as bandwidth allocation to the CPs need to be incorporated into the storage sharing problem to make the scheme more practical for implementation. 

In the scenario with multiple sellers and one buyer, the authors in \cite{mangili2016bandwidth} applied the reverse auction as presented in Section~\ref{subsec:Auction_forward_reverse_double_auction} to minimize the caching payment for the CP. The model consists of one CP as the buyer and multiple MNOs as the sellers. Each MNO owns one SBS. First, the MNOs submit their asks to the CP. Each MNO's ask specifies the price of storage that the MNO is willing to pay and the amount of storage that its SBS can provide. Based on this information, the CP calculates \textit{cache hit rate}, i.e., a probability that the requested content can be found in the cache, for each SBS. The CP calculates a ratio of the price submitted by each MNO to the cache hit rate of the MNO's SBS. The CP then sorts the SBSs in an ascending order of their ratios and selects a certain number of SBSs with the lowest ratios as the winning SBSs. Such a selection process is to minimize the caching payment for the CP. A price paid to each MNO of the winning SBS is determined based on the ratio of a \textit{critical} SBS, i.e., the first SBS which is not selected by the CP. Since the payments to the MNOs are larger than their submitted prices, the proposed scheme achieves the individual rationality, i.e., the non-negative utilities for the MNOs, and truthfulness. However, when the prices submitted by the MNOs are high, the CP may pay more to the MNOs. In this case, the utility of the CP decreases while that of the MNOs increases, meaning that there is a transfer of utility from the CP to the MNOs. Thus the proposed scheme does not guarantee the social welfare maximization. 

The caching payment minimization and truthfulness can be guaranteed by a multi-item auction as proposed in \cite{hu2015small}. Here, items are the storages of SBSs owned by MNOs, and the bidders are the CPs which have different sets of contents. Based on the storage prices offered by the MNOs, the CPs bid the most preferred storages of the contents using the preferred-storage graph \cite{demange1986multi}. The Breadth-First-Search (BFS) algorithm \cite{kurant2010bias} is applied to match each content and each storage. If there are some unmatched contents, the MNOs increase the prices of the unmatched storages uniformly until at least one CP changes its preferred-storages. If the lowest price is non-zero, the prices of all storages are deducted by the lowest price. This pricing strategy is to keep the prices of all storages as the lowest market-clearing price which achieves the truthfulness and reduces the caching payment for the CPs. Again, the preferred-storage graph is rebuilt until all the contents are matched. The final prices are the trading prices of the auction. 

In fact, the CPs can change their cached contents dynamically to adapt to the variation of the content popularity and the user preference. Since the new contents are transferred from core networks via the backhaul, there is additional instantaneous traffic load which burdens the backhaul and increases the average delay. To improve the revenue for the MNOs and to reduce the content replacement rates of the CPs, the authors in \cite{Hu2016caching} introduced an \textit{additional price} to charge the CPs for replacing the original contents. The additional price is generally determined based on the traffic load of the backhaul and the content replacement rates by the CPs. As shown in the simulation results, a high additional price reduces sharply the content replacement rate by the CPs. However, the high additional price also reduces the utility of the CPs and may discourage them to use the caching services.  
 
\subsubsection{Profit maximization}
\label{caching_profit}

To reduce heavy traffic load during peak time periods and hence improve QoS for users, the CP can proactively serve its users' future content requests during the off-peak time periods. To achieve this goal, the CP needs to track and predict the content demand of the users, e.g., using machine learning tools.

In this context, the authors in \cite{tadrous2016joint} adopted the profit maximization problem to maximize the CP's profit while improving the QoS for the users. The model consists of one CP as a seller and one user as a buyer. The CP owns several data items such as movies. First, the CP constructs a demand profile for the user at every time slot by using machine learning tools such as collaborative filtering. The demand profile at a time slot includes elements, each of which is the probability that the user requests a data item at the time slot. The CP's problem is to determine (i) data items which will be proactively served at every time slot and (ii) the prices for the data items to maximize the CP's profit, i.e., the difference between its revenue and cost. The CP solves the profit maximization problem by running an iterative algorithm. By showing that the CP's profit increases after each iteration and has an upper bound, the algorithm is proved to converge to a globally optimal solution of the prices and the data items to be served. However, no specific method is given for the convergence proof.

The proposed scheme in \cite{tadrous2016joint} is the pioneer work which discusses the proactive content provision in wireless environments. However, more general scenarios, e.g., multiple CPs and multiple users or one CP and multiple users, need to be investigated. For the scenario with multiple CPs and multiple users, each user can select its preferred CP based on the content freshness and the prices given by the CP. Assigning each user to each CP can be performed by, e.g., the double auction or matching theory, to achieve desired economic properties. For the scenario with one CP and multiple users, to guarantee the profit improvement for the CP, the profit maximization in \cite{tadrous2016joint} can be used again. However, in this case, the iterative algorithm with two steps should be executed in a distributed manner to reduce the centralized complexity at the CP. More specifically, the first step can be run by the CP to set prices so as to maximize its profit while the second step is executed by the users to determine data items to be proactively served so as to minimize the users' own payments. Nevertheless, it is challenging to prove the convergence of the algorithm in this case.


\subsubsection{Contract theory}
\label{sec:App_Small_Cell_Caching_Contract}

Each CP serves directly requests of its users, the CP is thus aware of content popularity as well as preferences of the users while the MNO may not have this information. Therefore, there is an \textit{information asymmetry} between the MNO and the CP. The content popularity creates so-called \textit{types} of each CP. The CP's type is high (low) when its content popularity is high (low). The CP may announce false information about its type to maliciously improve its utility as well as the performance of its users. For example, by claiming that its content popularity is higher than it actually is, the CP can enable the MNO to allocate more storage space. The CP then pays lower prices while also lowering the interference experienced by its users. As a result, the CPs may have an incentive to not reveal their correct types so as to pay lower prices to the MNO. 

To address this issue, the authors in \cite{hamidouche2016breaking} adopted the contract theory to construct storage-price bundles. The model consists of one MNO, i.e., seller, and multiple CPs, i.e., buyers. The MNO's problem is to determine the prices charged to the CPs based on their types and the required amounts of storage space to be allocated to the CPs so as to maximize the total utility of the CPs. Each CP's utility is the difference between its valuation and the price that the CP pays the MNO for the allocated storage. The optimization problem is NP-hard, and the matching theory and the swap-based deferred acceptance algorithm \cite{resende2007fast} are adopted to assign the storages of the SBSs of the MNO to the CPs. Then, the MNO determines the price charged to each CP by taking into account the impact of the CP on the utility of other CPs. Generally, a higher price will be charged to the CP if a larger amount of storage space is allocated to the CP. The simulation results show that the proposed scheme improves the utility of the CPs up to 140\% compared with the equal storage allocation model. The reason is that the proposed scheme allocates to all CPs only the amounts of storage space that they need. 


The optimization problem and model in \cite{hamidouche2016breaking} were also considered in \cite{liu2017resource}. However, a parameter which represents the content popularity is incorporated in the optimization problem. Depending on the value of the popularity parameter, the MNO checks the individual rationality constraint of the contract or uses the standard Lagrangian method to determine the optimal contract, i.e., the optimal prices and storages, for types of CPs. The simulation results show that the profits of both the MNO and CPs are convex functions which decrease with the increase of the popularity parameter.

\subsection{Energy Efficiency Optimization Though Offloading}
\label{sec:App_Caching_Offloading_Offloading}
In 5G HetNets, MBSs consume the major part of the network energy, and thus the minimization of the consumed energy
is needed. This goal can be achieved by mobile data offloading schemes which use the resources of SBSs to offload traffic of users from MBSs. Since the SBSs may be owned by third-parties, e.g., Femto Holders (FHs), the profit improvement for the third-parties needs to be guaranteed. Pricing models based on auctions or game theory are developed to provide an optimal solution that minimizes the energy consumption and maximizes the third-party's profit.  

 
To achieve the objectives while guaranteeing the truthfulness, the VCG auction can be used as proposed in \cite{bousia2016auction}. The model includes multiple MNOs as buyers and one FH as the seller, i.e., the auctioneer. Each MNO owns one MBS, and the FH has multiple SBSs. First, each MNO submits a bid which specifies an SBS and a price that the MNO is willing to pay the FH. The problem is to select the MNOs to maximize the FH's profit while minimizing the number of active MBSs of the MNOs, i.e., minimizing the energy consumption for the MNOs. To achieve the objectives of all stakeholders simultaneously, the Integer Linear Programming (ILP) multi-objective optimization problem is adopted and solved for the winning MNO selection. Then, to guarantee the truthful bidding, the VCG payment policy is applied to determine the charge for each winning MNO. The simulation results show that the proposed scheme improves significantly the energy efficiency compared with the switching-off scheme \cite{paris2013bandwidth}. However, the proposed scheme has higher complexity due to the multi-objective problem. 

The approach based on the VCG auction for the energy efficiency maximization is also found in \cite{xu2016reverse}. In this model, the MBS is the buyer, and SBSs are the sellers. The SBSs submit their asks to the buyer. Each ask includes the information of the number of resource blocks and power units which the SBS can provide. The MBS selects the SBSs which can maximize the system energy efficiency, i.e., the ratio of the total throughput to the total transmit power. The WDP is considered to be multiple knapsack problem which is then solved by the dynamic programming method with KKT conditions. Similar to \cite{bousia2016auction}, the VCG payment policy is used to calculate the charge for each winner. As shown in the simulation results, the proposed scheme significantly improves the system energy efficiency compared with the traffic offloading scheme based on fractional frequency
reuse \cite{liu2013energy}. However, the proposed scheme's performance only increases slightly when the number of SBSs is large. The reason is that inter-cell interferences are more severe.

To maximize the utility of the MBS, the forward auction can be applied as proposed in \cite{basutli2017auction}. The considered model is similar to that in \cite{xu2016reverse}, but the MBS is the auctioneer, i.e., the seller, its users are commodities, and SBSs are bidders, i.e., buyers. Moreover, the MBS selects the SBSs which can provide the highest SINRs for offloading the MBS's users.

\begin{figure}[t!]
 \centering
\includegraphics[width=7.2cm, height = 4.7cm]{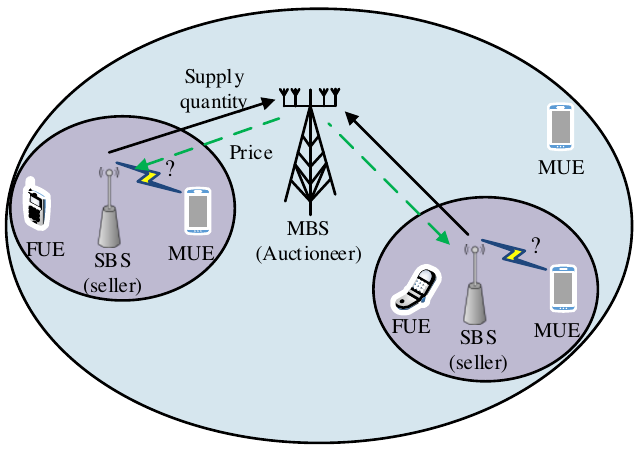}
 \caption{\small Auction-based mobile data offloading in HetNets.}
 \label{Multi_tier_mobile_data_offloading_auction}
\end{figure}
It can be seen that using the multi-objective optimization or multiple knapsack problem as proposed in \cite{bousia2016auction} or \cite{xu2016reverse} increases the computational complexity. Distributed auctions such as the ACA can be used as proposed in \cite{sken2016energy}. The model is shown in Fig.~\ref{Multi_tier_mobile_data_offloading_auction} with SBSs as the sellers and the MBS as the auctioneer. The SBSs compete with each other to offload users of the MBS. The MBS first announces an initial price that it is willing to pay to all the SBSs. Given the price, each SBS determines its supply quantity, i.e., the number of users that can be offloaded by the SBS, to maximize its profit. Meanwhile, the MBS calculates its demand quantity, i.e., the number of offloading users, to maximize its utility. Here, the MBS's utility is the difference between its revenue for saving power from the user offloading and the total price that the MBS pays all SBSs. If the total supply quantity is less than the demand quantity, the MBS increases the price to stimulate the SBSs to increase their supply quantities. The process continues until the total supply equals the demand. Since the SBSs calculate the supply quantities based only on their local information and the given price, the proposed scheme has small communication overhead, and thus reducing energy consumption of the network. However, how the MBS then assigns the exact users with the shortest distance to the corresponding SBS to minimize the total energy consumption is not considered in the proposed scheme. 

Apart from improving the energy efficiency, mobile data offloading in 5G HetNets aims to maximize the utility for the owners of MBSs and SBSs. One approach can be found in \cite{zhang2015cloud} in which one MNO as the buyer leases the SBSs of one FH, i.e., the seller, to offload the data of the MNO's MBSs. To maximize the utility of both the MNO and the FH, the Stackelberg game is adopted. First, the FH as the leader determines the price for offloading a unit of traffic to maximize its profit. Given the price, the MNO as the follower calculates the amount of traffic so as to maximize its utility, i.e., the difference between its satisfaction and the price paid to the FH for offloading. A more general scenario with multiple FHs and MNOs is then investigated, and the maximum weighted bipartite matching algorithm \cite{cheng1996maximum} is applied to find the optimal pairs of sellers and buyers. For the future work, the energy efficiency issue, e.g., the determination of inactive SBSs/MBSs, can be considered. 

Another approach using the Stackelberg game is proposed in \cite{rahmati2015price} in which the FH is the leader, and the users of MBSs, i.e., MUEs, are followers. First, given the FH's power prices, each MUE as a buyer determines its power demand to maximize its utility, i.e., the difference between its utility gain and the power price. The MUE's problem is a convex optimization which has a globally optimal solution. Given the MUEs' optimal power, the FH chooses optimal prices for the MUEs to maximize its revenue. When the total power demand of the MUEs is greater than the maximum power of the FH, the FH increases the prices to regulate the MUEs' demand. Then, the MUE's problem and FH's problem are solved again. This process repeats until the total power demand equals the maximum power. Based on the first order Taylor approximation, the algorithm is proved to converge to the unique Stackelberg equilibrium. 



\begin{table*}
\caption{Applications of economic and pricing models for wireless caching and mobile data offloading (HN: Heterogeneous Network).}
\label{table_caching_offloading}
\scriptsize
\begin{centering}
\begin{tabular}{|>{\centering\arraybackslash}m{0.6cm}|>{\centering\arraybackslash}m{0.4cm}|>{\centering\arraybackslash}m{1.4cm}|>{\centering\arraybackslash}m{0.65cm}|>{\centering\arraybackslash}m{0.7cm}|>{\centering\arraybackslash}m{1cm}|>{\centering\arraybackslash}m{8cm}|>{\centering\arraybackslash}m{1.3cm}|>{\centering\arraybackslash}m{0.7cm}|}
\hline
\multirow{2}{*} {\textbf{}} & \multirow{2}{*} {\textbf{Ref.}} & \multirow{2}{*} {\textbf{Pricing model}} & \multicolumn{3}{c|} {\textbf{Market structure}} & \multirow{2}{*} {\textbf{Mechanism}}& \multirow{2}{*} {\textbf{Solution}} & \multirow{2}{*} {\textbf{Network}} \tabularnewline
\cline{4-6} & & & \textbf{Seller} & \textbf{Buyer} & \textbf{Item} & &&\tabularnewline
\hline
\hline
\parbox[t]{2mm}{\multirow{9}{*}{\rotatebox[origin=c]{90}{ \hspace{-4cm}Profit maximization through caching}}}
& \cite{li2016efficient} &Stackelberg game& MNOs&Content Provider (CP)&SBSs&CP uses the conventional water-filling algorithm to decide optimal fractions of SBSs to rent. Given the optimal fractions, the MNOs determine their leasing prices via the non-cooperative game. &Stackelberg equilibrium&HN\tabularnewline \cline{2-9}

&\cite{shen2016stackelberg} &Stackelberg game& MNO&CPs&Caching service&CPs determine the optimal quantities of their caching files via the non-cooperative game. Then, the MNO determines the optimal service price.&Stackelberg equilibrium&HN\tabularnewline \cline{2-9}

&\cite{Zhang2016social} &Stackelberg game&Caching user&Peers&Caching service&Peers determine the optimal amounts of content for caching. Then, the caching user determines the caching prices for the peers taking into account the network and congestion effects. &Stackelberg equilibrium&HN\tabularnewline \cline{2-9}

&\cite{liu2016d2d} &Stackelberg game&Caching users&MBS&Power&Caching users determine the optimal prices, and then the MBS calculates its optimal power for each caching user. &Stackelberg equilibrium&HN\tabularnewline \cline{2-9}

&\cite{chen2016caching} &Stackelberg game&Caching users&MBS&Caching content&Caching users determine the optimal portions of contents by using the Schauder fixed-point theorem. Then, the MBS calculates the optimal price. &Stackelberg equilibrium&HN\tabularnewline \cline{2-9}

&\cite{you2017auction}&ACA&MNO&CPs&SBSs&The MNO adjusts the leasing price such that the total supply
quantity equals the demand quantity. &Walrasian
equilibrium&HN\tabularnewline \cline{2-9}

&\cite{xu2016self} &Double auction&MNO&CPs&Cache storages& Matching the VCPs' bids and the MNO's asks as well as payment determination for them are similar to those of the classical double auction.& Market
equilibrium &HN\tabularnewline \cline{2-9}

&\cite{mangili2016bandwidth} &Reverse auction&MNOs&CP&Cache storages&CP uses the greedy algorithm to select SBSs according to their cache hit rates. & Nash
equilibrium&HN\tabularnewline \cline{2-9}

&\cite{hu2015small} &Multi-item auction&MNOs&CPs&Cache storages&At each iteration, the CPs set prices for their storages. Given the prices, the MNOs bid the most preferred storages using the preferred-storage graph. The BFS algorithm is applied for the content-storage matching. &Nash equilibrium&HN\tabularnewline \cline{2-9}


&\cite{tadrous2016joint} &Profit maximization&CP&User&Data items&CP constructs demand profiles of the user. The CP runs an iterative algorithm to determine optimal data items to be served and their optimal prices. &Optimal
solution &HN\tabularnewline \cline{2-9}  

&\cite{hamidouche2016breaking} &Contract theory&MNO&CPs&Cache storages&MNO applies the matching theory and the swap-based deferred acceptance algorithm to assign storages to the CPs. The prices charged to the types of CPs are based on VCG auction payment. &Optimal
solution&HN\tabularnewline \cline{2-9}  

&\cite{liu2017resource} &Contract theory&MNO&CPs&Cache storages&MNO applies the standard Lagrangian method to determine the optimal prices and storages for types of CPs. &Optimal
solution&HN\tabularnewline \cline{2-9}  

\hline
\parbox[t]{2mm}{\multirow{9}{*}{\rotatebox[origin=c]{90}{ \hspace{-0.5cm}Energy efficiency optimization}}}
\parbox[t]{2mm}{\multirow{9}{*}{\rotatebox[origin=c]{90}{\hspace{-0.5cm} though offloading }}}

&\cite{bousia2016auction} &VCG auction&FH&MNOs&Offloading service&The ILP multi-objective optimization is used for selecting the winning MNOs. The winners are charged based on the VCG payment policy. &Bayesian Nash equilibrium&HN\tabularnewline \cline{2-9}

&\cite{xu2016reverse} &VCG auction&SBSs&MBS&RBs and power& The dynamic programming method with KKT conditions is used for selecting the winning SBSs. The winners are charged based on the VCG payment policy. &Bayesian Nash equilibrium&HN\tabularnewline \cline{2-9}


&\cite{sken2016energy} &ACA&SBSs&MBS&MUEs&MBS adjusts the offloading price such that the total supply quantity equals the demand quantity. &Walrasian equilibrium &HN\tabularnewline \cline{2-9}

&\cite{zhang2015cloud} &Stackelberg game&FH&MNO&Data traffic&FH determines the optimal offloading price, and then the MNO calculates the optimal amount of traffic. &Stackelberg equilibrium&HN\tabularnewline \cline{2-9}

&\cite{rahmati2015price} &Stackelberg game&FH&MUEs&Power&MUEs determine their optimal power demands. Then, the FH adjusts power prices such that the total power demand equals the maximum power of the FH. &Stackelberg equilibrium &HN\tabularnewline \cline{2-9}

\hline
\end{tabular}
\par\end{centering}
\end{table*}

\begin{table*}[h]
\caption{A summary of advantages and disadvantages of major approaches for wireless caching and mobile data offloading.}
\label{table_sum_advantage_wireless_caching}
\scriptsize
\begin{centering}
\begin{tabular}{|>{\centering\arraybackslash}m{2cm}|>{\centering\arraybackslash}m{7.8cm}|>{\centering\arraybackslash}m{6cm}|}
\hline
\cellcolor{myblue} &\cellcolor{myblue} &\cellcolor{myblue} \tabularnewline
\cellcolor{myblue} \multirow{-2}{*} {\textbf{Major approaches}} &\cellcolor{myblue} \multirow{-2}{*} {\textbf{Advantages}} &\cellcolor{myblue} \multirow{-2}{*}{\textbf{Disadvantages}} \tabularnewline
\hline
\hline
\cite{Zhang2016social} &\begin{itemize} \item Achieve win-win solution and consider network and congestion effects \end{itemize} & \begin{itemize}  \item Support only one caching user \end{itemize}\tabularnewline \cline{2-3}
\hline
\cite{hu2015small}&\begin{itemize} \item Support multiple CPs and multiple MNOs \end{itemize} & \begin{itemize}  \item Do not consider content replacement rates\end{itemize}\tabularnewline \cline{2-3}
\hline
\cite{tadrous2016joint} &\begin{itemize} \item Consider the proactive content provision\end{itemize} & \begin{itemize}  \item Support only one CP\end{itemize}\tabularnewline \cline{2-3}
\hline
\cite{bousia2016auction} &\begin{itemize} \item Achieve multiple objectives\end{itemize} & \begin{itemize}  \item Have high computational complexity\end{itemize}\tabularnewline \cline{2-3}
\hline
\end{tabular}
\par\end{centering}
\end{table*}

\textbf{Summary:} In this section, we have reviewed the applications of economic and pricing models for wireless caching and mobile data offloading in 5G HetNets. The related works are summarized in Table~\ref{table_caching_offloading}, and advantages and disadvantages of major approaches are given in Table~\ref{table_sum_advantage_wireless_caching}. As clearly shown in Table~\ref{table_caching_offloading}, economic and pricing approaches for the wireless caching receive more attentions. The reason may be that reducing network traffic and energy consumption over backhaul links in HetNets are crucial. However, some challenges for the wireless caching such as limited cache space of SBSs, high user mobility, and privacy concerns, need to be considered.

\section{Summary, challenges, and future research directions}
\label{sec:Open_issues}
\begin{figure}[t!]
    \begin{subfigure}[t]{0.5\linewidth}
        \centering
        \includegraphics[width=1.02\linewidth]{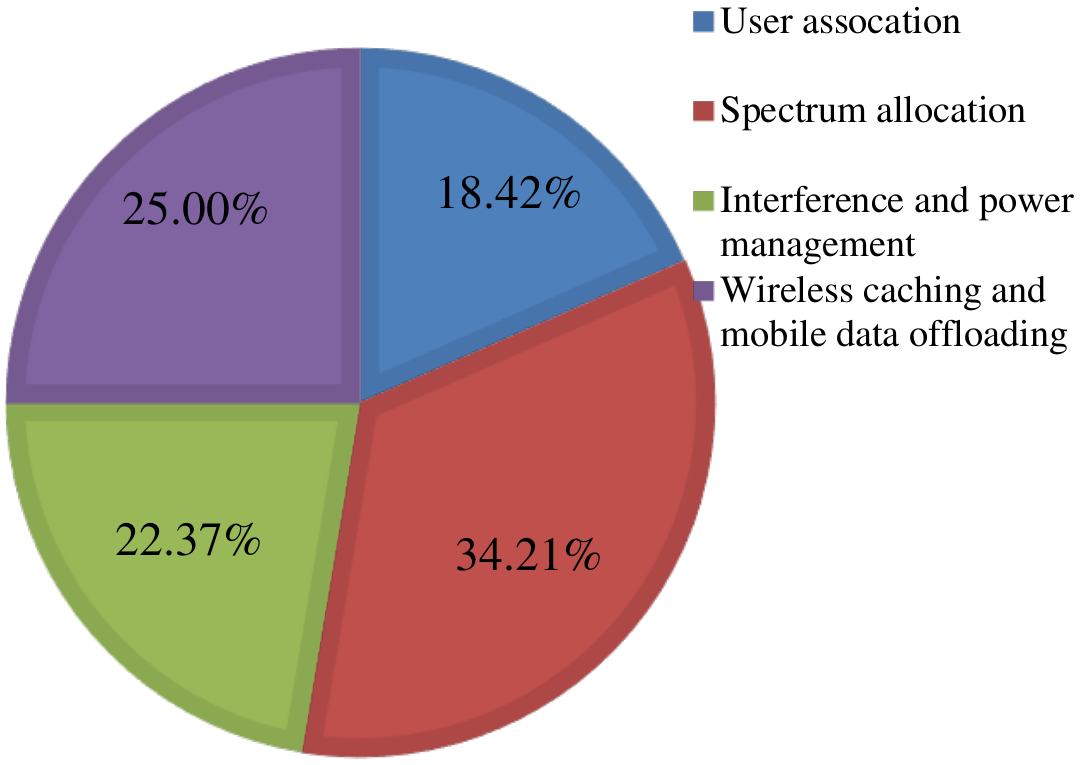}
        \caption{}
    \end{subfigure}%
 \hfill
    \begin{subfigure}[t]{0.5\linewidth}
        \centering
        \includegraphics[width=1.02\linewidth]{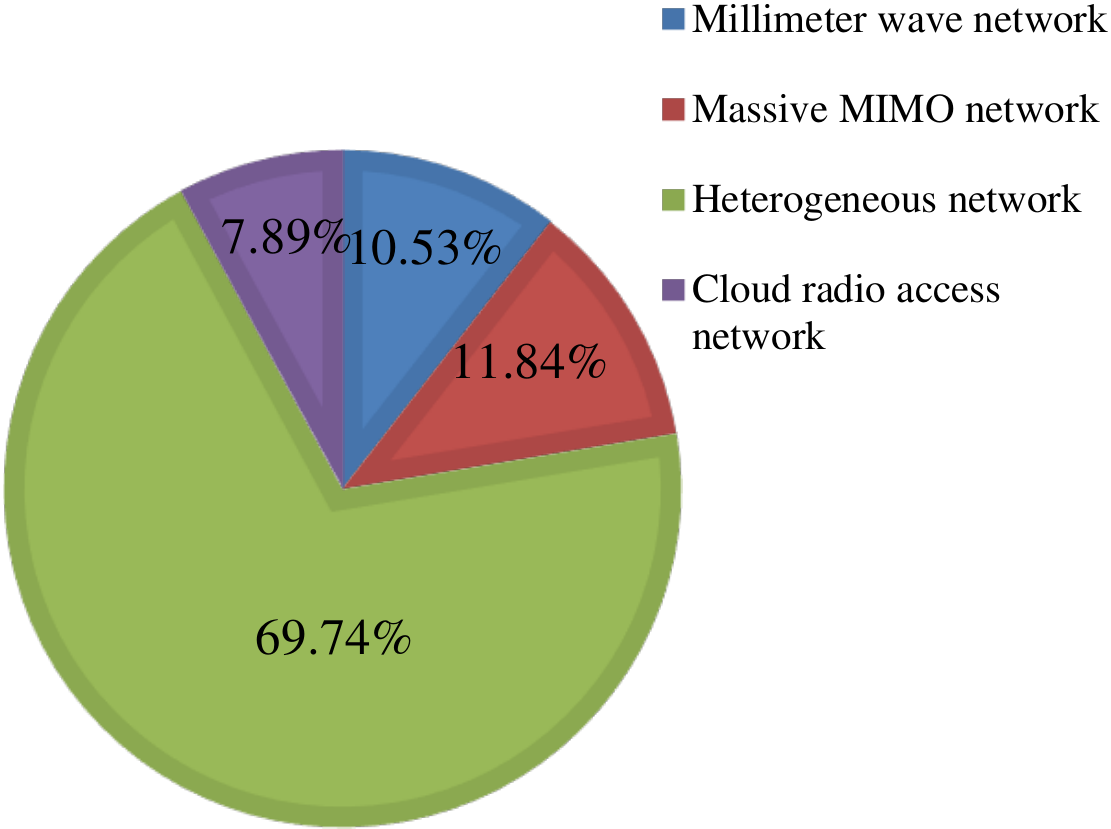}
        \caption{}
    \end{subfigure}
   \caption{\small Ratios of related work for (a) resource management issues and (b) 5G technologies.}
    \label{Summary_Figure}
\end{figure}

Different approaches reviewed in this survey evidently
show that economic and pricing models can effectively address resource management issues to meet the diverse requirements of the emerging 5G technologies. Fig.~\ref{Summary_Figure}(a) and Fig.~\ref{Summary_Figure}(b) show the ratios of economic and pricing approaches for different resource management issues and different 5G technologies respectively. From the figures, we observe that majority of the approaches are for HetNets while few approaches are for C-RANs. There are further challenges and new research directions as follows.

\subsection{Mobility of users}
\label{sec:Open_issues_Caching_relay}
In the spectrum allocation \cite{swainspectrum}, an offline auction is used in which mobile users are assumed to be stationary. They simultaneously submit required bandwidth and prices that they are willing to pay to the MNO. The MNO then determines the winner. In reality, the mobile users can move through multiple cells of different MNOs with high speed, e.g., a hundred km/h. In this case, the users may arrive in each cell one by one in a random order, and the offline auction may not be suitable. Alternatively, an online auction, e.g., as presented in \cite{ba2003building}, can be used which allows each MNO to decide whether the user wins or not immediately based on hitherto information available with the awareness of potential future bids.

\subsection{Security and privacy}
\label{sec:Open_issues_Security_Privacy}
When users move through multiple cells, the security and privacy issues arise due to the
possible involvement of untrusted or compromised network devices. Two common attacks that the users face are \textit{eavesdropping attack} and \textit{Distributed Denial-of-Service (DoS) attack}.   

In the eavesdropping attack, untrusted networks can eavesdrop the communication information between legitimate users and intended networks, e.g., small cell access points. To prevent the eavesdropping attack, cryptographic techniques are typically used. However, the techniques often require centralized authorities and additional secure channels for key exchanges. They thus have computation burden to both the legitimate users and the intended networks which are undesirable for 5G low-power network infrastructures. Alternatively, economic and pricing models can provide distributed solutions which maximize the secrecy
capacity for the legitimate communication links without requiring
additional secure channels for key exchanges \cite{luong2017applications}.

In the DDoS attack, a large number of compromised devices, also known as \textit{bots}, inside the network can be controlled by an external attacker to transmit radio jamming signals with high power to legitimate users and base stations. This attack makes the legitimate users and cells unavailable to respond to any service requests from the legitimate users. The DDoS attack is known as one of the most
severe attacks due to the use of thousands of bots distributed over the network. However, since the bots need to consume much network resources for their attack actions, they are subject to the resource prices. If the bots are rational, setting high resource prices discourages them to spend more power to perform the attack actions. Behavior-based pricing models such as Bayesian optimal pricing \cite{chorppathbayesian} can be adopted to easily achieve this goal. 

\subsection{Resource Management for Network Function Virtualization (NFV) and Software-Defined Networking (SDN)}
\label{sec:Open_issue_NFV_SDN}

NFV and SDN allow 5G to efficiently accommodate a wide range of services over a common network infrastructure. While the SDN decouples the network control and forwarding functions \cite{nguyen2017resource}, the NFV decomposes network functions from the physical network equipments. Such software-based solutions can provide the programmability, flexibility, and modularity that are required to create multiple logical/virtual networks. The virtual networks are namely \textit{network slices}, each of which serves a specific requirement of a user or a set of users. The NFV and SDN bring several benefits such as reducing the OPEX and CAPEX for the MNOs, facilitating the deployment of new services with increased agility and faster time-to-value, and achieving better system scalability according to user demands. However, implementing the NFV and SDN in 5G imposes several issues. 

For the NFV, the first issue is that resource sharing among multiple users can lead to congestion on the physical infrastructure as well as an unfair use of the resources. The second issue is how to allocate different physical resources, e.g., radio and storage capabilities, to the network slices so as to maximize the profit of MNOs while satisfying the requirements of network slices. To address the first issue, pricing models, e.g., smart data pricing, can be applied to offer incentives to users to use resources efficiently. For the second issue, the iterative auction as proposed in \cite{jiangnetwork} can be used to allocate different types of resources to the network slices. The profit maximization problem as discussed in \cite{tadrous2016joint} can be adopted to model and solve the second issue. Additionally, the combinatorial auction is a potential solution. However, these approaches are generally centralized mechanisms which have high computational complexity. 
\subsection{Incentive Mechanisms for Edge/Fog Computing}
\label{sec:Open_issue_Edge_Fog}
In the cloud computing, requests for cloud services such as computation and storage will go through a BS and core network to finally reach the cloud, i.e., remote data centers. However, given the massive dense users in 5G, such a paradigm faces issues such as long latency, high operational costs, and bandwidth bottlenecks at the BSs. To address the issues, the edge/fog computing paradigm is used. By using near-user edge devices such as fog servers, SBSs, e.g., picocells/femtocells, and user devices, the edge/fog computing pushes the frontier of cloud resources and services
away from the remote data centers to the periphery or edges of the network \cite{nguyen2017resource}. As a result, the edge/fog computing (i) significantly reduces the data traffic, cost, and latency, (ii) alleviates the major bottleneck and the risk of a potential point of failure, and (iii) provides high levels of scalability, reliability, and automation. However, to deploy the edge/fog computing infrastructure, the most important step is how to attract the owners of the edge devices to contribute their resources. Incentive mechanisms using pricing and payment strategies can be used to guarantee the stable scale of participants and QoS. 

\subsection{Power Allocation in Non-Orthogonal Multiple Access (NOMA) systems}
\label{sec:Open_issues_NOMA}
NOMA is expected to be a promising multiple access technique for 5G due to its high spectrum efficiency. Indeed, by exploiting the Successive Interference Cancellation (SIC) technique, multiple users can use the same frequency at the same time, but with different power levels. However, this raises the power allocation issue. For example, how a BS allocates power to the users to maximize the utilities of the BS and the users  subject to the transmit power constraint. To achieve the win-win solution, the pricing models, e.g., based on the Stackelberg game \cite{li2016price}, can be adopted. Note that the power allocated to one user affects not only the utility of that user but also the utilities of other users due to the interference and power constraint of the BS. Thus the interactions among the users can be modeled as the non-cooperative game. 
\subsection{User Assignment in Full-Duplex (FD) Cellular Networks}
\label{sec:Open_issues_Full_Duplex}
In-band FD is also one of key enabling technologies in 5G which can drastically increase the spectral efficiency. The FD allows a BS to transmit downlink traffic to a downlink user while simultaneously receiving uplink traffic from an uplink user using the same frequency. However, if the uplink user and the downlink user are not far enough from each other, the inter-user interference between them becomes too high as the same frequency is used. The inter-user interference reduces network throughput and spectrum efficiency. Therefore, the problem for the BS is to determine pairs of uplink users and downlink users and their power allocation to maximize the spectrum efficiency of all users in the network. Pricing models such as the forward auction \cite{basutli2017auction} and \cite{zhang2017auction} can be used. In this case, the uplink users, i.e., bidders, submit bids on the downlink users, i.e., commodities, to the BS, i.e., the auctioneer. The BS can match each uplink user with a downlink user such that the interference caused by the two users is minimized. 

In practice, in addition to bids, both uplink and downlink users are required to provide their location information to the BS for performing the user assignment. This naturally reveals the users' location information, and they have a high risk from physical attacks by adversaries. The combination of the forward auction and cryptographic algorithms can guarantee an efficient and privacy-preserving assignment.
\section{Conclusions}
\label{sec:Conclusion}
This paper has presented a comprehensive survey on the applications of economic and pricing theories to resource management in 5G wireless networks. Firstly, we have described key 5G technologies and resource management issues. Then, we have introduced and analyzed various economic and pricing models with the aim to understand the motivations of using these models in 5G wireless networks. Afterwards, we have provided detailed reviews, analyses, and comparisons of the economic and pricing approaches in solving a variety of resource management issues, i.e., user association, spectrum allocation, interference and power management, and wireless caching and mobile data offloading. Finally, we have outlined open issues as well as future research directions. 

\bibliographystyle{IEEEtran}
\bibliography{5GDatabase}{}


\end{document}